\documentclass[12pt]{article}
\usepackage{natbib}
\usepackage{authblk}

\usepackage{amsthm,amsmath,amsfonts,amssymb}

\RequirePackage[colorlinks,citecolor=blue,urlcolor=blue,backref=page,backref=page]{hyperref}
\RequirePackage{graphicx}
\usepackage{tabularx}
\usepackage{mathtools}
\mathtoolsset{showonlyrefs=true}
\usepackage{ulem}
\usepackage{nicefrac}
\usepackage{bbm,qtree,forest}
\usepackage{multicol}
\usepackage{multirow}
\usepackage{booktabs}
\usepackage{lscape}
\usepackage{subfig}
\usepackage{titlesec}

\makeatletter
\@addtoreset{section}{part}
\makeatother
\titleformat{\part}[display]
{\normalfont\LARGE\bfseries}{}{0pt}{}

\newcommand{\numberset}{\mathbb}
\newcommand{\N}{\numberset{N}}
\newcommand{\R}{\numberset{r}}

\newcommand{\E}[1]{\mathbb{E}\left[#1\right]}
\newcommand{\Var}[1]{\mathbb{V}ar\left(#1\right)}

\newcommand{\aoff}{\alpha^{\text{off}}}

\newcommand{\ton}{t_0^{\text{on}}}
\newcommand{\tong}{t_{0,g}^{\text{on}}}

\title{BayVel: A Bayesian Framework for  RNA Velocity Estimation in Single-Cell Transcriptomics}

\author[1]{Elena Sabbioni}
\author[2]{Enrico Bibbona}
\author[3]{Gianluca Mastrantonio}
\author[4]{Guido Sanguinetti}

\affil[1]{Politecnico di Torino, Dpt. of Mathematical Science, \texttt{elena.sabbioni@polito.it}}
\affil[2]{Politecnico di Torino, Dpt. of Mathematical Science, \texttt{enrico.bibbona@polito.it}}
\affil[3]{Politecnico di Torino, Dpt. of Mathematical Science, 
\texttt{Gianluca Mastrantonio}}
\affil[4]{International School of Advanced Studies (SISSA), \texttt{gsanguin@sissa.it}}

\makeatletter
\@addtoreset{section}{part}
\makeatother
 
\begin{document}

\maketitle
\begin{abstract}
RNA velocity is a model of gene expression dynamics designed to analyze single-cell RNA sequencing (scRNA-seq) data, and it has recently gained significant attention. However, despite its popularity, the model has raised several concerns, primarily related to three issues: its heavy dependence on data preprocessing, the need for post-processing of the results, and the limitations of the underlying statistical methodology.
Current approaches, such as \textit{scVelo}, suffer from notable statistical shortcomings. These include identifiability problems, reliance on heuristic preprocessing steps, and the absence of uncertainty quantification. To address these limitations, we propose \textit{BayVel}, a Bayesian hierarchical model that directly models raw count data. \textit{BayVel} resolves identifiability issues and provides posterior distributions for all parameters, including the RNA velocities themselves, without the need for any post-processing.
We evaluate \textit{BayVel}’s performance using simulated datasets. While \textit{scVelo} fails to accurately reconstruct parameters, even when data are simulated directly from the model assumptions, \textit{BayVel} demonstrates strong accuracy and robustness. This highlights \textit{BayVel} as a statistically rigorous and reliable framework for studying transcriptional dynamics in the context of RNA velocity modeling.
When applied to a real dataset of pancreatic epithelial cells previously analyzed with \textit{scVelo}, \textit{BayVel} does not replicate their findings, which appears to be strongly influenced by the postprocessing, supporting concerns raised in other studies about the reliability of \textit{scVelo}.
\end{abstract}
\part*{}
\section{Introduction}
\label{sec:introduction}
\indent Starting from a work by \cite{zeisel2011coupled}, RNA velocity was introduced by \cite{laManno2018rna} as a system of Ordinary Differential Equations (ODEs) that captures the processes of RNA transcription, splicing, and degradation. In this framework, unspliced messenger RNA (mRNA), which contains both exons (coding-part of mRNA) and introns (non-coding part), is processed into mature spliced mRNA, formed by exons, and the transcription process is either turned ON or OFF depending on the current level of development of the specific cell. This formulation allowed the problem to be reinterpreted as a linear regression, leading to the development of \textit{velocyto}, where RNA velocity is estimated as the ratio of unspliced to spliced mRNA. Building upon La Manno's model, \cite{bergen2020generalizing} proposes a more general statistical method that is able to handle experimental observations that correspond to the transient phase. An associated Python tool named \textit{scVelo} was released. This approach relies on an extensive pre-processing pipeline that converts discrete counts into continuous data and uses an expectation-maximization (EM) algorithm.\\
\indent RNA velocity has gained popularity because it adds a temporal interpretation to static single-cell RNA sequencing (scRNA-seq) snapshots, and many studies extending the \textit{scVelo} model appeared in the last years. Under a machine-learning framework, researches developed \textit{Velo-Predictor} \citep{VeloPredictor}, \textit{VeloAE} \citep{qiao2021veloAE}, \textit{CellDancer} \citep{li2024cellDancer}, \textit{DeepVelo} \citep{cui2024deepvelo}, \textit{SymVelo} \citep{xie2024symVelo}, \textit{NeuroVelo} \citep{kouadri2023neurovelo} and \textit{scTour} \citep{li2023sctour}. Many approaches, based on different mathematical methodologies, have been proposed, such as \textit{UniTVelo} \citep{gao2022unitvelo}, \textit{dynamo} \citep{qiu2022dynamo}, \textit{$\kappa$-velo} \citep{marot2022kvelo}, \textit{eco-velo} \citep{marot2022kvelo}, \textit{VeloCycle} \citep{lederer2024veloCycle}, \textit{Velde} \citep{jia2023velde} and \cite{su2024hodge}. A lot of works rely on Bayesian statistic, though many still in pre-print stages. Among them we find \textit{Pyro-Velocity} \citep{qin2022pyro}, \textit{cell2fate} \citep{aivazidis2023cell2fate}, \textit{VeloVAE} \citep{gu2022veloVAE}, \textit{VeloVI} \citep{gayoso2024veloVI} and \textit{ConsensusVelo} \citep{zhang2024consensusVelo}. The original RNA velocity model has also been extended to integrate various omics layers or gene regulatory patterns, such as in \textit{protaccel} \citep{gorin2020protaccel}, \textit{Multivelo} \citep{li2023multiVelo}, \textit{Chromatin Velocity} \citep{tedesco2022chromatinVelocity}, \textit{SIRV} \citep{abdelaal2024sirv}, \textit{TFvelo} \citep{li2024tfvelo}, \textit{scKINETICS} \citep{burdziak2023sckinetics}, \cite{qiu2020massively} and \cite{peng2023storm}, and to incorporate RNA velocity directly into cell trajectory inference, as in \textit{CellPath} \citep{CellPath}, \textit{VeTra} \citep{weng2021vetra}, \textit{DeepCycle} \citep{riba2022deepCycle}, \textit{CytoPath} \citep{gupta2022cytoPath}, \textit{RNA-ODE} \citep{liu2022rnaOde}, \textit{CellRank} \citep{lange2022cellrank}, and \textit{Velorama} \citep{singh2024velorama}. Alternative visualization tools and simulation frameworks (e.g., \textit{VeloViz} \citep{atta2022veloviz}, \textit{VeloSim} \citep{zhang2021velosim}) have also been proposed.\\
\indent Despite the significant success of \textit{velocyto} and \textit{scVelo}, several studies have raised concerns about RNA velocity’s reliability. For example, variability in measuring unspliced and spliced mRNA, due to differences in computational tools, can lead to biased estimates \citep{soneson2021preprocessing}. Biological systems may have sub-populations following different kinetics and only a subset of genes may conform to the kinetic models assumed by RNA velocity \citep{bergen2021rna}. Moreover the data may only capture a portion of the dynamic process, leading to potential erroneous predictions of developmental trajectories \citep{qin2022pyro}. Several studies \citep{bergen2021rna, gorin2022rna} have also presented datasets where RNA velocity estimates are highly distorted, with phase portraits that fail to reflect meaningful transitions and that contradict biologically expected outcomes. Many of the assumptions underlying \textit{scVelo}, such as data normality, are unrealistic for scRNA-seq data, which is characterized by low-copy numbers. Moreover, \textit{scVelo} relies heavily on projection techniques that can introduce artificial distortions \citep{zheng2023pumping}. For example, RNA velocity is typically visualized by representing the high-dimensional data into a UMAP embedding \citep{mcinnes2018umap} using the \textit{transition probability} method. It remains unclear under which conditions UMAP faithfully represents biological reality and \cite{zheng2023pumping} show that \textit{transition probability} method introduces a post-processing of velocity that tends to interpolate the embedding space, making a cell's velocity largely dependent on the gene expression of its nearest neighbors rather than uncovering novel developmental trajectories. Furthermore, the low-dimensional projection is highly sensitive to the k-nearest neighbors (k-NN) graph used during pre-processing. As noise levels increase, the k-NN graph can become distorted, leading to inaccurate RNA velocity estimates and a poor correspondence between the high-dimensional velocities and their low-dimensional representations. From a statistical perspective, \textit{scVelo} is also limited by its reliance solely on point estimates, without any measures of uncertainty, which restricts the confidence in its conclusions.\\
\indent In response to these criticisms, we propose a new estimation framework, denoted as \textit{BayVel}, designed to address all the statistical limitations of \textit{scVelo} by providing mathematically and biologically justified solutions.
Instead of relying on preprocessed data, we work directly with the original count data, removing the distortion introduced by \textit{scVelo} pre-processing steps. To derive the data distribution, we model the process as a stochastic representation of cellular dynamics based on a chemical reaction network (CRN), using a product of Poisson distributions for the initial state. This ensures that the theoretical distribution at any given time remains a product of Poisson distributions \citep{Jahnke2007}.
By introducing a cell-specific capture efficiency parameter and a gene-specific overdispersion term, we account for variability in transcript detection across cells, as well as additional sources of measurement noise. These deviations from the pure Poisson assumption naturally lead to a Negative Binomial distribution, which forms the basis of our model.
We also relax the unrealistic assumption that all cells share the same dwell time in the induction phase by allowing the parameters to vary across subpopulations of cells. In parallel, we allow groups of cells to share the same elapsed time since activation, which facilitates the estimation of kinetic parameters.
To visualize the results, we avoid using UMAP embedding in favor of a more robust principal component analysis (PCA) projection. Thanks to its linear nature, PCA does not distort the data structure, allowing for direct interpretation of the estimated velocities in the high-dimensional space.
Moreover, our approach is grounded in Bayesian statistics, and parameter estimation is carried out via Markov Chain Monte Carlo (MCMC), providing a natural way to quantify uncertainty. We validate \textit{BayVel}’s performance through an extensive simulation study and application to real data, comparing its results with those obtained using \textit{scVelo}.\\
\indent The remaining parts of this paper are organized as follows. In Section \ref{subsec:modeling}, we present the mathematical model underlying RNA velocity, while Section \ref{subsec:scvelo} describes the nature of the data and reviews the \textit{scVelo} algorithm and its limitations. In Section \ref{sec:bayVel}, we introduce our new framework, \textit{BayVel}, detailing its model innovations, data distribution assumptions, and prior specifications. Section \ref{sec:simRNAvel} presents our simulation studies, comparing \textit{scVelo} and \textit{BayVel}, followed by the application to real data in Section \ref{sec:real}. Finally, Section \ref{sec:conclusionRNA} concludes the paper and outlines future research directions. The Supplementary material \citep{supplementary}, available on the journal website, contains additional details on the mathematical model in Section 1, Section 2 and 3 provides further information about the statistical approach. Section 4 presents additional details on the simulation strategy. Section 5 and 6 contains respectively additional results on the simulated and real data. 

\section{Background material}\label{sec:background}
\subsection{Modeling scRNA-seq data and RNA velocity}\label{subsec:modeling}

In this section, we present the two mathematical models describing the dynamics of the expression levels of a gene. For notational convenience, and to facilitate the description of the models, we will show them for a single gene. Additionally, we introduce the related concept of RNA velocity.

\begin{figure}[t]
	\centering
	\subfloat[\label{fig:structureAlphaUS}]{\includegraphics[scale = 0.075]{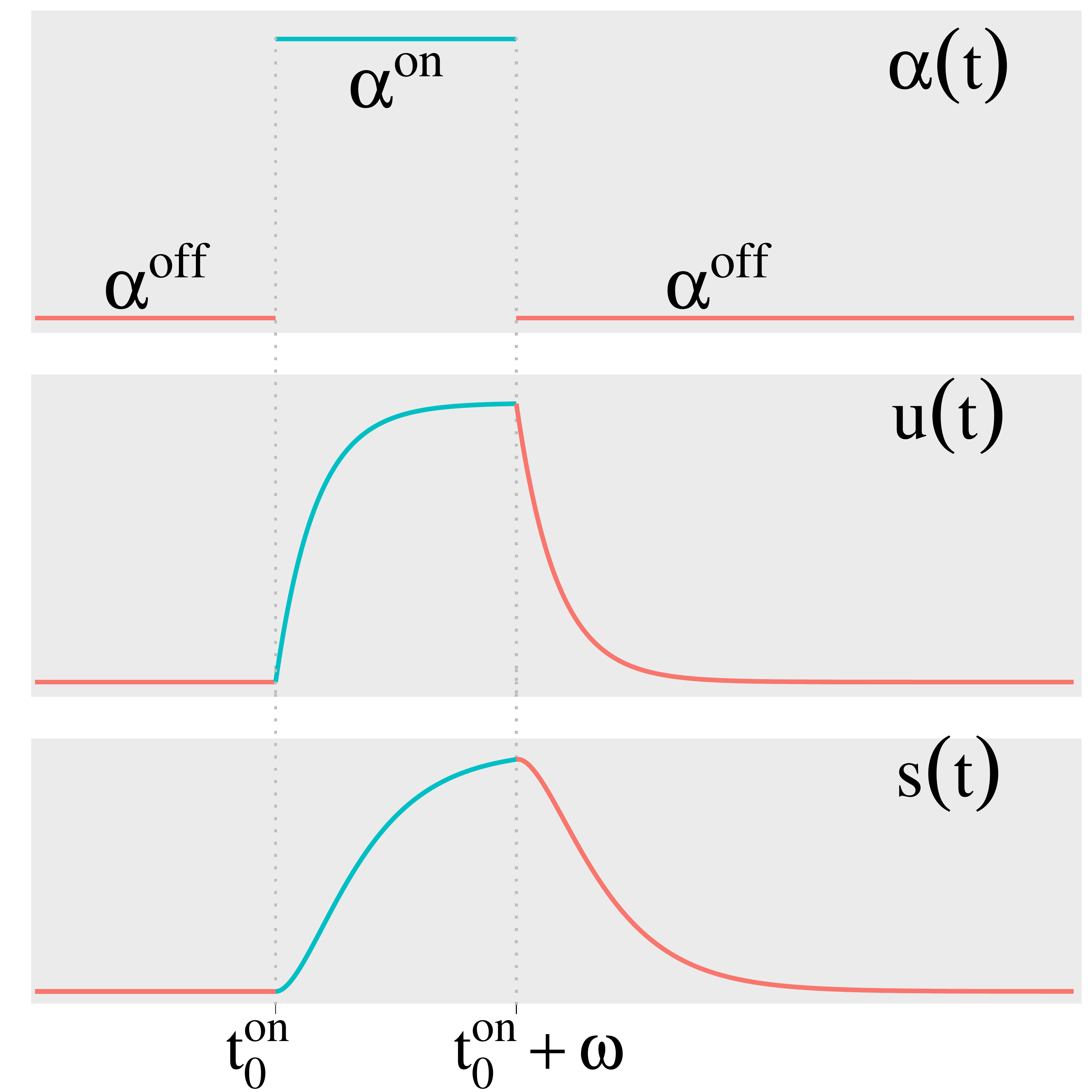}}
	\subfloat[\label{fig:geneDyn}]{\includegraphics[scale = 0.075]{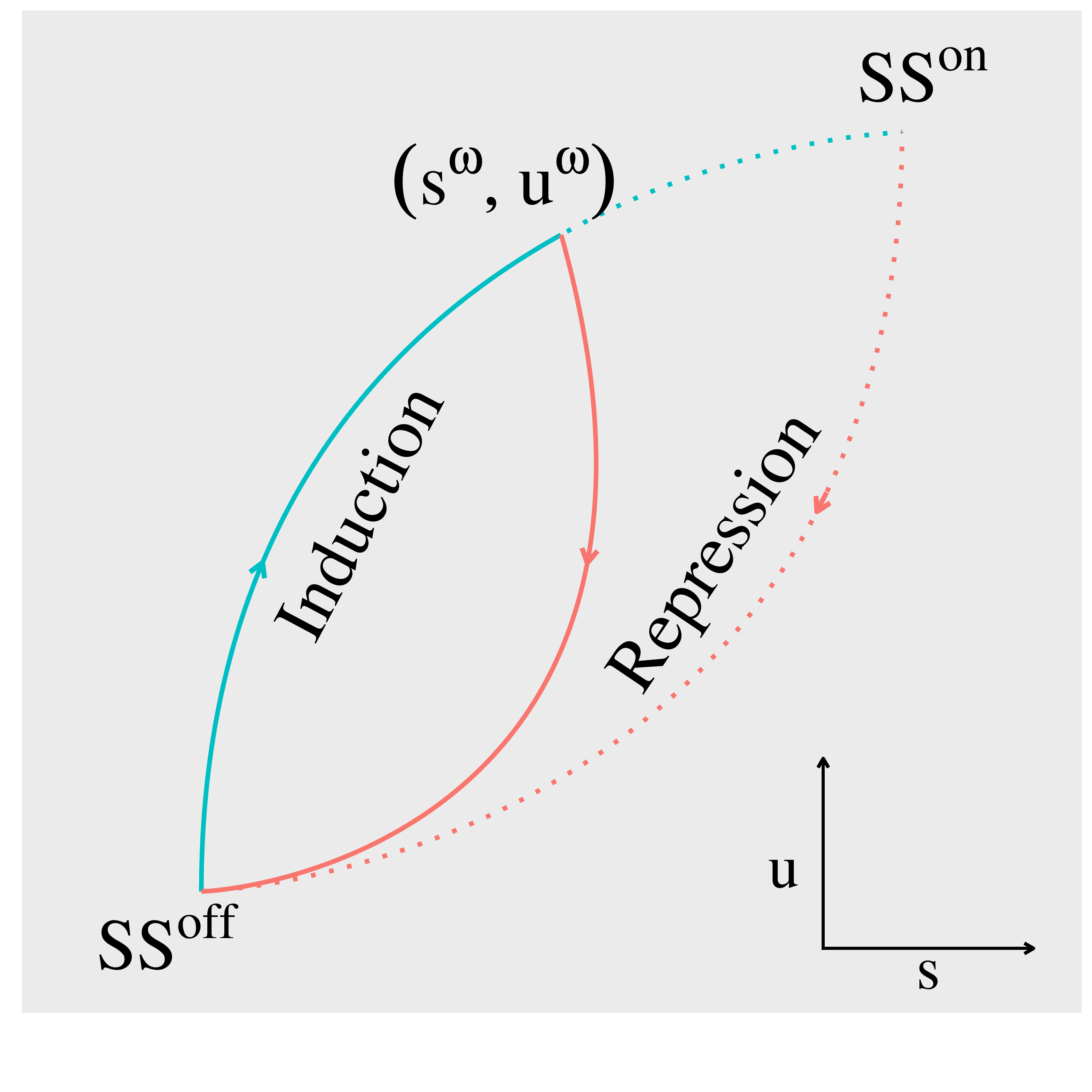}}
	\subfloat[\label{fig:dynamicDiffSwitch}]{\includegraphics[scale = 0.075]{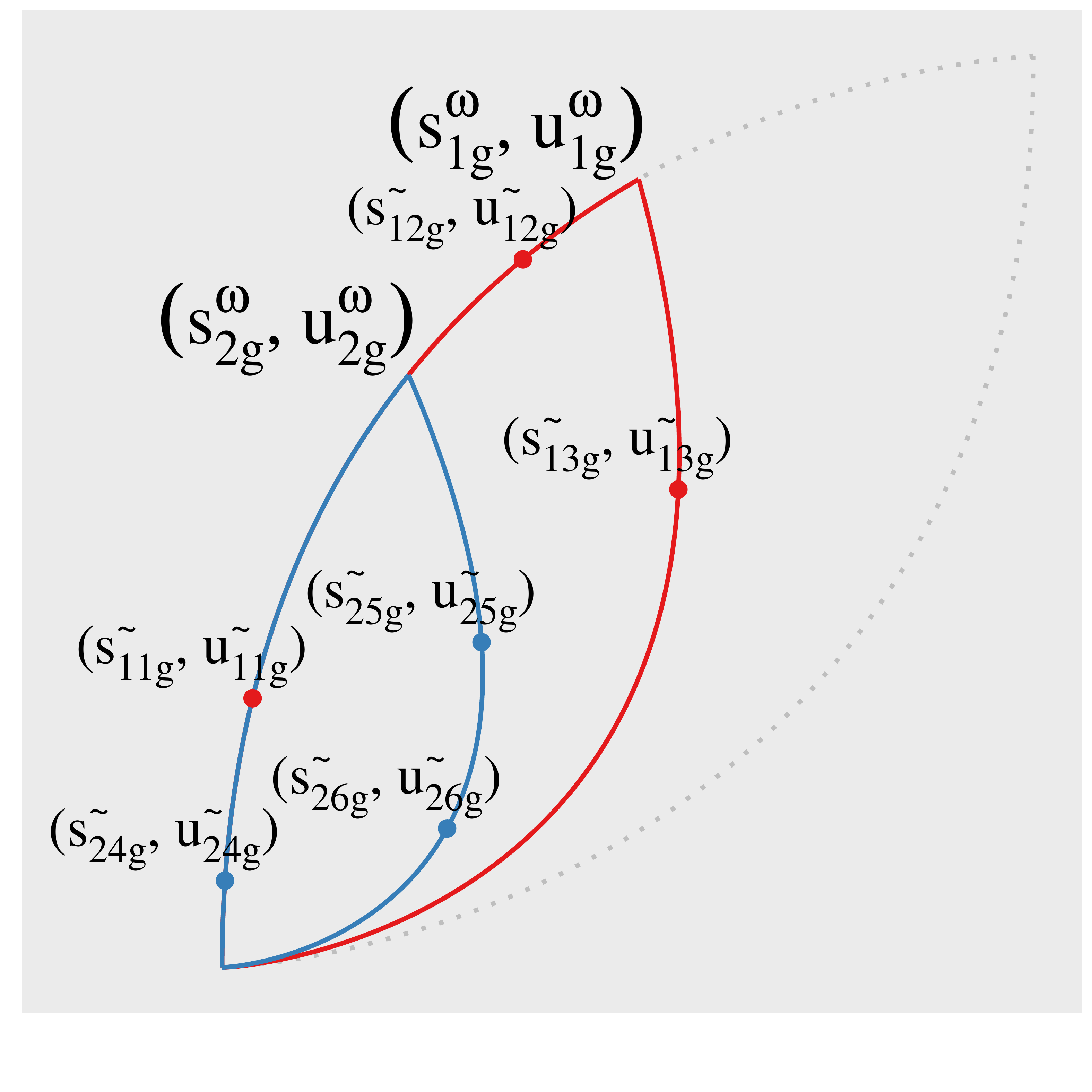}}
	\caption{Left: Time evolution of the transcription rate $\alpha(\cdot)$ (first row), of the associated unspliced $u(\cdot)$ (second row) and of spliced $s(\cdot)$ (third row) dynamic. Center: Evolution of the solution of the ODE system (\ref{eq:systemODE}) in the $(s,u)$ plane, with dotted line representing the potential behavior when $\omega\rightarrow + \infty$ and $\text{\textbf{SS}}^{\text{on}}$ is reached, while solid line depicts the dynamic when $\omega < + \infty$ and a switch occurs at $\left(s^\omega, u^\omega \right)$. Right: Representation of the group-subgroup structure modeled by \textit{BayVel}. For a fixed gene $g$, we illustrate the \emph{almond} structure with two distinct group-specific switching points, $(s_{1g}^\omega, u_{1g}^\omega)$ and $(s_{2g}^\omega, u_{2g}^\omega)$. The dotted line represents the unique potential dynamic governed by the rate parameters $\boldsymbol{\theta}_{g}^{\text{B2}}$. Subgroups $1, 2, 3$ (red dots), which belong to group $1$, are characterized by the positions $(s_{1rg}^\sim, u_{1rg}^\sim)$ for $r = 1, 2, 3$. Similarly, subgroups $4, 5, 6$ (blue dots), associated with group $2$, are described by the coordinates $(s_{2rg}^\sim, u_{2rg}^\sim)$ for $r = 4, 5, 6$. }
	\label{fig:geneDynamicTimePlane}
\end{figure}

\paragraph{Deterministic model} Following \cite{bergen2020generalizing}, we assume that the biological mechanism of gene expression can be described by the following CRN.
\begin{equation} \label{eq:CRN}
	\emptyset\stackrel{\alpha(t)}{\longrightarrow}\text{u-RNA}  \stackrel{\beta}{\longrightarrow} \text{s-RNA} \stackrel{\gamma}{\longrightarrow}  \emptyset.
\end{equation} 
The network models three processes: transcription of DNA into unspliced mRNA (u-RNA) at rate $\alpha(t)$, processing of unspliced mRNA into its spliced counterpart (s-RNA) at rate $\beta$, and degradation of spliced mRNA at rate $\gamma$. In the CRN \eqref{eq:CRN}, DNA is represented as $\emptyset$, as it is not consumed in the transcription mechanism and u-RNA is similarly created out of nothing. The rate of transcription $\alpha(t)$ is modeled as a piecewise constant function (cf. Fig \ref{fig:structureAlphaUS})
\begin{equation} \label{eq:alpha}
	\alpha(t) :=
	\begin{cases}
		\alpha^{\text{off}} & \text{ if }  0 \leq  t < t_{0}^{\text{on}},\\
		\alpha^{\text{on}} & \text{ if }   t_{0}^{\text{on}} \leq  t < t_{0}^{\text{on}} + \omega, \\
		\alpha^{\text{off}} & \text{ if }   t_{0}^{\text{on}} + \omega \leq t,
	\end{cases}
\end{equation} 
characterized by an OFF phase (repression), during which only a basal level of transcription is present ($\alpha^{\text{off}}$), followed by an ON phase (induction) with a higher transcription rate ($\alpha^{\text{on}} > \alpha^{\text{off}}$), which eventually transitions back to the OFF state. 
The parameters $\beta$ and $\gamma$, both belonging to $\mathbb{R}^+$, represent the splicing and degradation rate, respectively.
The initial ON phase begins at time $\ton$, with its duration denoted by $\omega$. The CRN \eqref{eq:CRN} is governed by mass-action kinetics, and it can be equally described by using the following system of ODEs, where $s(t, \dots)$ and $u(t, \dots)$ denote the amounts of spliced and unspliced mRNA at time $t$
\begin{equation}\label{eq:systemODE} 
	\begin{cases}
		\frac{d u(t,\ton,\omega, \boldsymbol{\theta})}{d t} = \alpha(t)  - \beta u(t,\ton,\omega, \boldsymbol{\theta}), \vspace{0.1cm}\\ 
		\frac{ds(t,\ton,\omega, \boldsymbol{\theta})}{dt} = \beta u(t,\ton,\omega, \boldsymbol{\theta}) - \gamma s(t,\ton,\omega, \boldsymbol{\theta}).
	\end{cases}
\end{equation}
with $\boldsymbol{\theta}:= \left(\alpha^{\text{off}}, \alpha^{\text{on}}, \beta, \gamma \right)$.
Assuming the initial conditions $s(0,\ton,\omega, \boldsymbol{\theta}) = {\alpha(0)}/{\gamma} = {\alpha^{\text{off}}}/{\gamma}$ and $u(0,\ton,\omega, \boldsymbol{\theta})= {\alpha(0)}/{\beta} = {\alpha^{\text{off}}}/{\beta}$, we can solve \eqref{eq:systemODE} in closed form.
If $\gamma \not = \beta$, the solution of \eqref{eq:systemODE} can be written as 
\small
\begin{equation}
	\label{eq:ussol}
		\left(s(t,\ton,\omega, \boldsymbol{\theta}), u(t,\ton,\omega, \boldsymbol{\theta}) \right) =
		\begin{cases}
			\left(\frac{\aoff}{\gamma}, 	\frac{\aoff}{\beta} \right) & \text{ if }  0 \leq  t < t_{0}^{\text{on}},\\
		\left(	s^\text{on}(t,\ton,\omega, \boldsymbol{\theta}), 	u^{\text{on}}(t,\ton,\omega, \boldsymbol{\theta})\right) & \text{ if }   t_{0}^{\text{on}} \leq  t \leq t_{0}^{\text{on}} + \omega, \\
			\left(s^\text{off}(t,\ton,\omega, \boldsymbol{\theta}) ,	u^\text{off}(t,\ton,\omega, \boldsymbol{\theta}) \right) &  \text{ if }   t_{0}^{\text{on}} + \omega < t,
		\end{cases} 
\end{equation}
\normalsize
where 
\begin{equation}
	\label{eq:solutionON} 	
	\begin{cases}
		s^\text{on}(t,\ton,\omega, \boldsymbol{\theta}):= \frac{\aoff}{\gamma} e^{-\gamma \tilde{t}(t)} \; + \frac{\alpha^{\text{on}}}{\gamma}\left[1 - \-e^{-\gamma \tilde{t}(t)} \, \right]+ \frac{\alpha^{\text{on}} - \aoff}{\gamma - \beta}\left[e^{-\gamma \tilde{t}(t)} - e^{-\beta\tilde{t}(t)}\right],\\
		u^\text{on}(t,\ton,\omega, \boldsymbol{\theta}) := \frac{\aoff}{\beta} e^{-\beta \tilde{t}(t)} + \frac{\alpha^{\text{on}}}{\beta}\left[1- e^{-\beta \tilde{t}(t)}\right], \\
		s^\text{off}(t,\ton,\omega, \boldsymbol{\theta})   := s^\text{on}\,(t_{0}^{\text{on}} + \omega,\ton,\omega, \boldsymbol{\theta}) e^{-\gamma \left[\tilde{t}(t) - \omega\right]}\; + \frac{\alpha^{\text{off}}}{\gamma}\left[1-e^{-\gamma \left[\tilde{t}(t) - \omega\right]} \right] + \\
		\qquad \qquad \qquad \qquad \qquad \qquad \qquad \;  + \frac{\alpha^{\text{off}} - \beta  u^\text{on}(t_{0}^{\text{on}} + \omega,\ton,\omega, \boldsymbol{\theta}) }{\gamma - \beta}\left[e^{-\gamma\left[\tilde{t}(t) - \omega\right]} - e^{-\beta\left[\tilde{t}(t) - \omega\right]}\,\right],\\
		u^\text{off}(t,\ton,\omega, \boldsymbol{\theta})  := u^\text{on}(t_{0}^{\text{on}} + \omega,\ton,\omega, \boldsymbol{\theta}) e^{-\beta\left[\tilde{t}(t) - \omega\right]} + \frac{\alpha^{\text{off}}}{\beta}\left[1- e^{-\beta\left[\tilde{t}(t) - \omega\right]}\,\right].\\
	\end{cases}
\end{equation}
and $\tilde{t}(t) := t - t_0^{\text{on}}$.
The solution for $\gamma = \beta$ can be derived as the limit of \eqref{eq:ussol} for $\gamma$ approaching $\beta$ and can be found in Section 1 of the Supplementary Material \citep{supplementary}. 
As $t \rightarrow \infty$ the dynamics converges to the following steady state
\begin{equation}
	\label{eq:SSoff}
		\text{\textbf{SS}}^\text{{off}} := \left(\sigma^{\text{off}}, \upsilon^{\text{off}}\right):= \lim_{t \rightarrow + \infty } \left( s(t,\ton,\omega, \boldsymbol{\theta}), u(t,\ton,\omega, \boldsymbol{\theta}) \right) = \left(\frac{\alpha^{\text{off}}}{\gamma}, \frac{\alpha^{\text{off}}}{\beta}\right),
\end{equation}
which corresponds to the initial conditions of the system.
If the ON phase lasts forever ($\omega = \infty$), the system \eqref{eq:systemODE} tends to a different steady state
\begin{equation}
		\label{eq:SSon}
		\text{\textbf{SS}}^\text{{on}} :=  \left(\sigma^{\text{on}}, \upsilon^{\text{on}}\right):= \lim_{t \rightarrow + \infty}  \left( s^{\text{on}}(t,\ton,\omega, \boldsymbol{\theta}), u^{\text{on}}(t,\ton,\omega, \boldsymbol{\theta}) \right)  = \left(\frac{\alpha^{\text{on}}}{\gamma}, \frac{\alpha^{\text{on}}}{\beta}\right).
\end{equation}
A full graphical description of Equations \eqref{eq:alpha} and \eqref{eq:ussol} is given in Figures \ref{fig:geneDynamicTimePlane}. 
Figure \ref{fig:structureAlphaUS} illustrates the time evolution of $\alpha(\cdot)$, $s(\cdot)$, and $u(\cdot)$, while Figure \ref{fig:geneDyn} depicts the trajectory of the solution \eqref{eq:ussol} in the $(s,u)$ plane. The shape that goes from $\text{\textbf{SS}}^\text{{off}}$ to $	\text{\textbf{SS}}^\text{{on}}$, and the back to $\text{\textbf{SS}}^\text{{off}}$, representing a dotted almond shape in Figure \ref{fig:geneDyn}, frames the dynamic and it will be from here on, referred to as the \emph{almond}. In general the system begins at the lower-left corner of the $(s, u)$ plane at $\text{\textbf{SS}}^\text{{off}}$ and remains there until time $\ton$. Subsequently, it moves along the
ON branch of the \emph{almond} towards $\text{\textbf{SS}}^\text{{on}}$, although it does not reach this point, as the system switches to the OFF branch at time $\ton + \omega$, attracted to $\text{\textbf{SS}}^\text{{off}}$ again. The coordinate of the switching point are $s^\omega := s(\ton + \omega,\ton,\omega, \boldsymbol{\theta})$ and $u^\omega := u(\ton + \omega,\ton,\omega, \boldsymbol{\theta})$. 
From Equations \eqref{eq:systemODE}, we can compute the time derivative of the spliced mRNA, $$v(t) :=
\frac{d s(t)}{dt} = \beta u(t,\ton,\omega, \boldsymbol{\theta})- \gamma s(t,\ton,\omega, \boldsymbol{\theta}).$$ This quantity, in \cite{laManno2018rna}, is called RNA velocity and it is the quantity of interest in the \textit{scVelo} procedures. It is important to note that the functions $s(t,\ton,\omega, \boldsymbol{\theta}) $ and $u(t,\ton,\omega, \boldsymbol{\theta})$ do not depend on both $t$ and $\ton $ individually, but only on their difference. Specifically, for each $a_1 \in \R$, we have the following invariance property
\begin{equation}
	s(t+a_1,\ton+a_1,\omega, \boldsymbol{\theta}) = s(t,\ton,\omega, \boldsymbol{\theta}),	\quad\quad u(t+a_1,\ton+a_1,\omega, \boldsymbol{\theta}) = u(t,\ton,\omega, \boldsymbol{\theta})\label{eq:invariance}
\end{equation}
 As observed in \cite{marot2022kvelo,zhang2024consensusVelo}, $s(t,\ton,\omega, \boldsymbol{\theta}) $ and $u(t,\ton,\omega, \boldsymbol{\theta})$ are also invariant under global rescaling of the rates, combined with inverse scaling of the time parameters. In particular, for any $a_2>0$,
\begin{equation}\label{eq:inv2}
	s\left(\frac{t}{a_2},\frac{\ton}{a_2},\frac{\omega}{a_2}, a_2\boldsymbol{\theta}\right) = s(t,\ton,\omega, \boldsymbol{\theta})\qquad u\left(\frac{t}{a_2},\frac{\ton}{a_2},\frac{\omega}{a_2}, a_2\boldsymbol{\theta}\right) = u(t,\ton,\omega, \boldsymbol{\theta}).
\end{equation}

\paragraph{Stochastic model}
The CRN \eqref{eq:CRN} can alternatively be modeled as a continuous-time Markov Chain. Therefore we introduce $\mathbf{Y}(t)= \left(S(t), U(t)\right) \in \N^2$, as the counts of spliced and unspliced mRNA at time $t$, respectively. Following stochastic mass action kinetics (cf. \cite{anderson2015stochastic}), the transition rates at time $t$ are defined as 
\begin{equation}
	\begin{split}
		& q_t\left(\left(s, u\right), \left(s, u + 1\right)\right) = \alpha(t),\; \quad q_t\left(\left(s, u\right), \left(s + 1, u - 1\right)\right) = \beta u,\\
		& q_t\left(\left(s, u\right), \left(s - 1, u \right)\right) = \gamma s, \qquad q_t\left(\left(s, u\right), \left(s', u' \right)\right) = 0 \qquad \text{ otherwise, }\\
	\end{split}
	\label{eq:stochProc}
\end{equation}
where $\alpha(t)$ is again given by equation \eqref{eq:alpha} and all other parameters retain the same interpretation as in the deterministic model.
By analogy with the initial conditions of the deterministic model, we assume that the initial distribution is a product of Poisson distributions, centered at $\text{\textbf{SS}}^{\text{off}} = (s(0,\ton,\omega, \boldsymbol{\theta}), u(0,\ton,\omega, \boldsymbol{\theta}))$, i.e. $\mathbf{Y}(0) \sim \text{Pois}(s(0,\ton,\omega, \boldsymbol{\theta}))\cdot\text{Pois}(u(0,\ton,\omega, \boldsymbol{\theta}))$. Proposition 2 of \cite{Jahnke2007} ensures that the transient distributions of our Markov Chain is again a product of Poisson distributions, centered at the solution of the ODE system \eqref{eq:systemODE}
\begin{equation}
	\label{eq:CME}
	\mathbf{Y}(t) \sim \text{Pois}(s(t,\ton,\omega, \boldsymbol{\theta}))\cdot\text{Pois}(u(t,\ton,\omega, \boldsymbol{\theta})).
\end{equation}
In this case the \emph{almond} in Figure \ref{fig:geneDyn} remains an appropriate illustration of the model since it plots the dynamics of the mean of the stochastic model. By stochastic CRN theory \cite{ACK2010,daniele_inv} we have that the initial distribution is stationary.

\subsection{The \textit{scVelo} tool and its methodological limitations }\label{subsec:scvelo}
RNA velocity modeling uses single-cell RNA sequencing data, which consists of gene expression levels for each cell at the time of sequencing.This technique provides read counts that are mapped to either intronic (unspliced) or exonic (spliced) regions, from which spliced and unspliced transcript counts are inferred using statistical methods (e.g., \citep{he2022alevin}, \citep{melsted2021modular}, \citep{putri2022analysing}). We assume this inference step has already been completed here. Thus, the input data consist of spliced and unspliced mRNA counts for each gene in each cell, denoted as $(y_{s, cg}, y_{u, cg}) \in \mathbb{N}_{\geq 0}^2$, where $g=1,\dots, G$ denotes the gene and $c=1, \dots, C$ the cell. In some cases, particularly when the data pertain to cell differentiation, each cell may be linked to a cell type annotation. A key limitation of such datasets is that the sequencing process destroys the sample, providing only a single static snapshot of all expression levels, without offering any information about temporal evolution. The tool \textit{scVelo} provides functions
which perform extensive preprocessing of the raw data, returning an approximately continuous version, $(m_{s, cg}, m_{u, cg})$, that are then used as input of \textit{scVelo}'s analysis. The preprocessing involves five main steps. Initially, raw counts are filtered to exclude genes with low expression levels. The data is then appropriately normalized and subsequently filtered again to
retain only the most highly dispersed genes. A logarithmic transformation follows, after which the values are smoothed and combined using PCA for dimensionality reduction. Based on this projection, \textit{scVelo} determines each cell’s $k$-nearest neighbors and replaces each cell’s data with the average values of its neighbors.\\
\indent The pre-processed data $\{m_{s, cg}, m_{u, cg}\}_{c,g}$ represent a snapshot of gene expression levels across numerous cells at the precise moment of sequencing. For each gene, these measurements are interpreted by \textit{scVelo} as noisy observations of the deterministic system \eqref{eq:systemODE}, characterized by gene-specific rate constants $(\alpha_g^{\text{off}}, \alpha_g^{\text{on}}, \beta_g, \gamma_g)$ and gene-specific duration $\omega_g$ of the ON phase. Additionally, the onset of the ON phase, $\tong$, is gene-dependent.
The instant at which sequencing occurs may correspond to different levels of maturity in the dynamics \eqref{eq:ussol} across genes and cells. This stage of maturity is denoted as $t_{cg}$, which is gene- and cell-specific, as detailed in \cite{bergen2020generalizing}. The invariance \eqref{eq:inv2} causes identifiability issues on this parameter \citep{marot2022kvelo, zhang2024consensusVelo}, that are not directly addressed in \cite{bergen2020generalizing}. The authors of \textit{scVelo} instead reduce the parameter space by assuming for every gene $\alpha_g^{\text{off}}=0$. Since this assumption suggests that at least some of the observed data values are close to zero, an additional transformation is applied when the data do not exhibit this behavior. Specifically the data $(m_{s, cg}, m_{u, cg})$ are translated as
\begin{equation}
	\label{eq:translation}  
	(m'_{s, cg}, m'_{u, cg}) = (m_{s, cg} - \min_c (m_{s, cg}), m_{u, cg} - \min_c (m_{u, cg})).
\end{equation} 
This step is performed before applying their method for further analysis. 
It is important to highlight that the three time-related parameters $(t_{cg}, \tong, \omega_g)$ cannot be estimated simultaneously. Specifically, if $t_{cg} \geq \tong$, the gene expression level, as described by equation \eqref{eq:ussol} and property \eqref{eq:invariance}, depends only on the difference $\tilde{t}_{cg} = t_{cg} - t_{0,g}^{\text{on}}$. This difference encapsulates dependencies on both the gene and the cell and can be interpreted as the time elapsed since the onset of the ON phase. Conversely, if for a given cell $c$ and gene $g$, the gene has not yet been activated at the time of sequencing ($t_{cg} < \tong$), the expression levels remain constant and independent of $t_{cg}$, thus provide no information about it. Hence, neither $t_{cg}$, nor $\tilde{t}_{cg}$ are identifiable when $t_{cg} < \tong$. As a consequence of these identifiability issues, the model parameters consist of the rate constants, which define the \emph{almond} in the $(s, u)$ plane and are specific to each gene, i.e. $\boldsymbol{\theta}_{g}^{\text{sc}} = (\alpha_g^{\text{off}}, \alpha_g^{\text{on}}, \beta_g, \gamma_g) = (0, \alpha_g^{\text{on}}, \beta_g, \gamma_g)$, as well as the time-related variables $\boldsymbol{\tau}_{cg}^{\text{sc}} = (\omega_g, \tilde{t}_{cg})$. As a result, we redefine the notation for the solutions of the ODE in equation \eqref{eq:ussol} as follows
\begin{equation}
	s(\boldsymbol{\tau}_{cg}^{\text{sc}}, \boldsymbol{\theta}_{g}^{\text{sc}}) = s(t_{cg},\tong,\omega_g, \boldsymbol{\theta}_{g}^{\text{sc}}), \qquad  
	u(\boldsymbol{\tau}_{cg}^{\text{sc}}, \boldsymbol{\theta}_{g}^{\text{sc}}) = u(t_{cg},\tong,\omega_g, \boldsymbol{\theta}_{g}^{\text{sc}}).  
\end{equation}
Furthermore, even for gene-cell pairs where $t_{cg} \geq \tong$, identifying $\tilde{t}_{cg}$ remains practically infeasible. This is because $\tilde{t}_{cg}$ is influenced by both the gene and the cell, yet only a single observation of the corresponding gene expression level is available to inform its estimation. This limitation, which was not addressed in \cite{bergen2020generalizing}, was explicitly highlighted in \cite{sisNonIdentifiability}. Each $\omega_g$ defines the point $(s_{g}^\omega, u_{g}^\omega)$ as
\begin{equation}  
	s_{g}^\omega = s(\tong + \omega_g,\tong,\omega_g, \boldsymbol{\theta}_g^{\text{sc}}), \qquad  
	u_g^\omega = u(\tong + \omega_g,\tong,\omega_g, \boldsymbol{\theta}_g^{\text{sc}}),  
\end{equation} 
which marks the transition to the OFF branch. 
In \cite{bergen2020generalizing}, parameter estimation is performed using maximum likelihood via the EM algorithm. The data distribution is defined in terms of the so-called Deming residuals
\begin{equation}
	e_{cg} := \text{sign}\left(m_{s, cg} - s(\boldsymbol{\tau}_{cg}^{\text{sc}}, \boldsymbol{\theta}_{g}^{\text{sc}})\right)
	\big\lVert (m_{s, cg}, m_{u, cg}) - (s(\boldsymbol{\tau}_{cg}^{\text{sc}}, \boldsymbol{\theta}_{g}^{\text{sc}}), u(\boldsymbol{\tau}_{cg}^{\text{sc}}, \boldsymbol{\theta}_{g}^{\text{sc}}))\big\rVert_2.
\end{equation}
The Deming residuals quantify the signed distances between the observed smoothed counts and the estimated mRNA levels, serving as a measure of how well the model approximates real data
They are assumed to be independent and normally distributed with gene specific variance, i.e. $e_{cg} \sim \mathcal{N}\left(0, \sigma^2_g\right)$. \\
\indent We summarize here various know issue of \textit{scVelo} methodologies, discussed in this paragraphs and in various literature papers. Firs of all, the data undergo extensive preprocessing, including both linear and nonlinear transformations, before being analyzed. The strong impact of this preprocessing has been examined in \cite{soneson2021preprocessing}. The output of \textit{scVelo} is further post-processed to compute the velocities. This post-processing critically affects the visualization of the results, as noted in \cite{zheng2023pumping}. Moreover, the parameters $\tilde{t}_{cg}$ exhibit weak identifiability, as discussed in \cite{sisNonIdentifiability} and imposing the constraint $\alpha_g^{\text{off}}=0$ as in \textit{scVelo} modify the parameters' interpretation due to translation \eqref{eq:translation}. As an additional point, the methodology does not provide any quantification of the uncertainty in the parameter estimates. As a result of all those issues, the estimate produced with the \textit{scVelo} tool is unreliable and we show some of its problem with a simulation study in Section \ref{sec:simRNAvel}.

\section{BayVel proposal} 
\label{sec:bayVel} 

We introduce a new Bayesian methodology, named \textit{BayVel}, to estimate the parameters of the models presented in Section \ref{subsec:modeling}. While preserving the core conceptual framework of RNA velocity, our approach directly addresses the statistical challenges inherent in \textit{scVelo} (see Section \ref{subsec:scvelo}).  Unlike \textit{scVelo}, we work directly with the raw data $\{y_{s,cg},y_{u,cg}\}_{c,g}$ and gene expression counts are modeled using discrete-space stochastic processes $\{\mathbf{Y}_{cg}\}$ that, for each cell $c\in \{1, \dots, C\}$ and gene $g \in \{1, \dots, G\}$, are independent copies of the process described in equation \eqref{eq:stochProc}, with specific
parameters shared across different cells and genes as detailed below. 

\subsection{The model}
\textit{BayVel} is based on \eqref{eq:stochProc}, where the gene-dependent rate parameters $\left(\alpha_g^{\text{off}}, \alpha_g^{\text{on}}, \beta_g, \gamma_g\right)$ are shared across cells. To resolve the identifiability issues arising from the invariance property \eqref{eq:inv2}, without loss of generality we fix $\beta_g=1$.
Regarding the time-related parameters, we adopt a different approach from \textit{scVelo}. First, \textit{scVelo} assumes that all cells share the same gene-specific duration $\omega_g$ of the ON phase. This assumption is overly restrictive, as it is common in many biological contexts to observe different cell types coexisting within the same sample. This coexistence introduces heterogeneity in expression profiles, potentially resulting from certain genes being switched ON or OFF at different times across various cell types. Conversely, \textit{scVelo} models the elapsed times since the start of the ON phase $\tilde t_{cg}$ as both gene- and cell-specific. This approach is excessively general and results in a lack of practical identifiability.
Our proposal instead leverages the fact that cells can be grouped based on similar expression profiles (e.g., cell types), allowing us to pool information from cells within the same group.
We assume then that each cell $c$ in the dataset carries a group label $k\in\{1,\dots,K\}$ with $K\leq C$. This label may correspond to a known cell-type annotation (if available) or be determined using a clustering method applied to expression profiles. Within this framework, we model both the ON-phase durations $\omega_{kg}$ and the elapsed times $\tilde t_{kg}$ as group- and gene-specific. This approach simultaneously addresses two key challenges: it relaxes the unrealistic assumption that all cells share the same ON-phase duration, and it improves identifiability by borrowing shared information across grouped cells.
The parameters of our first \textit{BayVel} model are $\boldsymbol{\theta}_{g}^{\text{B1}}= (\alpha_g^{\text{off}}, \alpha_g^{\text{on}}, \beta_g, \gamma_g)=(\alpha_g^{\text{off}}, \alpha_g^{\text{on}}, 1, \gamma_g)$, $\boldsymbol{\tau}_{kg}^{\text{B1}}=(\omega_{kg}, \tilde{t}_{kg})$. Compared to the original \textit{scVelo} approach, the proposed method offers several advantages but introduces an additional constraint. Specifically, for each fixed gene $g$, both time parameters $\omega_{kg}$ and $\tilde{t}_{kg}$ are group-dependent, preventing cells from sharing the same ON-phase duration while having different elapsed times since the start of the ON phase, a flexibility previously allowed in \textit{scVelo}. To restore this flexibility, we introduce a finer level of grouping, by allowing cells within each group $k$ to be assigned to different subgroups. Consequently, we introduce a subgroup label $r \in \{1, \dots, R\}$, where $K\leq R \leq C$. This additional label is obtained either by clustering the expression profiles of the cells with the same type annotation or by adopting a hierarchical clustering approach when such annotation is not available. It is assumed that cells corresponding to different groups cannot be in the same subgroup. The parameters of this more general \textit{BayVel} model are
\begin{equation}\label{B2}
		\boldsymbol{\theta}_{g}^{\text{B2}}= (\alpha_g^{\text{off}}, \alpha_g^{\text{on}}, \beta_g, \gamma_g)=(\alpha_g^{\text{off}}, \alpha_g^{\text{on}}, 1, \gamma_g),\qquad \boldsymbol{\tau}_{krg}^{\text{B2}}=(\omega_{kg}, \tilde{t}_{rg}).
\end{equation}
This model contains as subcases both the original \textit{scVelo} setting (where $K=1$ and each cell is considered as belonging to a separate subgroup) and the previous version $\boldsymbol{\theta}_{g}^{\text{B1}},\boldsymbol{\tau}_{kg}^{\text{B1}}$, considering $R=K$. Therefore, in the following part of this paper, we will always refer to the \textit{BayVel} model with the more general parameterization \eqref{B2}.\\
\indent For each cell $c$ belonging to group $k$ and subgroup $r$, the generative model is defined as
\begin{equation} \label{eq:pois}
	Y_{s,cg}\sim \text{Pois}(s(\boldsymbol{\tau}_{krg}^{\text{B2}} , \boldsymbol{\theta}_{g}^{\text{B2}})), \qquad   Y_{u,cg}\sim \text{Pois}(u(\boldsymbol{\tau}_{krg}^{\text{B2}} , \boldsymbol{\theta}_{g}^{\text{B2}})),
\end{equation}
with $Y_{s, cg} \perp Y_{s, c'g'}$, $Y_{u, cg} \perp Y_{u, c'g'}$ and $Y_{s,cg} \perp Y_{u,c'g'}$ for all $c,c' \in \{1,\dots , C\}$ and $g,g' \in \{1,\dots , G\}$.
It is reasonable to expect that additional noise sources may influence the measurement process, leading to deviations from the Poisson assumption. To address this, \textit{BayVel} relaxes the Poisson distribution assumption in favor of the more general Negative Binomial distribution, which accounts for overdispersion. We parameterize the Negative binomial distribution in terms of the mean $\mu$ and an overdispersion parameter $\eta$, which captures the excess variance beyond what is explained by the Poisson model. Specifically, if $Y\sim \text{NB}(\mu, \eta)$, then $\E Y = \mu$ and $\Var Y = \mu + \mu^2 \eta$. Since overdispersion may vary between genes, we assume a gene-specific parameter $\eta_g \in \mathbb{R}^+$.\\
\indent During scRNA-sequencing, only a fraction of the mRNA counts is successfully captured in each cell, as discussed in \cite{tang2023captureEfficiency}. \textit{scVelo} addresses this issue through a normalization step included in the preprocessing of the data. In contrast, in \textit{BayVel} we prefer to model this differences explicitly in the model, introducing a cell-specific ``capture efficiency'' parameter $\lambda_c \in (0,1]$.
With this refinement, the assumed data distribution becomes
\begin{equation} \label{eq:nb}
	Y_{s,cg}\sim \text{NB}(s(\boldsymbol{\tau}_{krg}^{\text{B2}} , \boldsymbol{\theta}_{g}^{\text{B2}}) \lambda_c, \eta_g), \qquad  Y_{u,cg}\sim \text{NB}(u(\boldsymbol{\tau}_{krg}^{\text{B2}} , \boldsymbol{\theta}_{g}^{\text{B2}}) \lambda_c, \eta_g),
\end{equation}
where the capture efficiency $\lambda_c$ scales the mean expression levels. However, this produces an additional identifiability issues since for any $a_3 \in \R^+$, we have
\begin{equation}
	\label{eq:invCapture}
	\begin{aligned}
		s(\boldsymbol{\tau}_{krg}^{\text{B2}} ,  (\alpha_g^{\text{off}}, \alpha_g^{\text{on}}, 1, \gamma_g)) \lambda_c &= s(\boldsymbol{\tau}_{krg}^{\text{B2}} ,  (a_3\alpha_g^{\text{off}}, a_3\alpha_g^{\text{on}}, 1, \gamma_g)) \frac{\lambda_c}{a_3},\\
		u(\boldsymbol{\tau}_{krg}^{\text{B2}} ,  (\alpha_g^{\text{off}}, \alpha_g^{\text{on}}, 1, \gamma_g)) \lambda_c &= u(\boldsymbol{\tau}_{krg}^{\text{B2}} ,  (a_3\alpha_g^{\text{off}}, a_3\alpha_g^{\text{on}}, 1, \gamma_g)) \frac{\lambda_c}{a_3}.
	\end{aligned}
\end{equation}
To resolve this issue, we impose the constraint that the average capture efficiency across all cells is fixed at 1. This constraint can be enforced through post-estimation rescaling (cf. Section \ref{sec:statistic}). 
As a final remark, to simplify the notation, we will hereafter denote $s^\sim_{krg} := s(\boldsymbol{\tau}_{krg}^{\text{B2}} , \boldsymbol{\theta}_{g}^{\text{B2}})$, and $u^\sim_{krg} := u(\boldsymbol{\tau}_{krg}^{\text{B2}} , \boldsymbol{\theta}_{g}^{\text{B2}})$.

\subsection{Statistical inference and prior definition}
\label{sec:statistic}

\begin{figure}
	\centering
	\includegraphics[scale = 0.22]{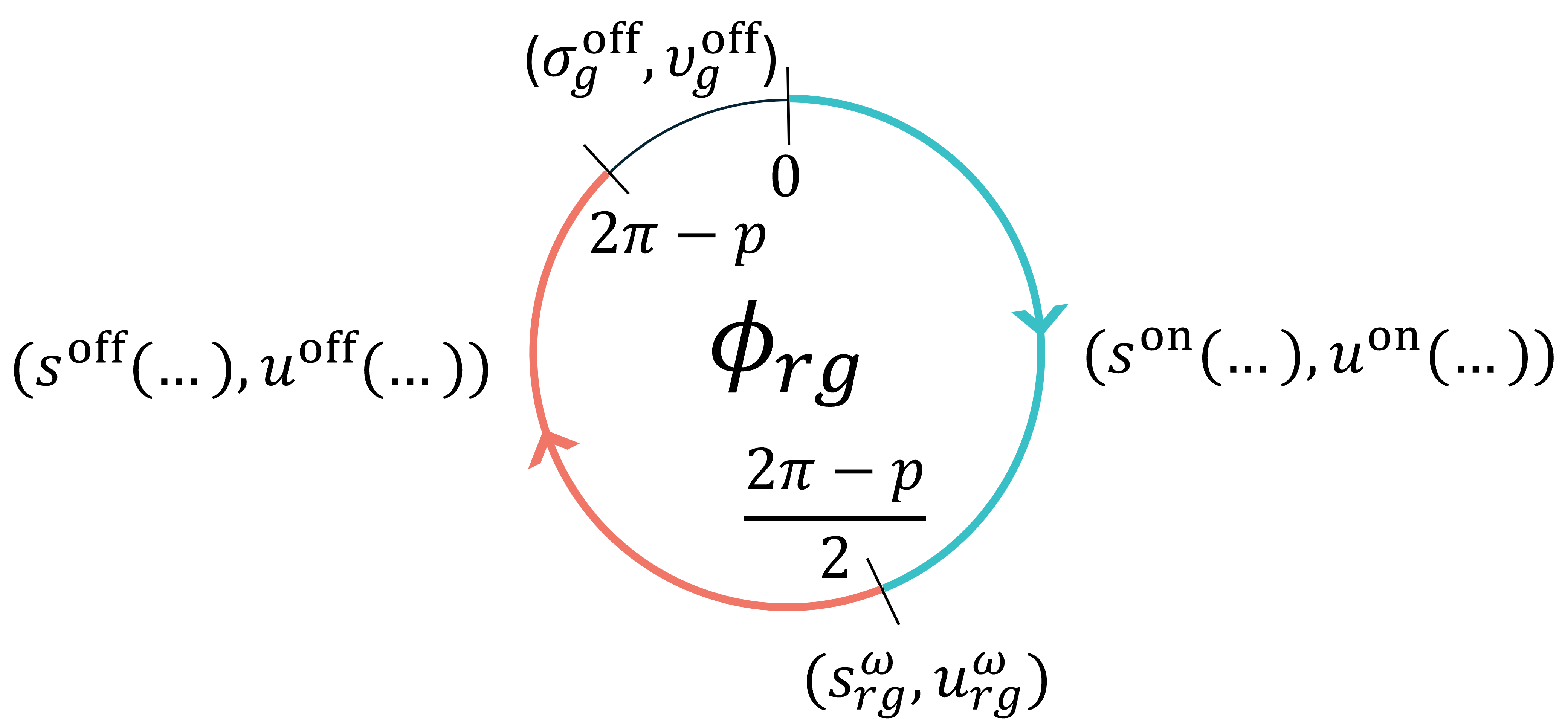}
	\caption{Graphical representation of the relationship between the parameter $\phi_{rg}$ and the \emph{almond}'s coordinates in the $(s, u)$ plane.}
	\label{fig:thetaTransform}
\end{figure}
The natural parameterization of the model is not the most convenient choice for the sampling algorithms. Indeed, for each gene and cell, our data consist of a pair that can be interpreted as a point in the $(s,u)$ space. The stochastic and deterministic models, along with the model parameters, define a "theoretical position" in this space. However, since the parameters enter nonlinearly into the solution \eqref{eq:ussol}, it is challenging to define a weakly informative prior over them. For example, consider the parameter combination $\gamma_g \tilde{t}_{rg}$ which appears as a negative exponent in equations \eqref{eq:solutionON}. When either $\gamma_g$ or $\tilde{t}_{rg}$ is very large,
the system reaches a steady-state region where even substantial changes in these parameters produce only negligible effects on the coordinates in the $(s,u)$ plane. Additionally, two points that are both close to the lower steady-state $\textbf{SS}_g^{\text{off}}$, but belonging to different branches, are close in the $(s, u)$ plane, but the corresponding time parameters $\tilde{t}_{rg}$ would be drastically different, one really small and the other one very large. This discrepancy makes it difficult for the sampling algorithm to move efficiently between branches when exploring the parameter space. To overcome these difficulties, we realized that working with coordinates in the $(s,u)$ plane
is a much more effective approach, as it allows for a more intuitive definition of weakly informative priors. As a first application of this principle, we use the steady-state coordinates $(\upsilon_g^{\text{off}}, \upsilon_g^{\text{on}}, \sigma_g^{\text{on}})$ as working parameters, rather than working directly with the reaction rates $\boldsymbol{\theta}_{g}^{\text{B2}}$. Moreover, to prevent the upper steady state from being searched in a parameter region that could lead to unreasonable high counts, we impose an arbitrary upper bound $a$ for both coordinates of the steady states, ensuring that $\upsilon_g^{\text{on}}, \sigma_g^{\text{on}} < a$. This assumption, which is also reasonable
from a biological viewpoint, allows use to define a weakly informative prior for the steady-state coordinates within the interval $[0,a]$.
Similarly, instead of using $\omega_{kg}$, we work with the $u$-coordinate of the switching point, $u^\omega_{kg}$. Regarding the time parameter $\tilde{t}_{rg}$, we replace it with an angular coordinate $\phi_{rg} \in [0, 2\pi]$, which, given the values of $(\upsilon_g^{\text{off}}, \upsilon_g^{\text{on}},
\sigma_g^{\text{on}})$ and of $u^\omega_{kg}$, determines the coordinates $(s(\boldsymbol{\tau}_{krg}^{\text{B2}}, \boldsymbol{\theta}_{g}^{\text{B2}}), u(\boldsymbol{\tau}_{krg}^{\text{B2}} , \boldsymbol{\theta}_{g}^{\text{B2}}))$. Figure \ref{fig:thetaTransform} shows the relationship between $\phi_{rg}$ and the coordinates on the $(s, u)$-plane (see Section 2 of Supplementary Material \citep{supplementary} for the analytic form). Assigning an entire circular sector to the lower-steady state enables us to plaice a point mass at this steady state. \\

\begin{equation}
	\label{eq:priorDir}
	\left(
	{\upsilon_g^{\text{off}}},{\upsilon_g^{\text{on}} - \upsilon_g^{\text{off}}}, a - {\upsilon_g^{\text{on}}}
	\right)/a \sim \text{Dirichlet}(1, 1, 1),
\end{equation}
which corresponds to a uniform on the symplex, and hence we have that $\upsilon_g^{\text{on}}/a \sim \text{Beta}(2,1)$.
Similarly, for the only free parameter for the $s$ component, we assume the following prior $\sigma_g^{\text{on}} /a\sim \text{Beta}(2,1)$. Since $\upsilon_g^{\text{off}}\leq  u_{kg}^\omega\leq\upsilon_g^{\text{on}} $, and $0\leq\phi_{rg} <2\pi$, a natural choice for a weakly informative prior is the Uniform distribution
\begin{equation}
		\label{eq:priorSwitch}
		u_{kg}^\omega | \upsilon_g^{\text{off}}, \upsilon_g^{\text{off}} \sim \mathcal{U}\left(\upsilon_g^{\text{off}}, \upsilon_g^{\text{on}}\right),\qquad
		\phi_{rg}\sim \mathcal{U}\left(0, 2\pi\right).
\end{equation}
The priors \eqref{eq:priorSwitch} induce conditional densities over the original time parameters $\boldsymbol{\tau}_{krg}^{\text{B2}}=(\omega_{kg}, \tilde{t}_{rg})$, that can be computed in closed form and are shown in Section 3 of the Supplementary Material \citep{supplementary}. The priors for the overdispersion and capture efficiency parameters are $\eta_g \sim \mathcal{N}_{[0, +\infty)}\left(0, 10000^2\right)$ and $\lambda_c \sim \mathcal{U}(0, 1)$, where $\mathcal{N}_{[a,b]}(\mu, \sigma^2)$ denotes a truncated Normal distribution support in $[a, b]$. While not explored here, alternative priors, such as mixtures with mass at $0$, could be used, promoting shrinkage toward a Poisson model for genes with minimal extra noise. \\
\indent Under this setting, no full conditional distributions for the distributions' parameters are available in closed form, and all parameters are sampled in the MCMC using Metropolis steps. In more detail, for each $g$, the parameters $\upsilon_g^{\text{off}}, \upsilon_g^{\text{on}}, \sigma_g^{\text{on}}$ are proposed jointly using a multivariate normal distribution with an adaptive covariance matrix, following \cite{andrieu2008adaptiveMCMC} (Algorithm 4), while the other parameters are proposed independently with an adaptive variance, as suggested in \cite{Robert}. The results presented in the next sections are obtained using $250{,}000$ iterations, a burn-in of $200{,}000$, and a thinning of $25$, yielding $2{,}000$ posterior samples. \textit{BayVel} MCMC algorithm is implemented in \texttt{Julia} \citep{bezanson2017julia} and run in parallel on HPC systems. A key drawback of \textit{BayVel} is its high computational cost, since it requires 1–3 days, depending on $K$ and $R$.

\section{Simulation study} \label{sec:simRNAvel}
We conduct a simulation study to comprehensively evaluate and compare the performance of \textit{scVelo} and \textit{BayVel}. Although both methods are theoretically applicable to raw count data, \textit{BayVel} operates directly on unprocessed counts, whereas \textit{scVelo} requires a preprocessing step. Consequently, the parameter estimates obtained from the two methods on the same sample may differ significantly. Additionally, the underlying assumptions of the two methods are not identical.
To ensure a fair comparison, we first identify a parameter setting in which both methods are expected to perform well. Using these shared parameters, we then generate separate datasets according to each method’s respective statistical model. Additionally, since \textit{BayVel} operates under less restrictive assumptions, we create a few specific datasets to further benchmark our proposed approach.
For both model, then, we simulate $G = 2000$ independent genes and $C = 3000$ cells, fixing $K = 1$ (all the cells belong to the same group, as in \textit{scVelo}) and compare the estimated values with the known ground true. We explore different datasets, varying $R \in \{10, 30, 100\}$. The number of genes, cells, as well as the number of subgroups is chosen to mimic the real dataset presented in Section \ref{sec:real}. 
As an additional benchmark, we generate three additional \textit{BayVel} dedicated datasets, setting $K = 10$ with $R \in \{10, 30, 100\}$, where the hierarchical subdivision in groups and subgroups is fully exploited. A detailed description can be found on the Supplementary Material \citep{supplementary}, Section 4.\\
\indent Once the parameters are simulated, we can generate data $\{y_{s, cg}, y_{u, cg}\}_{c,g}$ for \textit{BayVel} by sampling from the Negative Binomial. On the other hand, care must be taken when simulating data from \textit{scVelo}. Although the package provides a built-in simulation routine (\texttt{scvelo.datasets.simulation}), this does not produce data following the \textit{scVelo} statistical model in Section \ref{subsec:scvelo}. Instead, for each gene $g$ and cell $c$, it samples the observed values $m_{s, cg}$ and $m_{u, cg}$ independently from Normal distributions centered at the ODE solution \eqref{eq:ussol}, i.e. $m_{s, cg} \sim \mathcal{N}\left(s(\boldsymbol{\tau}_{cg}^{\text{sc}}, \boldsymbol{\theta}_{g}^{\text{sc}}), \sigma^2_{s}\right)$ and $m_{u, cg} \sim \mathcal{N}\left(u(\boldsymbol{\tau}_{cg}^{\text{sc}}, \boldsymbol{\theta}_{g}^{\text{sc}}), \sigma^2_{u}\right)$ (see Section 4.2 of Supplementary Material \citep{supplementary} for details about the variances used in the \textit{scVelo} simulation procedure).
To take into account the possible effect caused by this discrepancy, we adopt two alternative approaches for simulating \textit{scVelo} data $\{m_{s, cg}, m_{u, cg}\}_{c,g}$: i) we generate data accordingly to the \textit{scVelo} build-in function, we refer to the data simulated in this way as \textit{Independent-Normal-data} (\textit{IN}); ii) using the same parameter settings, we instead sample $(m_{s, cg}, m_{u, cg})$ according to the likelihood in Section \ref{subsec:scvelo}, assuming Normally distributed Deming residuals; this data are denoted as \textit{Deming-residuals-data} (\textit{Dem}).
These two simulation strategies allow us to obtain a comprehensive evaluation of \textit{scVelo}'s performance. A description on how to simulate from the Deming residuals is shown in Section 4.3 of Supplementary Material \citep{supplementary}.

\subsection{The performance of \textit{scVelo}}
\label{sec:perfScVelo}

\begin{table}[t]
	\small
	\centering
	\begin{tabular}{ll|rrrrrrrr}
		\hline
		& & $\upsilon_{g}^{\text{off}}$ & $\sigma_{g}^{\text{off}}$ & $\upsilon_{g}^{\text{on}}$ & $\sigma_{g}^{\text{on}}$ & $u_{g }^\omega$ & $s_{g}^\omega$ & $u_{cg}^{\sim}$ & $s_{cg}^{\sim}$ \\
		\hline
		\multirow{3}{*}{\textit{IN}} & $K = 1, R = 10$ & 0.794 & 1.000 & 0.474 & 0.485 & 0.136 & 0.138 & 0.059 & 0.301   \\
		&$K = 1, R = 30$ &0.848 & 1.000 & 0.463 & 0.489 & 0.136 & 0.125 & 0.071 & 0.295 \\
		&$K = 1, R = 100$ &0.846 & 1.000 & 0.473 & 0.474 & 0.123 & 0.128 & 0.068 & 0.290 \\
		\hline
		\multirow{3}{*}{\textit{Dem}} & 	$K = 1, R = 10$ & 0.802 & 0.865 & 0.466 & 0.462 & 0.127 & 0.113 & 0.046 & 0.286 \\
		&$K = 1, R = 30$ &  0.857 & 0.879 & 0.467 & 0.471 & 0.124 & 0.103 & 0.053 & 0.278 \\
		&$K = 1, R = 100$ & 0.847 & 0.879 & 0.467 & 0.479 & 0.112 & 0.111 & 0.051 & 0.271 \\
		\hline
	\end{tabular}
	\caption{Median relative error for the parameters estimated by \textit{scVelo}, when the simulated data are sampled from two independent Normal distributions (\textit{IN}) or according to \textit{scVelo} likelihood (\textit{Dem}).}
	\label{tab:error_scVelo}
\end{table}
\normalsize
Before inference, see Section \ref{subsec:scvelo}, \textit{scVelo} applies the translation in \eqref{eq:translation} to shift the system into a reference frame where $\alpha_g^{\text{off}} = 0$. 
Once inference is complete, the estimated parameters are used to reconstruct the gene-specific \emph{almond} shape. Then, the inverse translation is applied to the entire dynamic to bring back the system to its original position, with the lower steady state to $\left(\min_c(m_{s, cg}), \min_c(m_{u, cg})\right)$, and all other coordinates translated by the same amount. As a consequence, a direct comparison of \textit{scVelo}'s inferred parameters with the original simulated values is not feasible, since the parameters have a different interpretation in the two settings. Instead, we evaluate \textit{scVelo}'s performance by analyzing how well the estimated \emph{almond} structure in the $(s, u)$-plane aligns with the true simulated dynamics. Specifically, we
compare the coordinates of the inferred steady states $(\sigma_g^{\text{off}}, \upsilon_g^{\text{off}})$ and $(\sigma_g^{\text{on}}, \upsilon_g^{\text{on}})$, of the switching points $(s^\omega_{1g}, u^\omega_{1g})$ and of the positions $(s(\boldsymbol{\tau}_{cg}^{\text{sc}}, \boldsymbol{\theta}_{g}^{\text{sc}}), u(\boldsymbol{\tau}_{cg}^{\text{sc}}, \boldsymbol{\theta}_{g}^{\text{sc}}))$ with the corresponding simulated values; all these quantities are invariant under transformation \eqref{eq:inv2}. 
Figure \ref{fig:phasePlotSIM} shows the output of \textit{scVelo} (in blue) for three arbitrarily selected genes, compared to the known ground truth (in red) used in the simulations. For genes 1 and 3, no translation was needed since $\min_c(m_{s, cg})= \min_c(m_{u, cg})=0$. In contrast, for gene 12, a translation was applied, and the \emph{almond} is then shifted away from the origin. A clear discrepancy exists between the estimated and true values, which we further analyze by calculating the relative error $\delta = \left|({\hat{\psi} - \psi})/{\psi}\right|$, where $\psi$ is a generic parameter.
This metric provides a standardized measure of deviation, allowing for a fair assessment of the estimation quality across different genes and conditions. For each type of parameter, the relative errors computed for all the genes and cells are aggregated to provide a concise summary of the estimation accuracy. Specifically, we report the median relative error over the varying parameters, ensuring a robust measure that is less sensitive to extreme values; these results are shown in Table \ref{tab:error_scVelo}.
 The results indicate that \textit{scVelo} struggles to accurately reconstruct the \emph{almond} shape across all simulated settings. As expected, varying the number of subgroups does not influence the results, since \textit{scVelo} is not influenced by this hierarchical structure in the data. Notably, for $\sigma^\text{off}_g$ the median relative error under the \textit{IN} data is consistently 1. This occurs because, for many genes $\min_c(m_{s, cg}) = 0$, forcing the estimated $s$-coordinate of the lower steady state $\text{\textbf{SS}}^\text{{on}}$ to remain at $0$. Additionally, we do not observe significant differences in the errors obtained under the \textit{IN} and \textit{Dem}, with the error values remain consistently high in both cases. In particular, we observe that the coordinates $(\sigma^{\text{off}}_g, \upsilon_g^{\text{off}})$ of the lower steady state are estimated extremely poorly, significantly impacting the positioning of the entire \emph{almond} structure. This misestimation propagates through the inferred dynamics, leading to distortions in the overall shape and alignment of the trajectory in the $(s, u)$-plane. The relative error in velocity estimation reflects the challenges in estimating the previous parameters and is therefore extremely high, ranging from $76.0\%$ ($K = 1, R = 30,$ \textit{IN}) to $80.2\%$ ($R = 1, K = 10$, \textit{Dem}). However, assessing velocity errors is inherently less reliable, as we will discuss in Section \ref{sec:perfBayVel}.\\
\indent This simulation study confirms that the methodological issues raised in Section \ref{subsec:scvelo} have a serious impact on the quality of the estimates even when \textit{scVelo} operates in the best possible scenario, that is when data are exactly simulated according to its underlying statistical model. 

\subsection{The performance of \textit{BayVel}}
\label{sec:perfBayVel}

\begin{figure}[t]
	\centering
	\subfloat{\includegraphics[scale = 0.15]{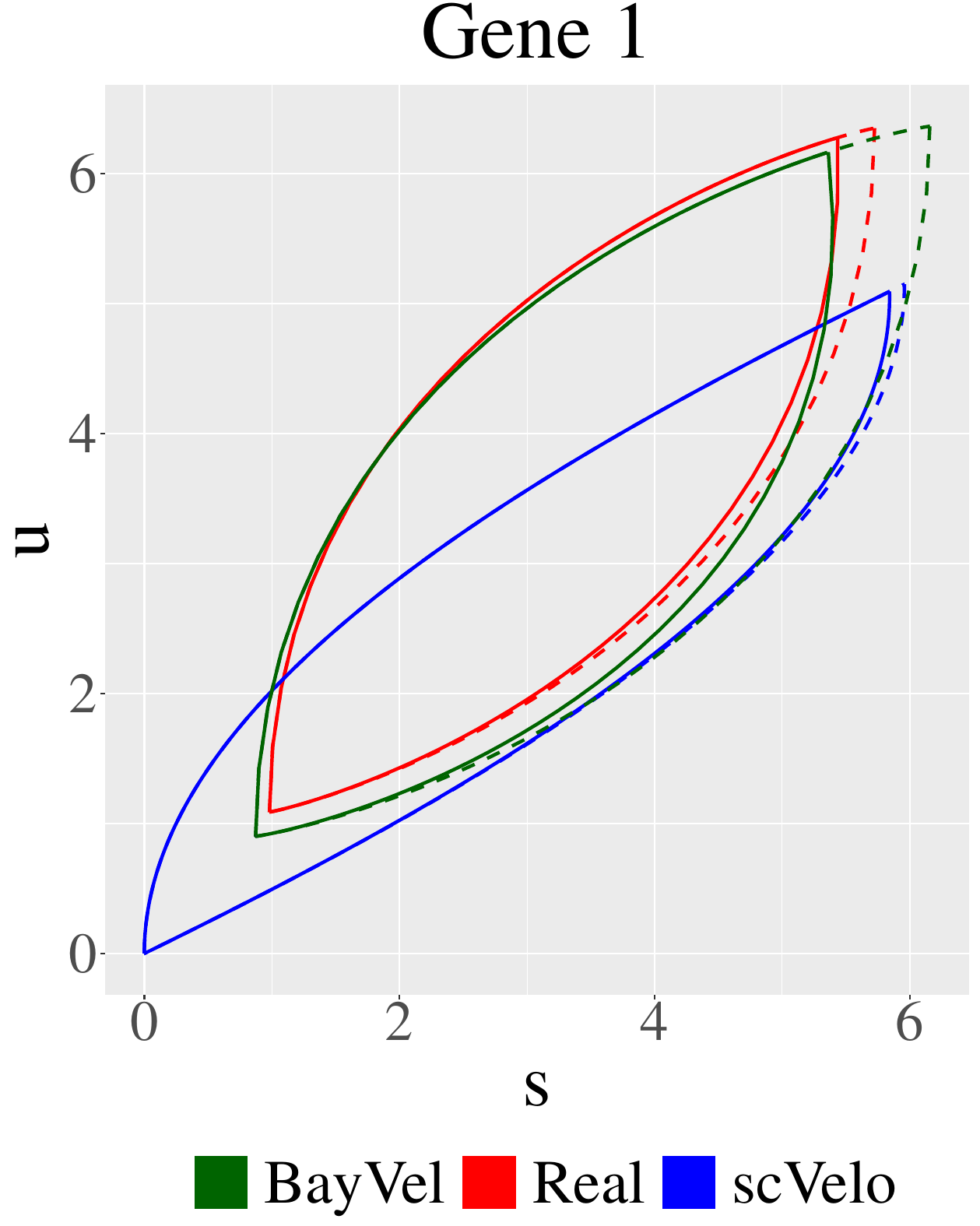}}
	\subfloat{\includegraphics[scale = 0.15]{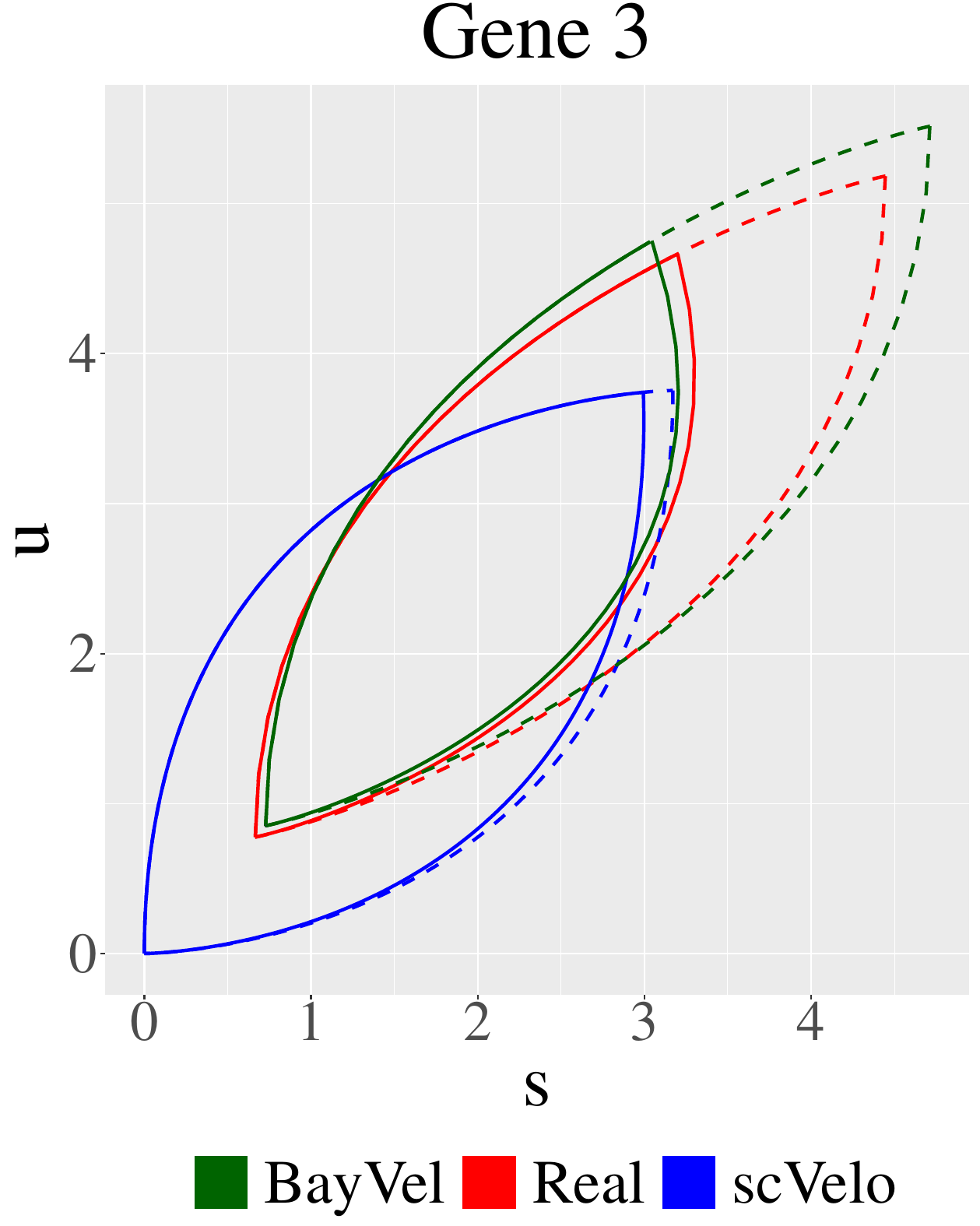}}
	\subfloat{\includegraphics[scale = 0.15]{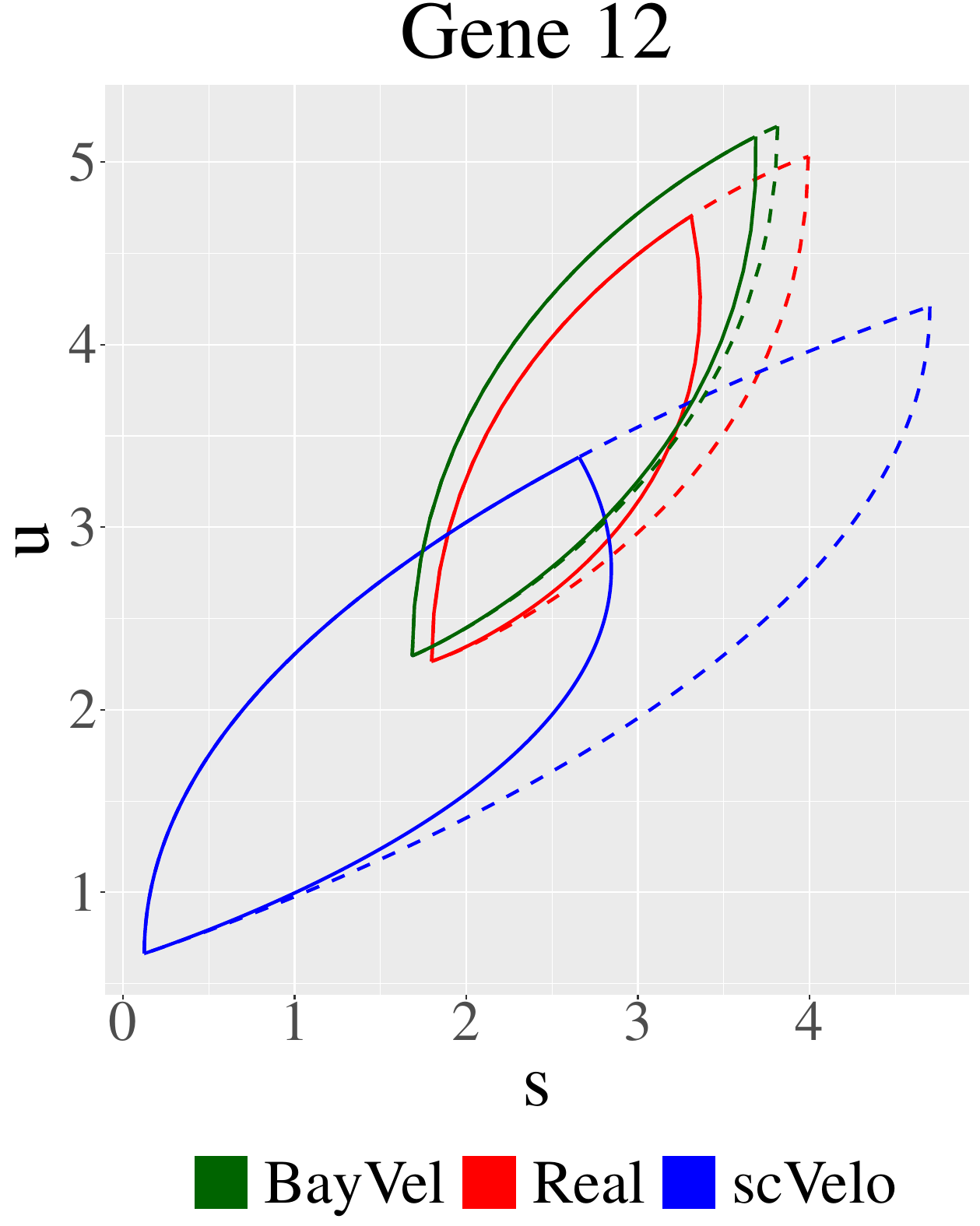}}
	\caption{Comparison of the estimation results for different genes on simulated data, obtained using \textit{scVelo} (blue line) and \textit{BayVel} (green line), when $K = 1, R = 10$. The phase plot for \textit{BayVel} is generated using the maximum a posteriori parameters. The red curve represents the ground truth dynamics, derived from the real parameters used to simulate the structure.}
	\label{fig:phasePlotSIM}
\end{figure}

\begin{table}[t]
	\small
	\centering
	\begin{tabular}{l|rrrrrrrrr}
		\hline
		&  $\upsilon_{g}^{\text{off}}$ & $\sigma_{g}^{\text{off}}$ & $\upsilon_{g}^{\text{on}}$ & $\sigma_{g}^{\text{on}}$ & $u_{kg }^\omega$ & $s_{rkg}^\omega$ & $u_{krg}^{\sim}$ & $s_{krg}^{\sim}$ \\
		\hline		
		$K = 1, R = 10$ & 0.048 & 0.037 & 0.109 & 0.151 & 0.036 & 0.043 & 0.034 & 0.030  \\
		$K = 1, R = 30$ & 0.054 & 0.032 & 0.094 & 0.140 & 0.031 & 0.043 & 0.043 & 0.037 \\ 
		$K = 1, R = 100$ & 0.071 & 0.045 & 0.118 & 0.192 & 0.035 & 0.051 & 0.071 & 0.058  \\ 
		$K = 10, R = 10$ & 0.046 & 0.038 & 0.172 & 0.229 & 0.103 & 0.140 & 0.031 & 0.031  \\
		$K = 10, R = 30$ & 0.045 & 0.035 & 0.143 & 0.203 & 0.076 & 0.109 & 0.047 & 0.043 \\ 
		$K = 10, R = 100$ & 0.054 & 0.052 & 0.161 & 0.235 & 0.082 & 0.102 & 0.074 & 0.064  \\ 
		\hline
	\end{tabular}
	\caption{Median relative error for the parameters estimated by \textit{BayVel} on different simulated scenarios, comparing the median of the estimated posterior chain with the true simulated values of the parameters.}
	\label{tab:relErroBayVel}
\end{table}
\normalsize
\begin{figure}[t]
	\centering
	\subfloat{\includegraphics[scale = 0.15]{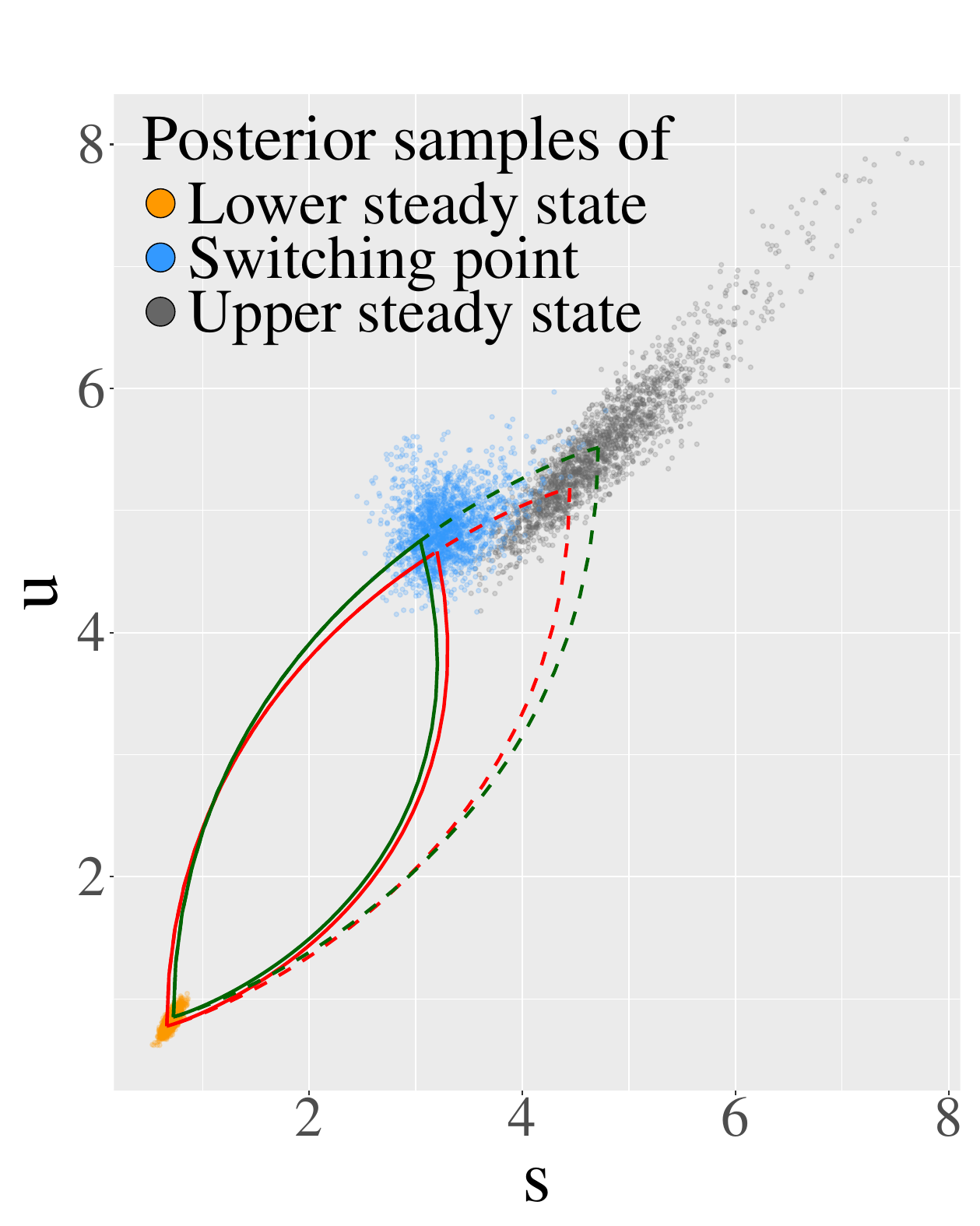}}
	\subfloat{\includegraphics[scale = 0.15]{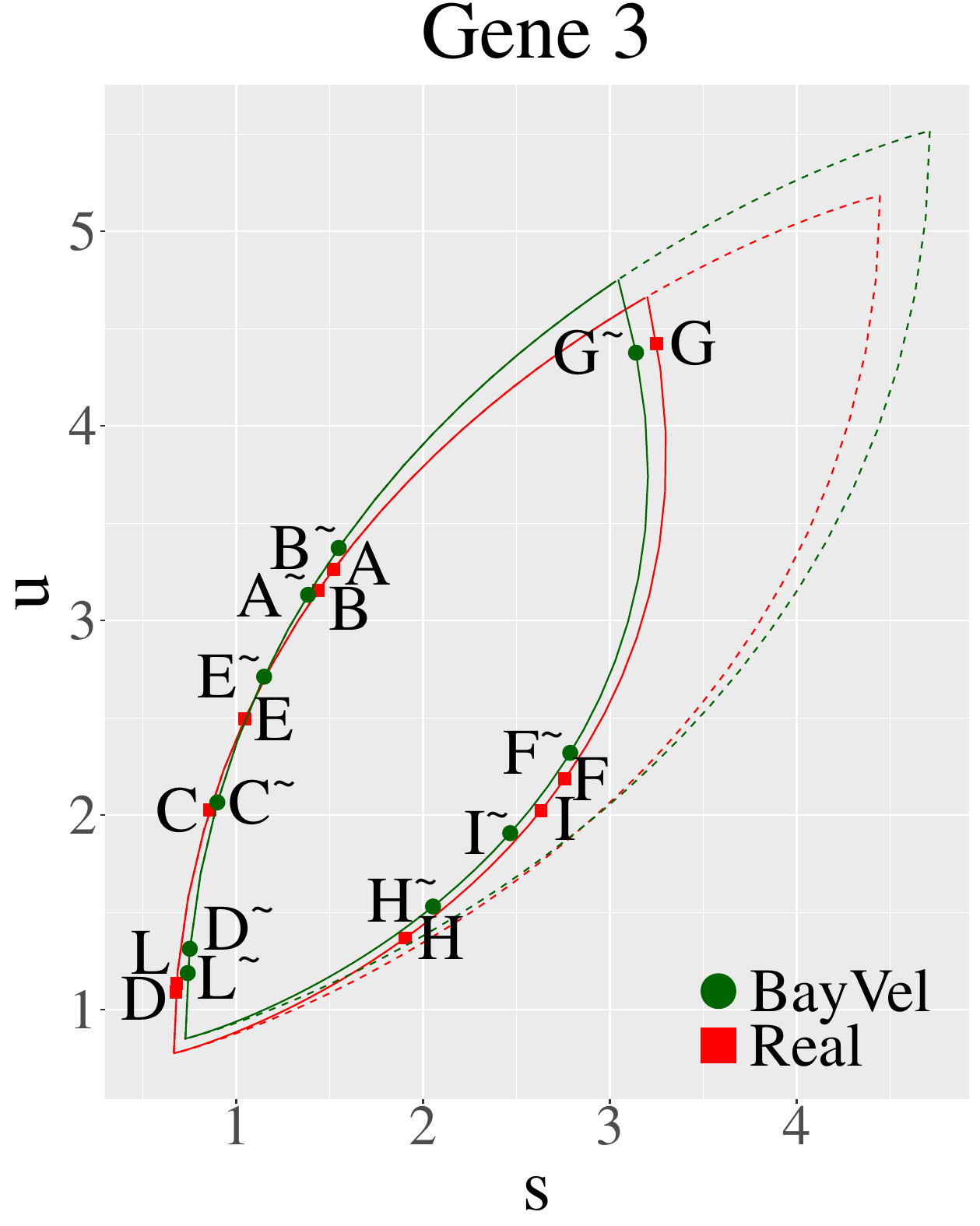}}
	\subfloat{\includegraphics[scale = 0.15]{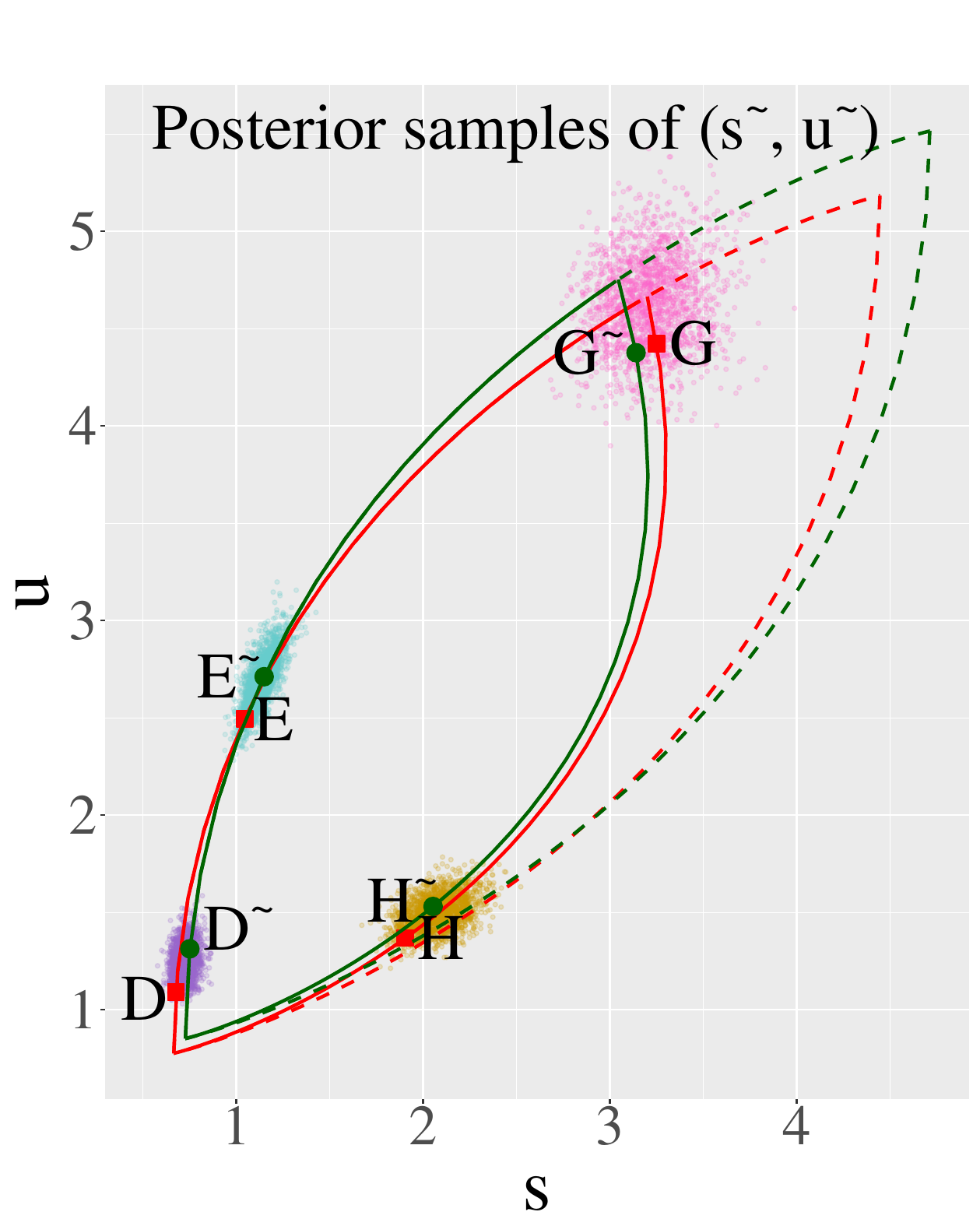}}\\
	\subfloat{\includegraphics[scale = 0.15]{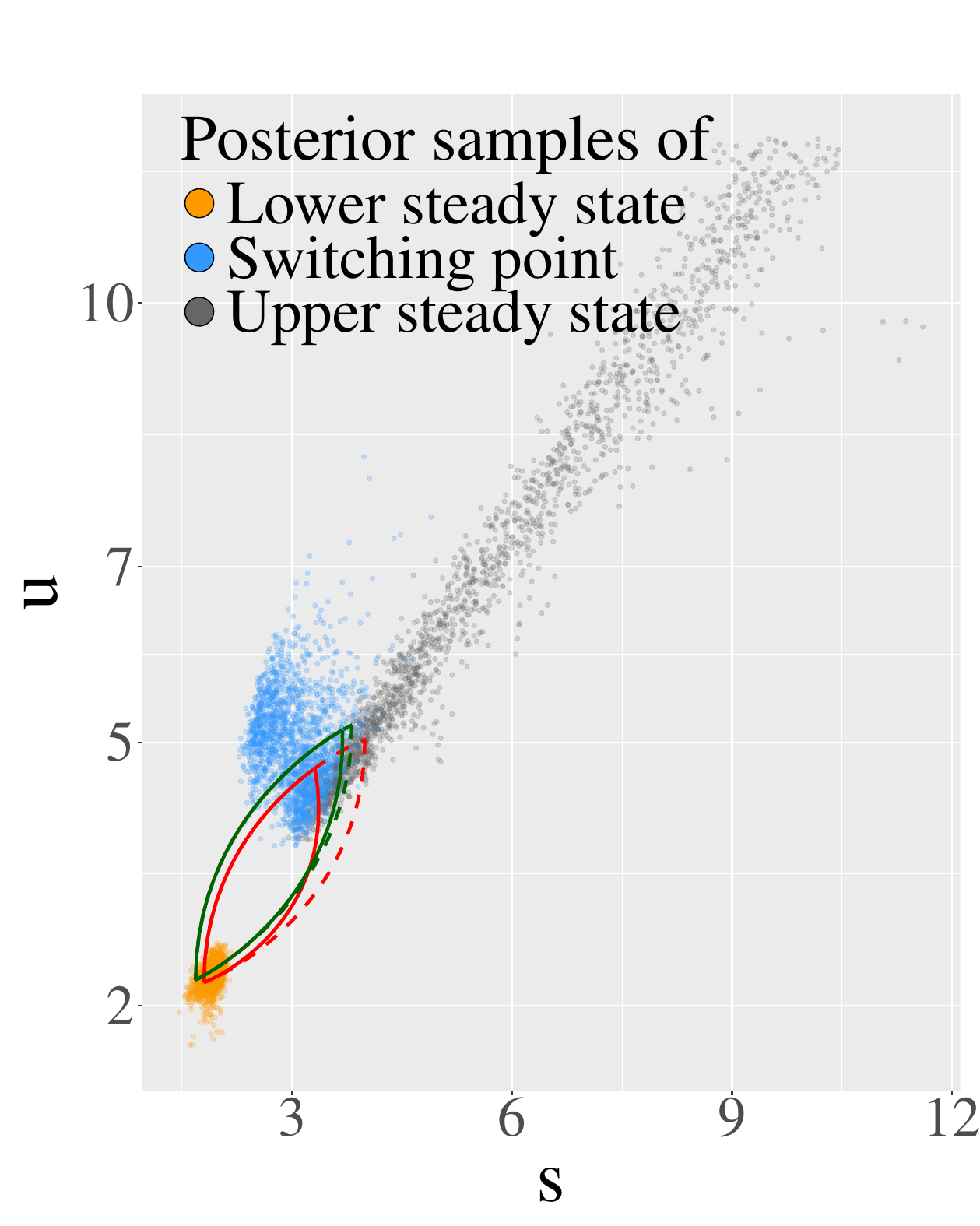}}
	\subfloat{\includegraphics[scale = 0.15]{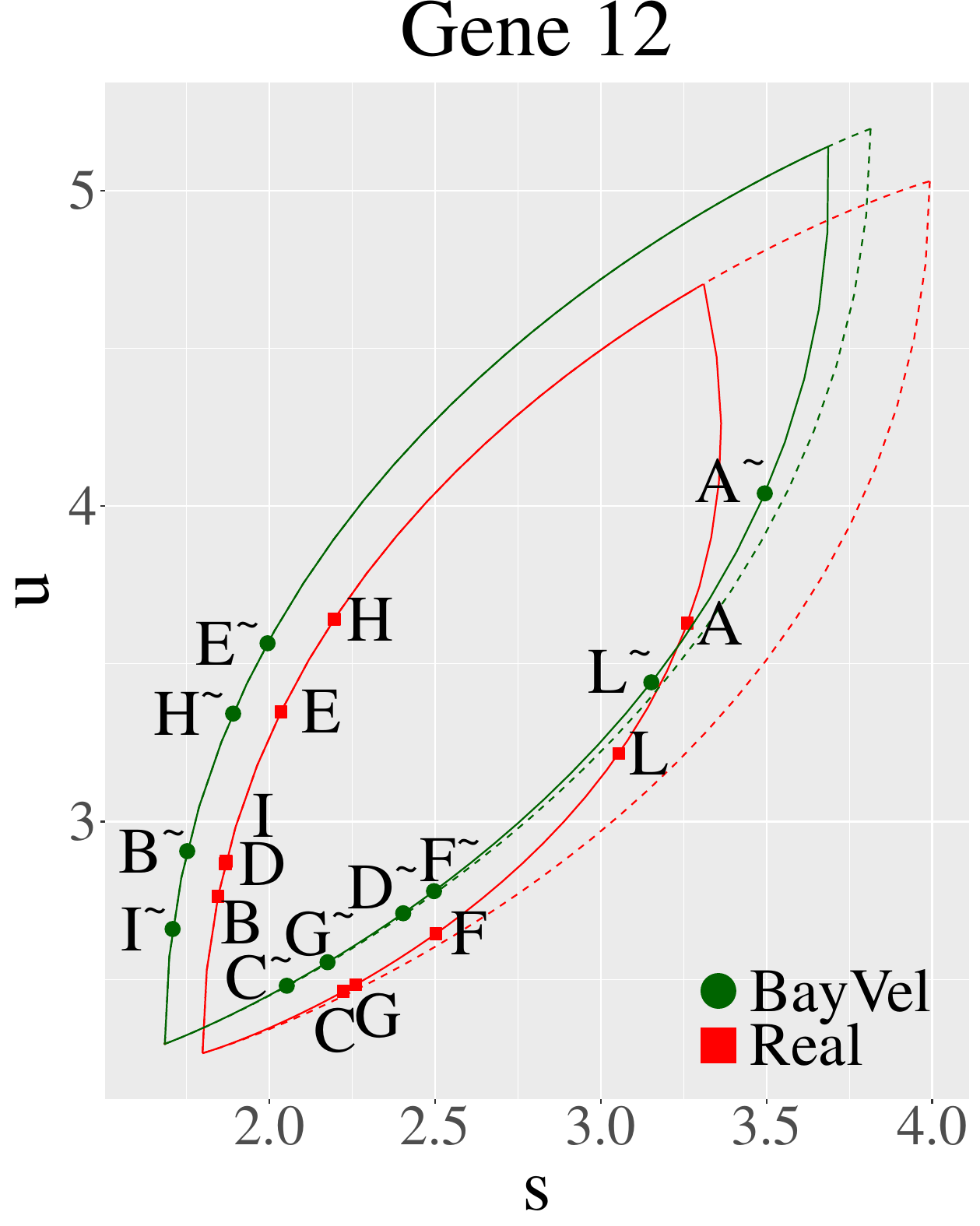}}
	\subfloat{\includegraphics[scale = 0.15]{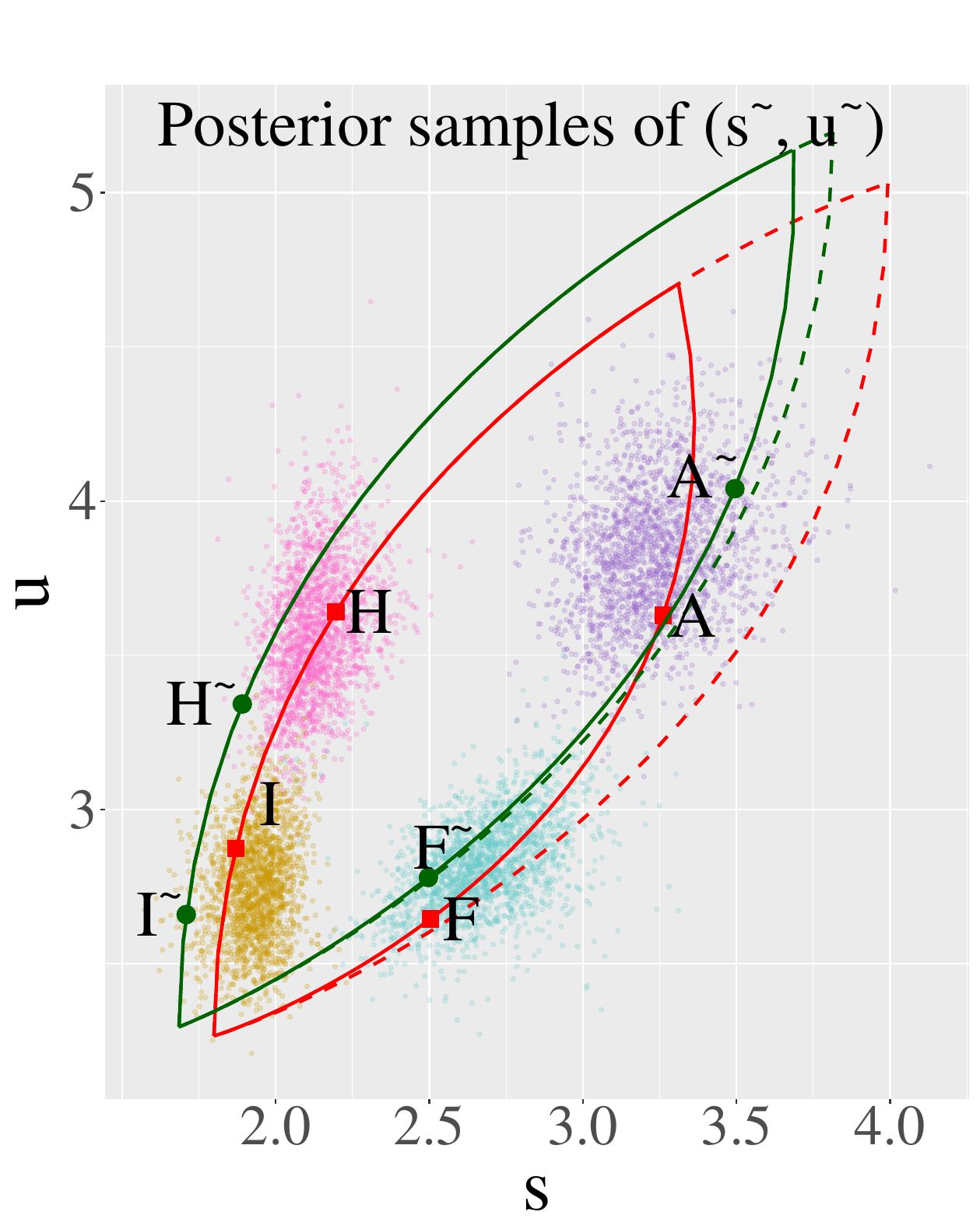}}
	\caption{\textit{BayVel} estimation of \emph{almond} for two simulated genes in the setting where $K = 1, R = 10$. In all the plots, the green \emph{almond} is plotted using the MAP estimates of \textit{BayVel}'s parameters to visualize the estimated dynamic, while the red curve represents the true simulated \emph{almond}, plotted using the true simulated parameters. Left column: variability of the coordinates of the lower (orange) and upper (grey) steady states and of the switching point (blue). Each dot represents a posterior sample. Central column: estimated positions of all subgroups, with green points corresponding to the MAP estimates from \textit{BayVel} and red dotted points representing the simulated true subgroup positions. Right column: variability of four selected subgroup-specific coordinates, where each dot represents a posterior sample. 
	}
	\label{fig:variability}
\end{figure}

\begin{table}[t]
	\small
	\centering
	\begin{tabular}{l|cccccccccccc}
		\hline
		& $\upsilon_{g}^{\text{off}}$ & $\sigma_{g}^{\text{off}}$ & $\upsilon_{g}^{\text{on}}$ &  $\sigma_{g}^{\text{on}}$ &  $u_{k g}^{\omega}$ & $s_{k g}^{\omega}$ 	& $u_{krg}^\sim$ & $s_{krg}^\sim$ & $v_{kr g}$  & $\eta_{g}$ &  $\lambda_{c}$\\
		\hline
		$K = 1, R = 10$ & 0.94 & 0.92& 0.84  &  0.82   & 0.92 & 0.89 & 0.95 & 0.93 & 0.90& 0.94 & 0.95 \\
		$K = 1, R = 30$ & 0.78 & 0.91& 0.78  & 0.77   & 0.87 & 0.83 & 0.95 & 0.93 & 0.90 & 0.95 & 0.95 \\ 
		$K = 1, R = 100$ & 0.62 & 0.80& 0.68  & 0.65   & 0.79 & 0.73  & 0.95 & 0.92 & 0.89 & 0.95 & 0.94 \\
		$K = 10, R = 10$ & 0.94 & 0.89& 0.53  &  0.45  & 0.93  & 0.93 & 0.95 & 0.94& 0.91 & 0.95 & 0.95 \\
		$K = 10, R = 30$ & 0.85 & 0.85& 0.41  &  0.31  & 0.82 & 0.81 & 0.95 & 0.93 & 0.88& 0.95 & 0.94 \\
		$K = 10, R = 100$ & 0.73 & 0.68& 0.30  &  0.21   & 0.67 & 0.67 & 0.93 & 0.90 & 0.86& 0.94 & 0.94 \\
		\hline
	\end{tabular}
	\caption{Percentage of true parameters that fall within the $95\%$ credible interval, as obtained from the posterior distributions estimated by \textit{BayVel} on simulated data.
	}
	\label{tab:percPRESI}
\end{table}

\begin{table}[t]
	\small
	\centering
	\begin{tabular}{l|rrrrr}
		\hline
		$K = 1, R = 10$ & $\upsilon_{g}^{\text{off}}$ & $\sigma_{g}^{\text{off}}$ & $\upsilon_{g}^{\text{on}}$ & $\sigma_{g}^{\text{on}}$ & $v_{rkg}$\\
		\hline
		\textit{BayVel} & $0.103$ & $0.080$ & $0.660$ & $0.963$ & $0.168$ \\
		\textit{scVelo} & $2.031$ & $1.998$ & $5.411$ & $5.624$ & $0.540$ \\
		\hline
	\end{tabular}
	\caption{Median absolute error for \textit{BayVel} and \textit{scVelo} for some the estimated parameters in the $K = 1, R = 10$ simulated scenario.}
	\label{tab:absErr}
\end{table}

\normalsize
\textit{ScVelo} offers only point estimates of the quantities of interest, without providing any measure of uncertainty. In contrast, \textit{BayVel}, operating within a Bayesian framework, provides posterior distributions for all parameters, enabling a two-level discussion of our findings. First, point estimates such as posterior medians or means are extracted and compared to the known true parameters using the relative error, allowing for a direct comparison of \textit{BayVel}’s accuracy relative to \textit{scVelo}. Second, the posterior variability offers valuable insight into the robustness and reliability of the estimates. At the point-estimate level, Figure \ref{fig:phasePlotSIM}
visually evaluates \textit{BayVel}’s ability to reconstruct the \emph{almond} dynamic for three randomly genes. The green structure, derived from maximum a posteriori (MAP) estimates obtained via \textit{BayVel}, closely matches the true \emph{almond} trajectory (in red). Table \ref{tab:relErroBayVel} quantifies these results by reporting median relative errors computed from the posterior medians of \textit{BayVel}'s chains. Overall, the errors remain consistently low, with most cases exhibiting deviations around or below $5\%$, demonstrating that \textit{BayVel} produces estimates that closely approximate the true values. However, a notable exception arises in the coordinates of the upper
steady state, $(\sigma_g^{\text{on}}, \upsilon_g^{\text{on}})$, where the error is higher. The underlying causes of this increased error and its impact on the reconstruction of the \emph{almond} shape are explored in the following discussion. As expected, increasing $K$ reduces precision in the estimation of the switching points $(s^\omega_{kg}, u^\omega_{kg})$, as fewer data points contribute to each parameter. Similarly, when increasing $R$ from $10$ to $100$, the sample size per subgroup decreases from $300$ to $30$, leading to a decline in performance, in line with known identifiability issues \citep{sisNonIdentifiability}. For
\textit{BayVel}-specific parameters, the median relative error for the overdispersion parameter ranges from $1.9\%$ in the simplest setting ($K = 1, R = 10$) to a maximum of $2.4\%$ (for $K = 1, R = 30$ and $K = 10, R = 300$). The capture efficiency remains consistently well estimated, with a median error of $0.9\%$ across all settings, except for $K = 1, R = 10$, where it reaches $1.1\%$. \\
\indent One of the key advantages of \textit{BayVel} over \textit{scVelo} is its ability to generate posterior sampels for all parameters, providing a measure of variability in the estimates. Figure \ref{fig:variability} qualitatively assesses this variability for the coordinates of the steady states and the switching point for two genes (named 3 and 12) in the setting $K = 1, R = 10$. Each point represents a posterior sample, with the green \emph{almond} based on MAP estimates and the red curve representing the true simulated dynamic (as in Figure \ref{fig:phasePlotSIM}). The left panel focuses on the variability of the lower (orange) and upper (grey) steady states, $\mathbf{SS}_g^{\text{off}}$ and
$\mathbf{SS}_g^{\text{on}}$, respectively, as well as the coordinates of the switching point $(s_{kg}^{\omega}, u_{kg}^{\omega})$ (blue). The central panel displays the estimated positions of all subgroups, with green points corresponding to the MAP estimates from \textit{BayVel} and red dotted points representing the true subgroup positions. Finally, the right panel highlights the variability of four selected subgroup-specific coordinates. 
 While the lower steady state is consistently well estimated, the upper steady state of Gene 12 shows greater variability, reflecting the complexity of capturing its dynamics.
For Gene 3 (top row), instead all the other estimates are highly accurate, and therefore also the lower-steady state variability is lower. To explain this behavior we should notice that, while for Gene 3 the subgroup positions span the entire gene’s dynamic range, enabling a more reliable reconstruction of the \emph{almond}, for Gene 12 (bottom row), the true subgroup positions cover only the lower part of the dynamic range. As a result, estimating the remaining portion becomes more challenging, leading to greater uncertainty, particularly in the upper part of the dynamic and in the steady state coordinates. Shifting the upper steady state along the distribution of grey points significantly affects the upper dynamic but has little impact on the lower part. Despite variations in the upper region, subgroup positions in the $(s, u)$-plane remain well estimated, with only a slight increase in variability. The behavior of genes like gene $12$ explains the increased relative error in $(\sigma_g^{\text{on}}, \upsilon_g^{\text{on}})$. Table \ref{tab:percPRESI} reports the percentage of true values falling within \textit{BayVel}'s $95\%$ credible intervals (CI), complemented by Table 2 in the Supplementary Material \citep{supplementary}, which presents the median CI length. Notably, subgroup-specific positions $(s^\sim_{krg}, u_{krg}^\sim)$ fall within the CI in most cases and settings, supporting
\textit{BayVel}’s reliability. The coverage and the median length for the CI of both $s_{kg}^{\omega}$ and $u_{kg}^{\omega}$ clearly reflect the behavior discussed previously and the effect of genes similar to Gene $12$. As expected, increasing the number of subgroups $R$ raises the number of estimated parameters while reducing subgroup sample size, impacting lower steady-state and switching point coordinates but not significantly affecting subgroup-specific positions. \\
\indent A specific comment is needed for the estimation of the RNA velocity. The median relative error in Table \ref{tab:relErroBayVel} ranges from $12.5\%$ in the simplest model ($K = 1, R = 10$) to
$55.8\%$ in the most complex model ($K = 10, R = 100$). While these errors still indicate an improvement over \textit{scVelo}, they are slightly higher than those observed for the other parameters. A contributing factor to this increase is the propagation of the errors from the individual parameters. However, the primary reason lies elsewhere: the true simulated velocity values tend to be smaller on average (for $K = 1, R = 10$, median: $0.289$, $(q_{0.25}; q_{0.75})$: $(-0.400$; $0.9208)$) compared to other parameters. As a result, the relative error raises immediately. Table \ref{tab:absErr} compares the median absolute error of velocity with that of other estimated parameters
($\sigma_g^{\text{off}}, \upsilon_g^{\text{off}}, \sigma_g^{\text{on}}, \upsilon_g^{\text{on}}$) in the setting $K = 1, R = 10$. The absolute error of velocity for \textit{BayVel} ($0.168$) is similar in magnitude to that of the $u$-coordinate of the lower steady state $\upsilon_g^{\text{off}}$ ($0.103$) and significantly smaller than that of $\upsilon_g^{\text{on}}$ ($0.660$). Nonetheless, the relative errors for both $u$-coordinates ($0.048$ and $0.109$ for lower and upper steady state respectively, see Table \ref{tab:relErroBayVel}) are lower than the relative error for velocity. The coverage percentages for velocity, shown in Table \ref{tab:percPRESI}, ranging from
$86\%$ to $91\%$, support the ability of our proposal to recover the true parameters.

\section{Analysis of real data}\label{sec:real}
We analyze a real dataset of pancreatic epithelial cells collected during mouse embryonic development at stage E15.5 \citep{bastidas2019edocrinogenesisDataset}. This dataset, available in the Gene Expression Omnibus (GEO) under accession number GSE132188, has also been studied in \cite{bergen2020generalizing} (accessible through the \textit{scVelo} package via \texttt{scvelo.datasets.pancreas)}. It consists of $C = 3696$ cells, each assigned to one of eight distinct cell types: \textit{Alpha}, \textit{Beta}, \textit{Delta}, \textit{Ductal}, \textit{Epsilon}, \textit{Ngn3 high EP}, \textit{Ngn3 low EP}, and \textit{Pre-endocrine}. We apply both \textit{BayVel} and \textit{scVelo} to this dataset. We first filter out low-expression genes using \textit{scVelo}'s \texttt{scv.pp.filter\_genes} function, obtaining the discrete counts $\{(y_{s, cg}, y_{u, cg})\}_{c, g}$ that are used as input by \textit{BayVel}. To apply \textit{scVelo}, we then follow the standard \textit{scVelo} pipeline and apply the remaining pre-processing steps, getting $\{(m_{s, cg}, m_{u, cg})\}_{c, g}$. In both cases, we retain $G = 2000$ genes for analysis. To evaluate \textit{BayVel}, we explore different numbers of groups and subgroups, consistent with our simulation structure and presented in Section \ref{sec:resBayVel}. \\
\indent Before analyzing performances of \textit{BayVel} and \textit{scVelo} on this real dataset, it is essential to discuss how the results are visualized. In \cite{bergen2020generalizing}, visualization primarily relies on representing gene expression of each cell into a low-dimensional UMAP embedding, though the authors note that their approach is also compatible with other dimensionality reduction techniques such as t-SNE \citep{van2008tsne} and PCA. Within this framework, estimated quantities are visualized in the embedded space, and velocity is represented using the \textit{velocity transition probability} method. However, several critiques \citep{zheng2023pumping} have been raised regarding both the use of UMAP for RNA velocity visualization and the general plotting approach employed by \textit{scVelo}. A key issue is that UMAP is a non-linear transformation, meaning it distorts the relationships between points in the original high-dimensional space when embedding them into a lower-dimensional representation. This distortion becomes particularly problematic when visualizing RNA velocity, which is defined as a first-order derivative. Specifically, the estimated future state of a cell, denoted as $s^{\sim *}_{krg}$, for a given gene $g$, cell group $k$, and subgroup $r$, can be approximated as $s^{\sim *}_{krg} = s^{\sim}_{krg} + v_{krg}dt$.
Here, $dt$ represents the time interval between the estimated current position $s^\sim_{krg}$ and the projected future position $s^{\sim *}_{krg}$. Note that, while this equation is written with \textit{BayVel} notation, it generalizes naturally to the \textit{scVelo} framework.
When a non-linear method such as UMAP is used, the choice of $dt$ has a strong influence on the location of the projected future positions \citep{zheng2023pumping}. Since the UMAP transformation alters distances and relationships between points, even a small change in $dt$ can significantly affect the direction of velocity vectors in the low-dimensional space. Therefore the inferred cell transitions may not reliably reflect the actual developmental dynamics, making velocity plots in UMAP embedding potentially misleading. To mitigate these issues, a linear projection method such as PCA is preferable, since it ensures that different choices of $dt$ only affect the scaling of
projected velocity vectors, without altering their directions. Additionally, Zheng et al. \citep{zheng2023pumping} demonstrate that \textit{scVelo}'s \textit{velocity transition probability} approach further distorts velocity directions. 
As a consequence, the directions of velocity vectors in the low-dimensional space may not accurately reflect the underlying biological dynamics, but instead be artifacts of the embedding itself. This raises additional concerns about the reliability of the evolutionary trajectories presented in \cite{bergen2020generalizing}.\\
\indent To ensure more objective and reliable conclusions, all plots presented in this section are generated using PCA projections and directly projecting the velocity vectors, without the post-processing steps applied by the \textit{velocity transition probability} method. Since RNA velocity is defined as the derivative of the spliced data, the PCA projections will be obtained using just eh spliced values, without directly including the unspliced component.

\subsection{Critical analysis on \textit{scVelo} results}
\begin{figure}[t]
	\centering
	\subfloat[\label{fig:posScVelo}]{\includegraphics[scale = 0.15]{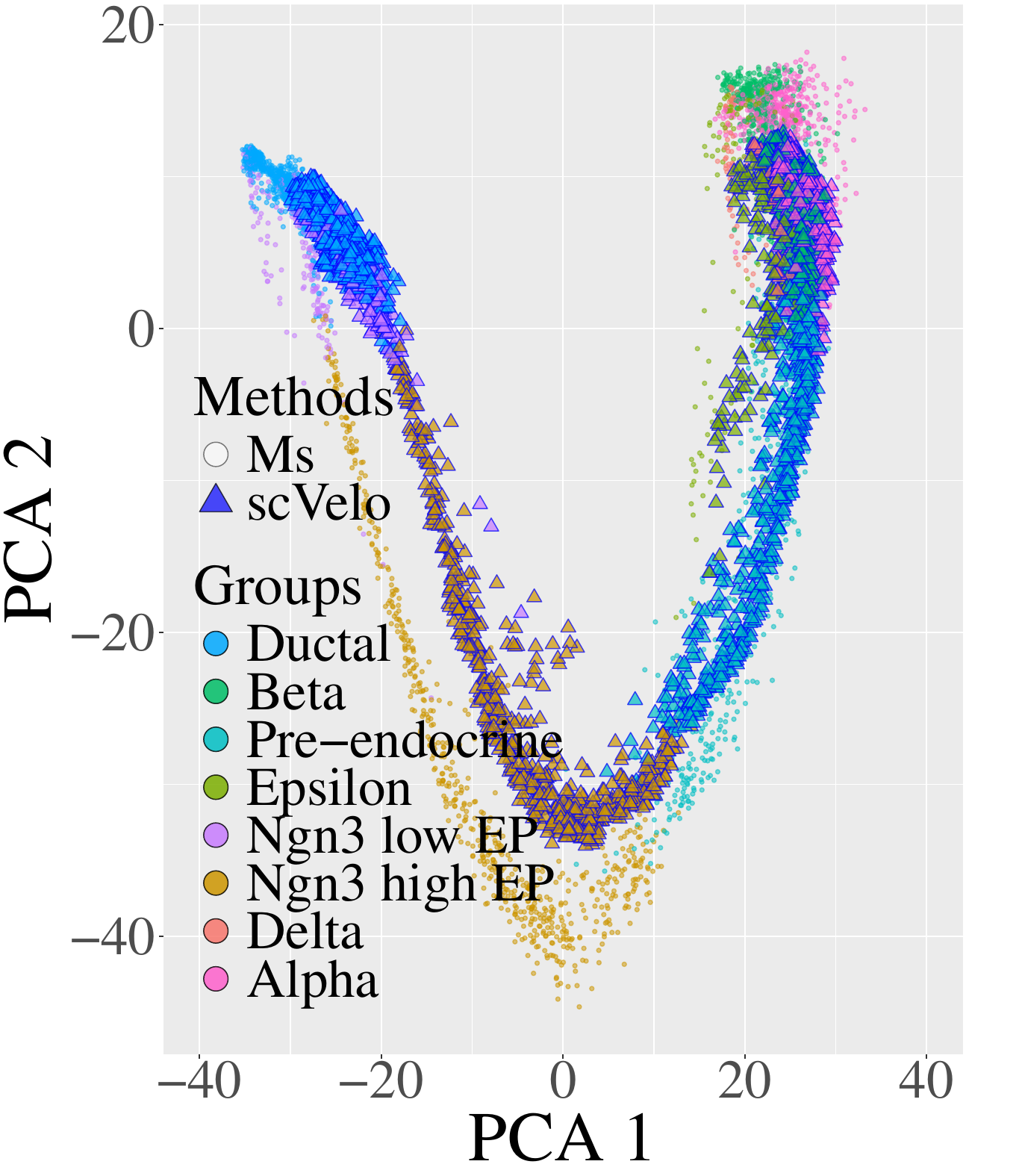}}
	\subfloat[\label{fig:velScVelo}]{\includegraphics[scale = 0.15]{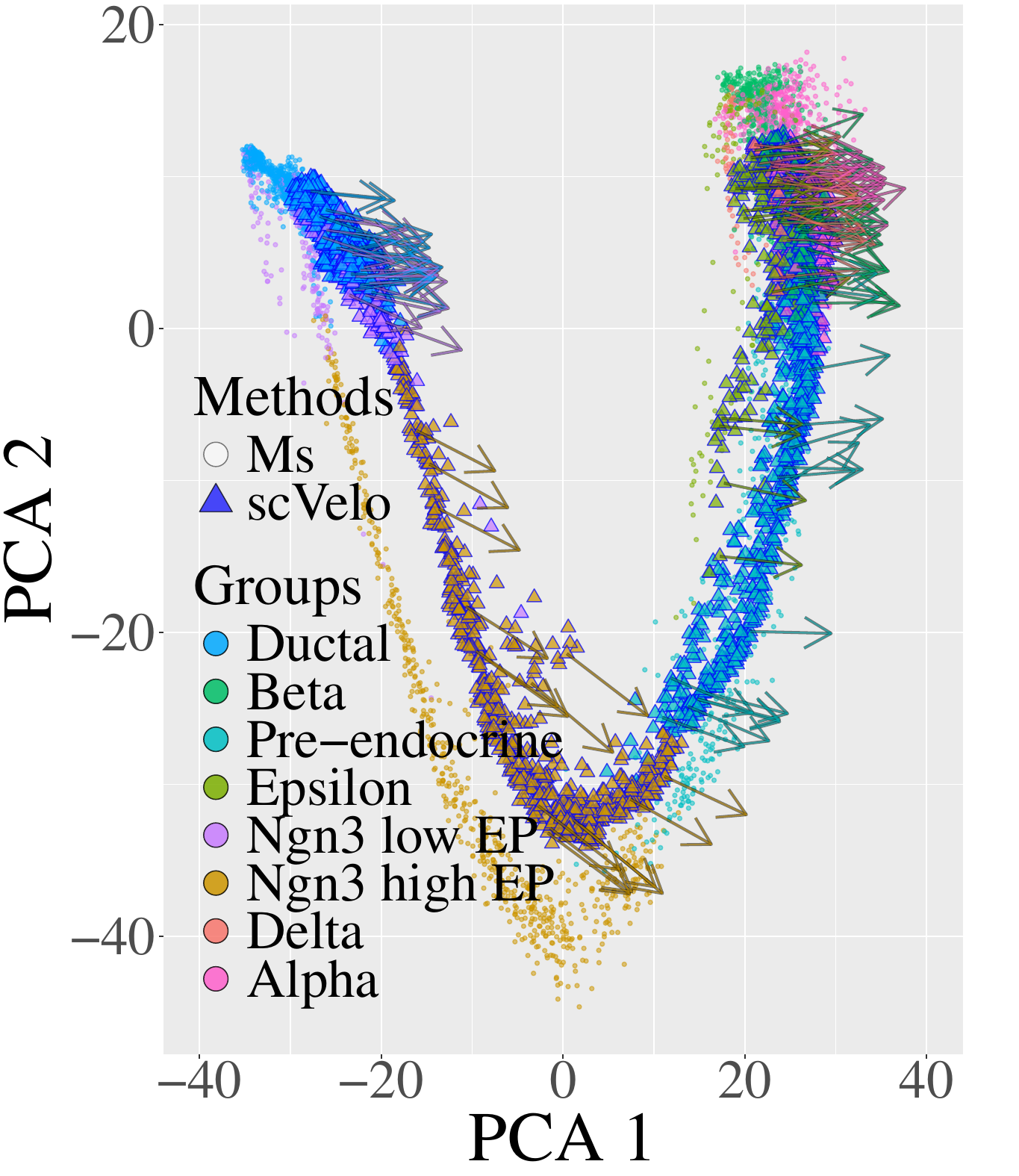}}
	\subfloat[\label{fig:umapScVelo}]{\includegraphics[scale = 0.3]{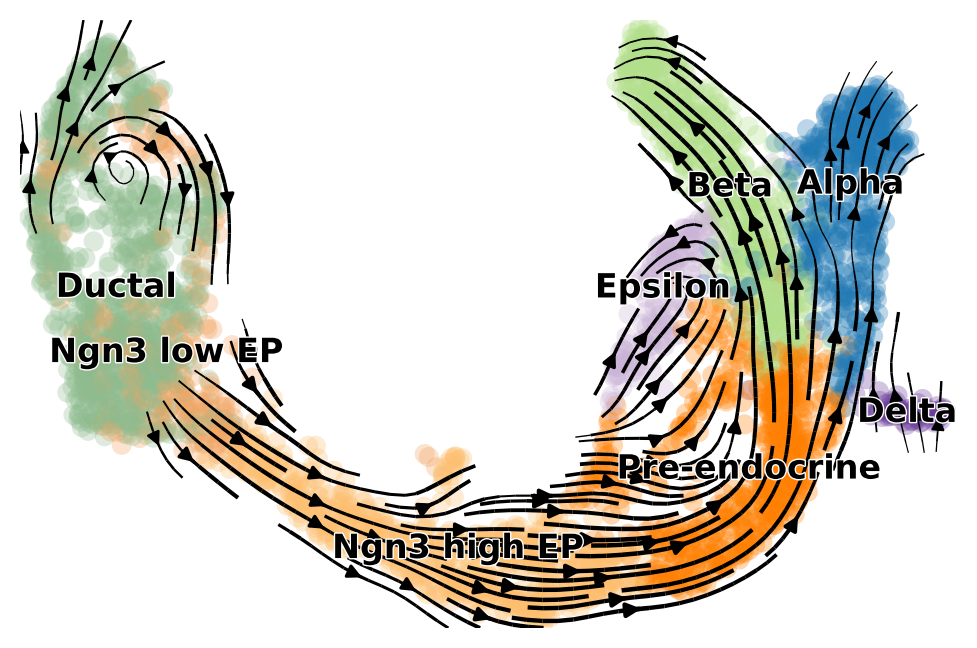}}	
	\caption{Left and Centre: PCA projection along the first and second components of the real pancreatic pre-processed spliced data and the corresponding \textit{scVelo} estimates of cells $s$-coordinates (left) and RNA velocity (centre). We first project the pre-processed spliced observations, $\{\boldsymbol{m}_{s, c} = (m_{s, c1}, \dots, m_{s, cG})\}_{c=1, \dots, C}$, from the pancreatic dataset into a low-dimensional space through PCA, with small dots representing individual cells and colors indicating different cell types. The estimated $s$-coordinates from \textit{scVelo} are projected onto the same PCA space and shown as colored triangles. To visualize velocity, we compute the future state $\tilde{s}^*_{cg}$ of each cell $c$ with $dt = 0.001$ and project it onto the PCA space. To enhance clarity, velocity arrows are normalized to a common magnitude. Given the large number of cells in the dataset, we randomly sample 5\% of the velocity vectors for each cell type to improve visualization. Right: RNA velocity estimated by \textit{scVelo} on the pancreatic dataset, projected with the \textit{velocity transition probability} method on the UMAP low dimensional embedding. The figure is obtained running \textit{scVelo} notebook at \cite{scvelo_notebooks}.}
\end{figure}

When \textit{scVelo}'s results are visualized using linear projection methods and without any post-processing of the velocity vectors, they differ significantly from those presented in \cite{bergen2020generalizing}. We first analyze the cells-specific $s$-coordinates $s(\boldsymbol{\tau}_{cg}^{\text{sc}}, \boldsymbol{\theta}_{g}^{\text{sc}})$.
Figure \ref{fig:posScVelo} presents \textit{scVelo}'s estimation of $s(\boldsymbol{\tau}_{cg}^{\text{sc}}, \boldsymbol{\theta}_{g}^{\text{sc}})$. The figure is generated by projecting the pre-processed spliced expression levels, $\{\boldsymbol{m}_{s, c} = (m_{s, c1}, \dots, m_{s, cG})\}_{c=1, \dots, C}$, onto a low-dimensional space using PCA. We plot the first principal component against the second, with small dots representing observed cells and colors indicating different cell types. The estimated $s$-coordinates from \textit{scVelo} are projected onto the same PCA space and depicted as colored triangles. According to \textit{scVelo}'s model, the estimated positions should align closely with the pre-processed data. However, substantial deviations emerge between the observed data points (dots) and the estimated positions (triangles), particularly in the left and lower regions of Figure \ref{fig:posScVelo}, where the estimated positions fail to overlap with the observed data projection. \\
\indent In terms of estimated RNA velocity, to facilitate comparison with the plots in the original paper, where velocities were projected using UMAP and the \textit{velocity transition probability} method, we show Bergen et coauthors' plot in Figure \ref{fig:umapScVelo}.
In our visualization approach, in Figure \ref{fig:velScVelo}, we compute the future state $\tilde{s}^*_{krg}$ for each cells $c$ with $dt = 0.001$ and then projected onto the PCA space of spliced pre-processed data. To improve clarity, velocity arrows are normalized to a common magnitude. 
Comparing Figures \ref{fig:velScVelo} and \ref{fig:umapScVelo}, it is evident that the velocity arrow directions differ substantially between the two plots. While \cite{bergen2020generalizing} emphasizes that the velocity trajectories mirror biologically expected evolution, this biological sequence is not apparent once the artificial post-processing artifacts of \textit{scVelo} are removed. In particular, the cycling dynamics of \textit{Ductal} cells, identified in Figure \ref{fig:umapScVelo}, are no longer evident in \ref{fig:velScVelo}. While \textit{Pre-endocrine} cells are expected to differentiate into the four terminal cell types (\textit{Alpha}, \textit{Beta}, \textit{Delta}, and
\textit{Epsilon}), this expected trajectory is not reflected in Figure \ref{fig:velScVelo}, as the velocity vectors of \textit{Pre-endocrine} cells do not consistently point toward these terminal states. These discrepancies align with the criticisms raised by \cite{zheng2023pumping}.

\subsection{Results of \textit{BayVel}}\label{sec:resBayVel}
\begin{figure}[t]
	\centering
	\subfloat[\label{fig:r8}]{\includegraphics[scale = 0.15]{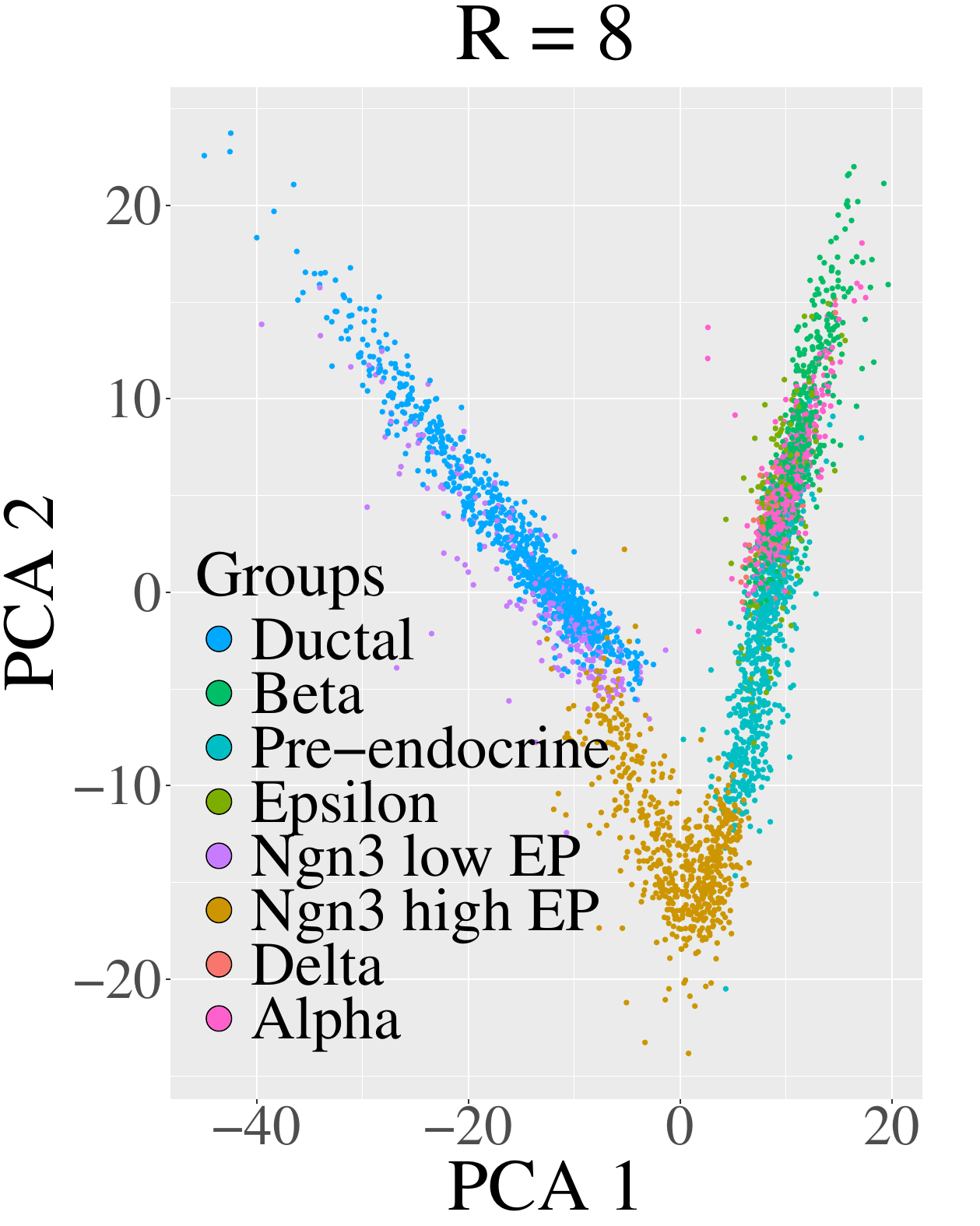}}
	\subfloat[\label{fig:r9}]{\includegraphics[scale = 0.15]{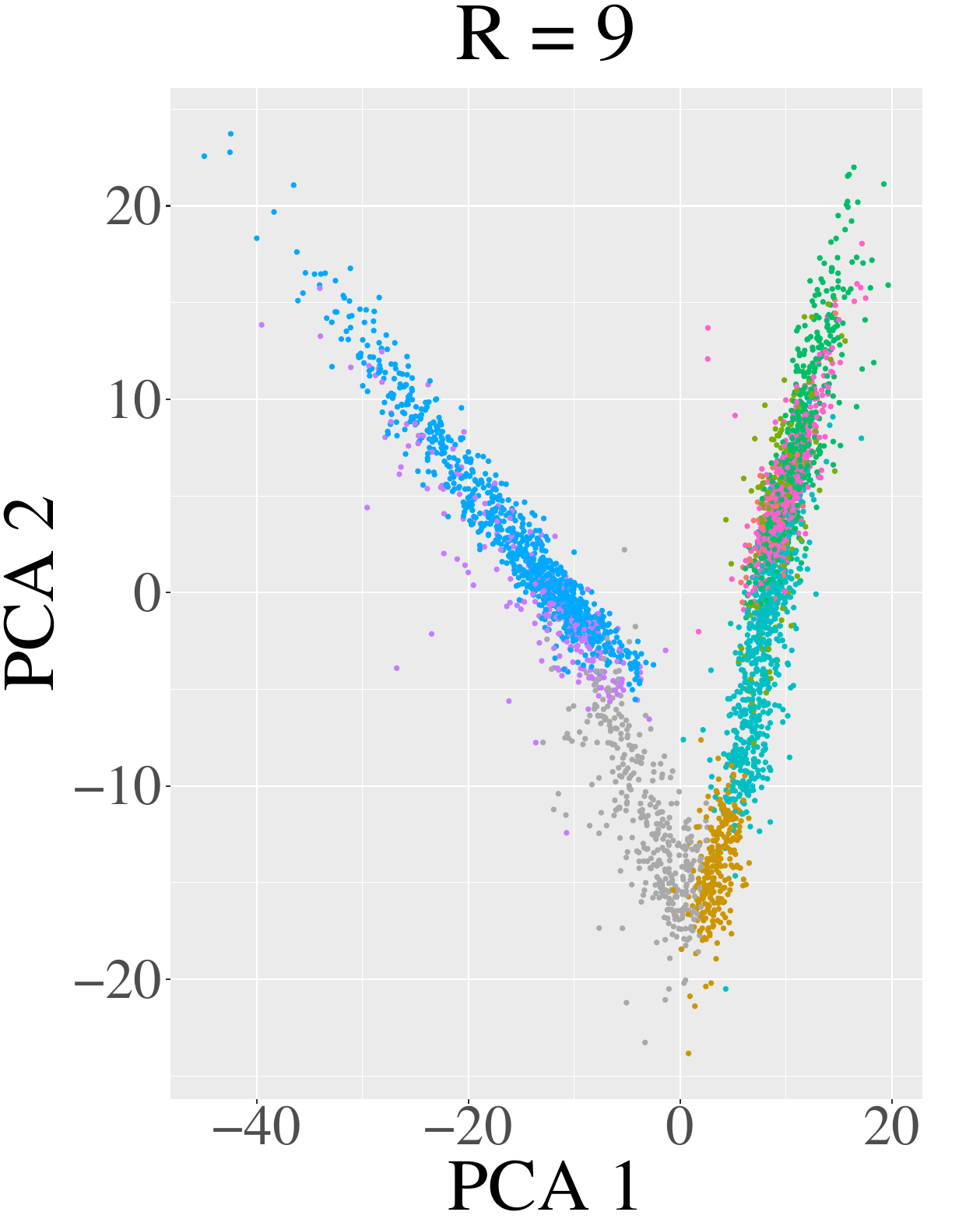}}
	\subfloat[\label{fig:r38}]{\includegraphics[scale = 0.15]{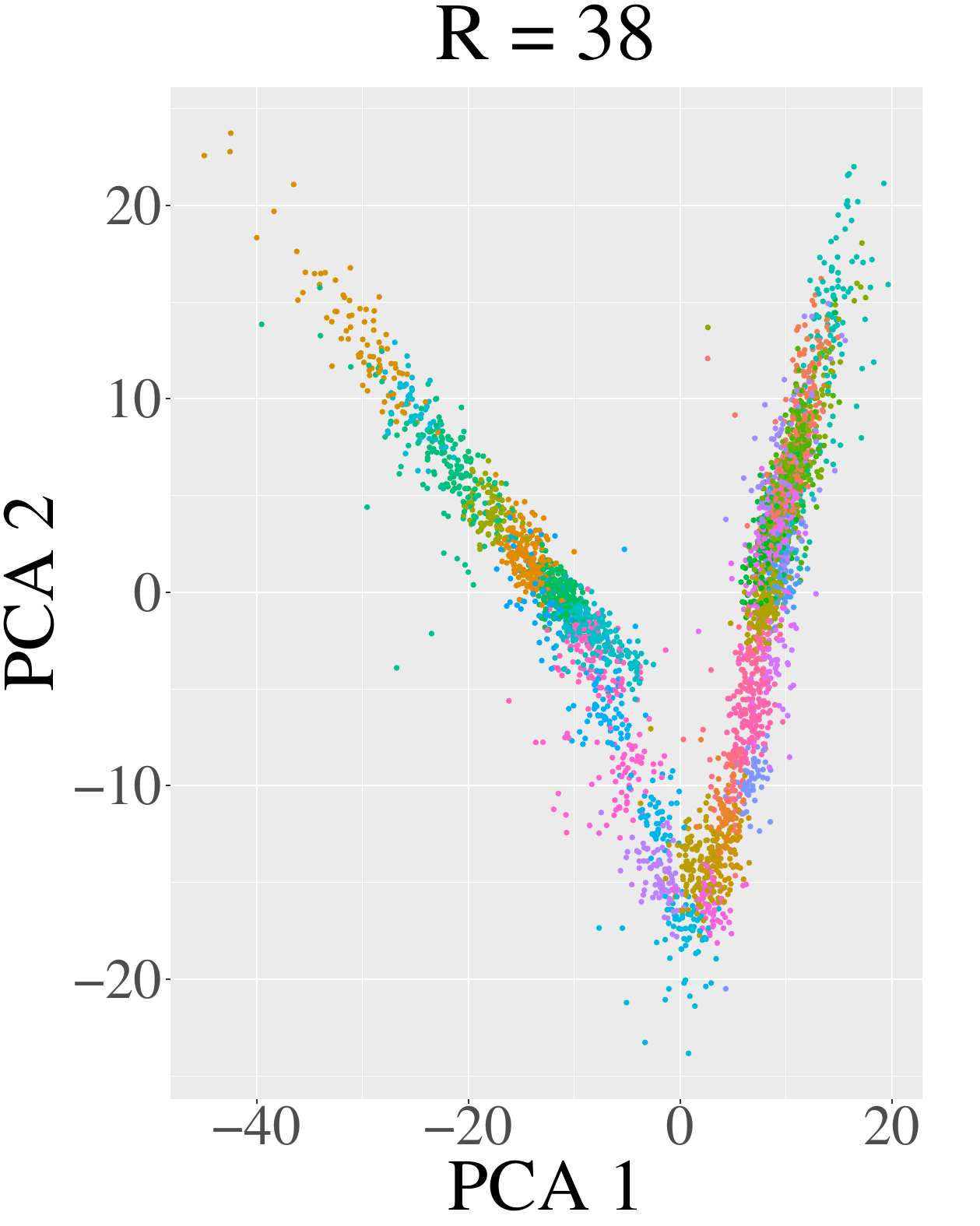}}
	\caption{Subgroups' structure assumed on the real pancreatic data for the different scenarios where \textit{BayVel} is applied. Different colors correspond to different subgroups, while the dots represent the PCA projection of the observed spliced counts $\{y_{s, cg}\}_{c, g}$. The first principal component is plotted against the second. The left panel corresponds to the setting with $R = 8$ subgroups, the central panel to $R = 9$, and the right panel to $R = 38$.}
	\label{fig:subgroups}
\end{figure}

\begin{table}[t]
	\small
	\centering
	\begin{tabular}{l|c}
		\hline
		& WAIC \\
		\hline
		$K = 1, R = 8$   & 15135112.942 \\
		$K = 1, R = 9$   & 15078383.755 \\
		$K = 1, R = 38$  & 15025752.306\\       
		$K = 8, R = 8$  & 15077787.934 \\       
		$K = 8, R = 9$  & 15026882.065 \\ 
		$\mathbf{K = 8, R = 38}$ & \textbf{14959815.879} \\  
		\hline
	\end{tabular}
	\caption{WAIC for the different \textit{BayVel} models on the real pancreatic dataset. The model with the lowest (and better) WAIC ($K = 8, R = 38$) is highlighted in bold.}
	\label{tab:waic}
\end{table}

\normalsize
\begin{figure}[t]
	\centering
	\subfloat[\label{fig:posBayVelGS1-T1-D4-12}]{\includegraphics[scale = 0.15]{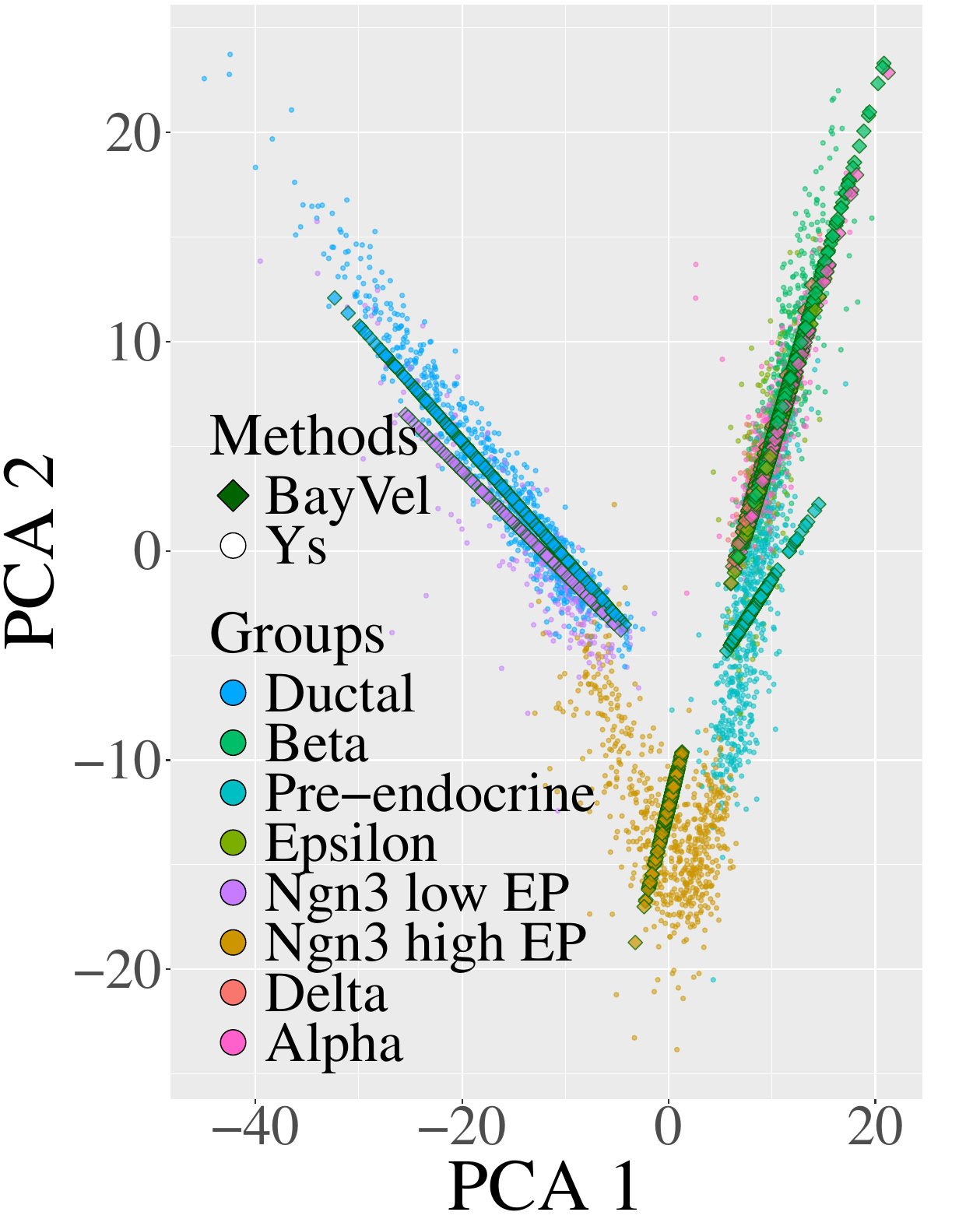}}
		\subfloat[\label{fig:velBayVelGS1-T1-D4-12}]{\includegraphics[scale = 0.15]{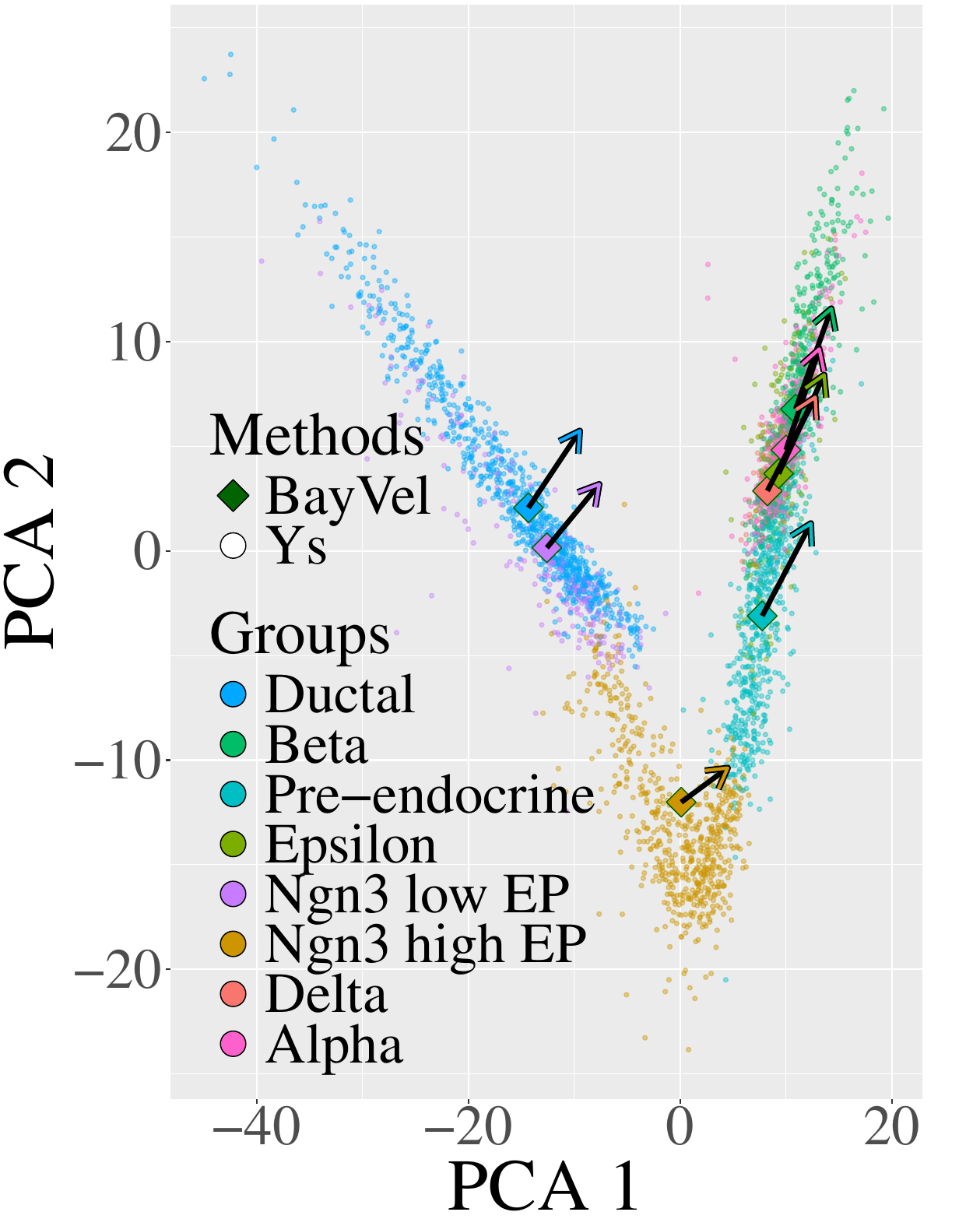}}
	\subfloat[\label{fig:posBayVelGS2-T3-D4-12}]{\includegraphics[scale = 0.15]{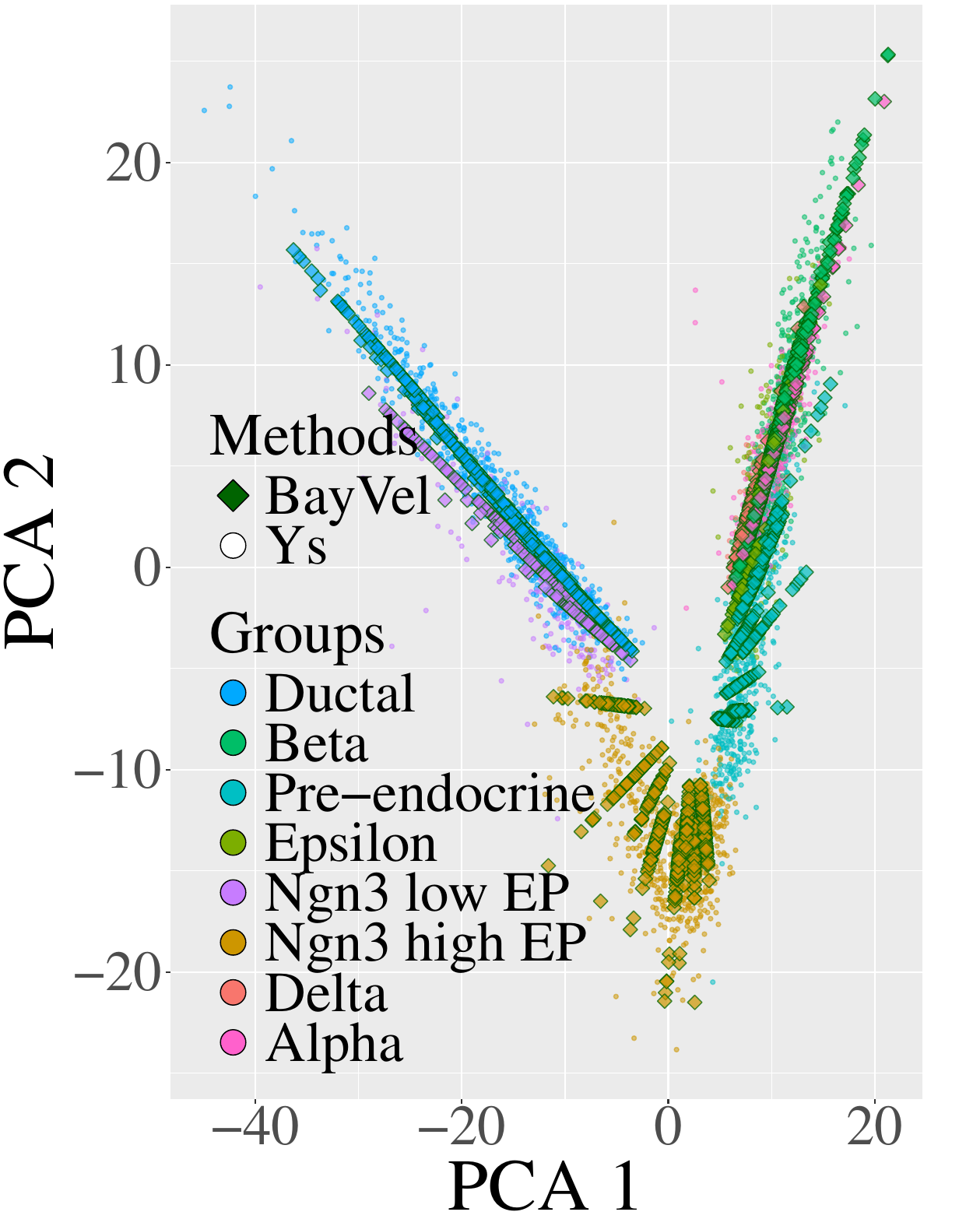}}
	\subfloat[\label{fig:velBayVelGS2-T3-D4-12}]{\includegraphics[scale = 0.15]{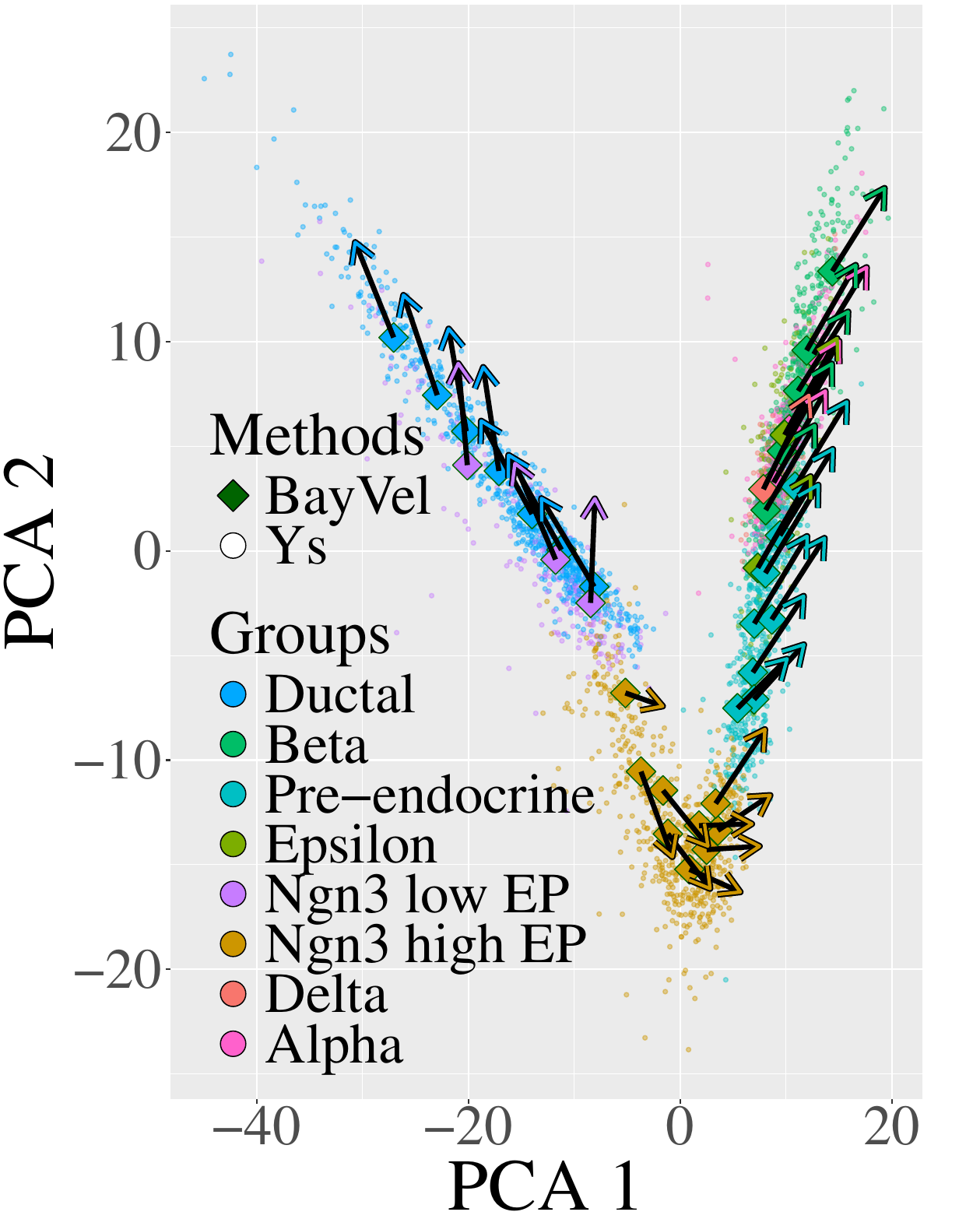}}
	\caption{PCA projection along the first and second components of the real pancreatic raw spliced counts and the corresponding \textit{BayVel} estimates of spliced expected counts (\ref{fig:posBayVelGS1-T1-D4-12}, \ref{fig:posBayVelGS2-T3-D4-12}) and of RNA velocity (\ref{fig:velBayVelGS1-T1-D4-12}, \ref{fig:velBayVelGS2-T3-D4-12}) for model with $K = 1, R = 8$ (\ref{fig:posBayVelGS1-T1-D4-12}, \ref{fig:velBayVelGS1-T1-D4-12}) and for the best WAIC-selected model with $K = 8, R = 38$ (\ref{fig:posBayVelGS2-T3-D4-12}, \ref{fig:velBayVelGS2-T3-D4-12}). We first project the discrete spliced count, $\{\boldsymbol{y}_{s, c} = (y_{s, c1}, \dots, y_{s, cG})\}_{c=1, \dots, C}$, from the pancreatic dataset into a low-dimensional space through PCA, with small dots representing individual cells and colors indicating different cell types. Then we project on the PCA space the MAP estimate of the expected mean of spliced count,
	$\lambda_c s^\sim_{krg}$, obtained with \textit{BayVel} (rectangles). Velocity vectors are computed plotting the future states using MAP estimates of \textit{BayVel} parameters. Due to the linear effect of capture efficiency, we display only one representative velocity arrow for subgroup.}
	\label{fig:pcaPosBayVel}
\end{figure}

\begin{figure}[t]
	\centering 
	\subfloat[\label{fig:putativeScVelo}]{\includegraphics[scale = 0.3]{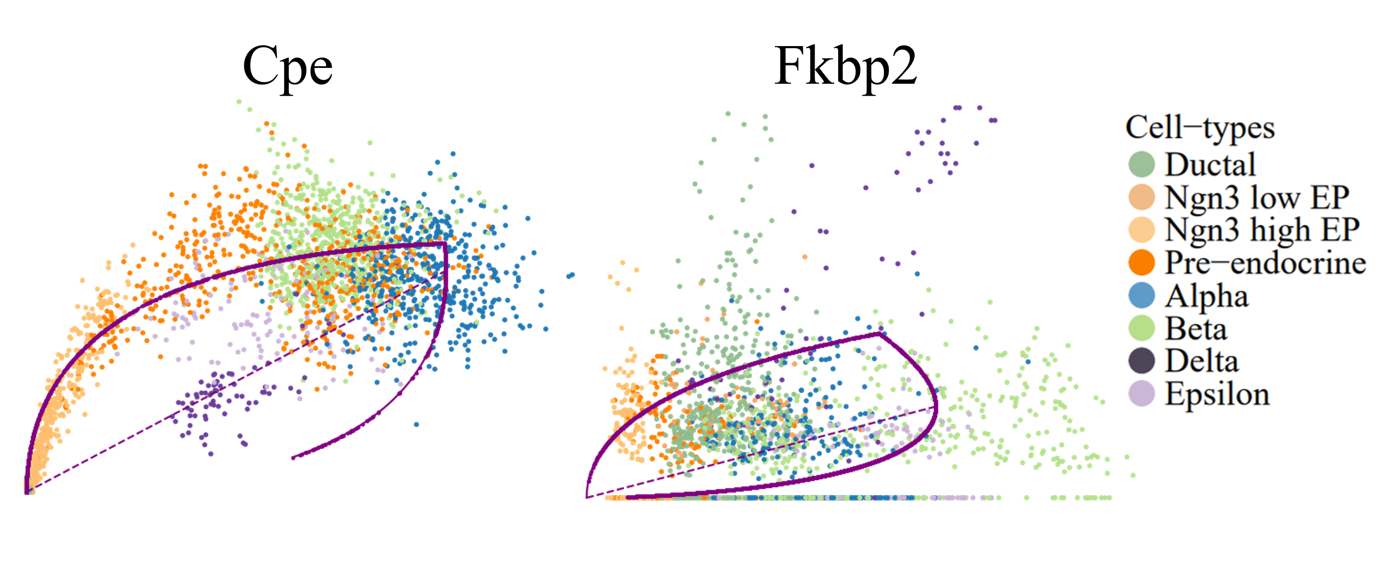}}
	\subfloat[\label{fig:putativeCpeBayVel}]{\includegraphics[scale = 0.15]{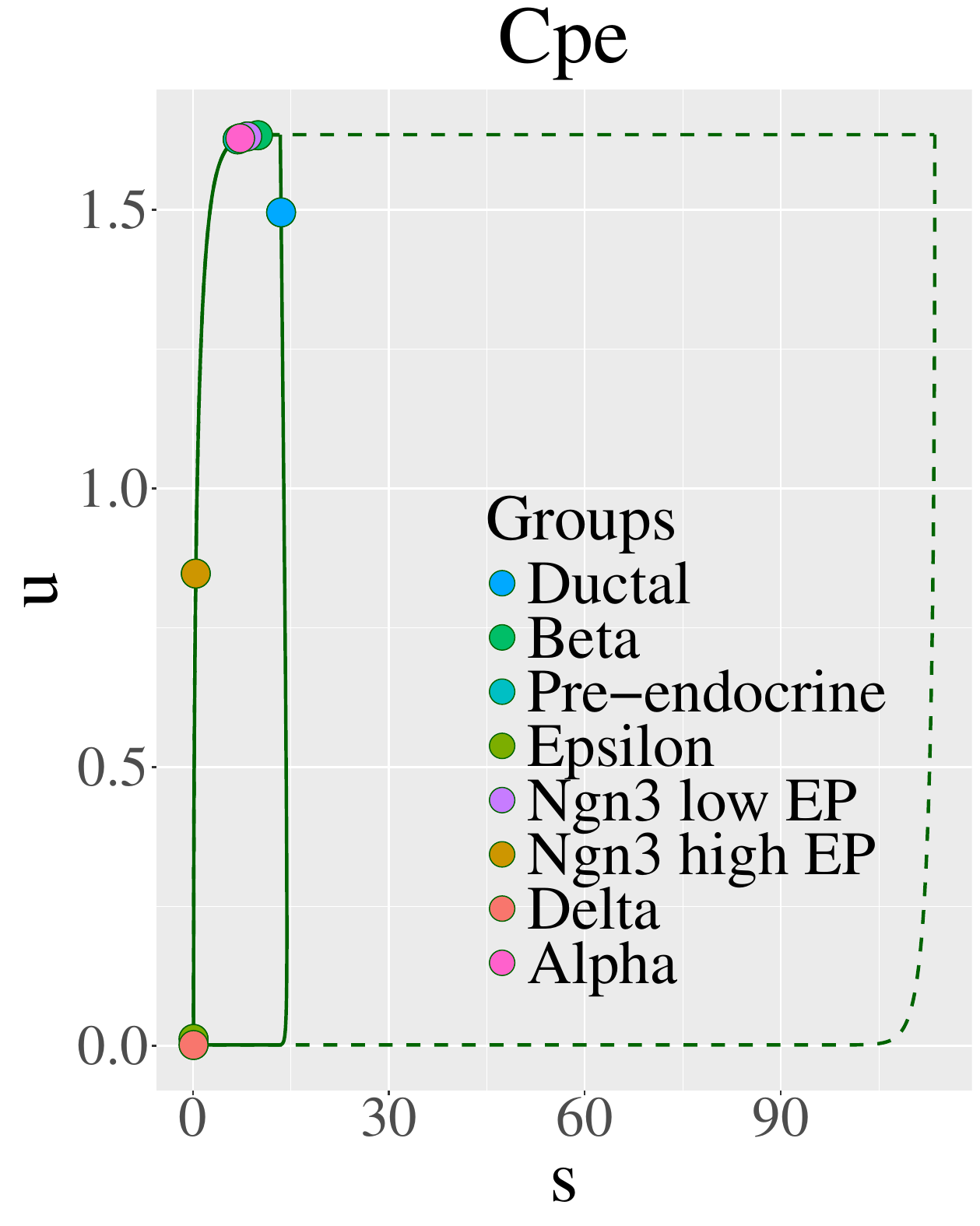}}
	\subfloat[\label{fig:putativeFkbp2BayVel}]{\includegraphics[scale = 0.15]{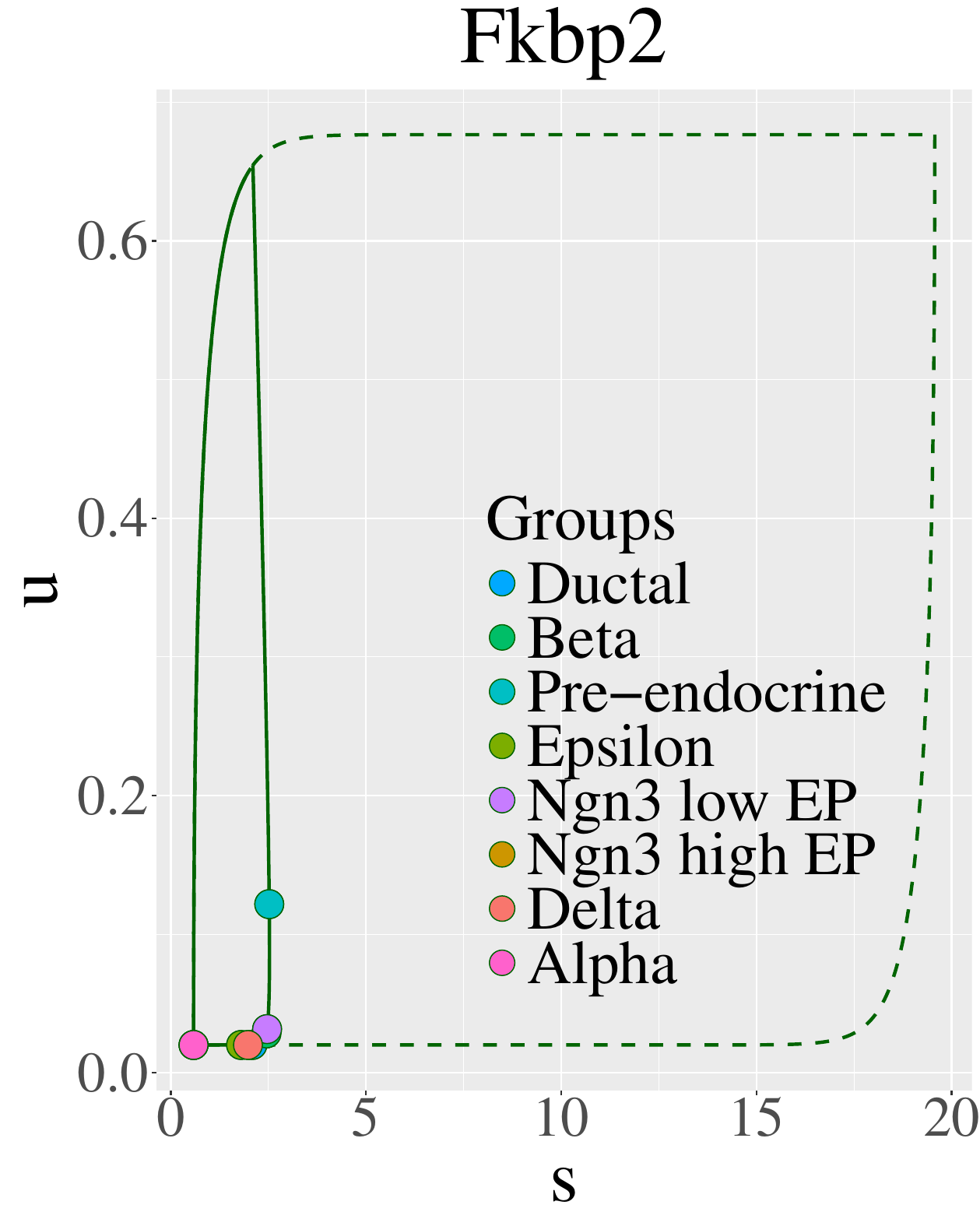}}
	\caption{Gene specific \emph{almonds} for genes Cpe and Fkbp2 estimated by \textit{scVelo} (\ref{fig:putativeScVelo}) and \textit{BayVel} (\ref{fig:putativeCpeBayVel} and \ref{fig:putativeFkbp2BayVel}). The results of \textit{scVelo} are reproduced using the notebook at \cite{scvelo_notebooks}. The colored dots in this figure represents the observed pre-processed data for these genes, the gene dynamic represent \textit{scVelo} estimates. \textit{BayVel}'s \emph{almonds} are obtained using the MAP estimates of the parameters, with points representing the estimated subgroup specific positions.}
	\label{fig:putative}
\end{figure}

\textit{BayVel} is applied on the real dataset in various scenario, exploring different values for the group and subgroup numbers. 
We consider two grouping possibilities, $K = 1$, where all cells belong to the same group, and $K = 8$, where groups correspond to cell-type annotations. For the number of subgroups $R$, we start with a simple setting where all cells of the same type form a single subgroup ($R = 8$). To further refine subgroup definitions, we analyze the distribution of spliced counts ${y_{s, cg}}$ in a low-dimensional space via PCA (Figure \ref{fig:r8}). The curvature in the PCA projection may stem from the "horseshoe effect," a known artifact in scRNA-seq data  \citep{shah2024systematic}. While this can distort true relationships, in our case, PCA results align with those from UMAP, supporting their validity. We therefore proceed with PCA here, though further investigation of this effect may offer additional insights.
 \textit{Ngn3 high EP} cells (shown in ocher) display distinct patterns, living along two separate lines, one with a negative slope and the other with a positive slope (Figure \ref{fig:r9}). This suggests a natural
subdivision of the \textit{Ngn3 high EP} population into two subgroups, leading to a second setting with $R = 9$. Finally, we test \textit{BayVel} in a more granular scenario with $R = 38$ subgroups (Figure \ref{fig:r38}). Here, subgroups are determined via hierarchical clustering: for each cell type, we project the corresponding cells onto the first two principal components, apply clustering, and cut the hierarchy when any subgroup contains fewer than 30 cells. This threshold ensures sufficient data to estimate the model parameters reliably. In total, we apply \textit{BayVel} under six different settings, varying $K \in \{1, 8\}$ and $R \in \{8, 9, 38\}$ and maintaining the same
MCMC-specific parameters as in the simulations (cf. Section \ref{sec:simRNAvel}). To compare these models, we compute the Watanabe–Akaike information criterion (WAIC) \citep{watanabe2010asymptotic}, a generalization of the Akaike information criteria, applicable in the Bayesian framework. 
The model with the lowest WAIC is considered the best. Table \ref{tab:waic} summarizes WAIC scores across different \textit{BayVel}
configurations. The results indicate that the model with $K = 8, R = 38$ provides the best fit, suggesting that this level of granularity captures the data structure most effectively. Additionally, given the same number of subgroups $R$, the settings with group-specific switching points ($K = 8$) consistently achieve lower WAIC values. \\
\indent To visually compare \textit{BayVel}'s results with the real data, we project the discrete counts $\{\boldsymbol{y}_{s, c} = \left(y_{s, c1}, \dots, y_{s, cG}\right)\}_{c = 1, \dots, C}$ onto a low-dimensional space (Figure \ref{fig:pcaPosBayVel}, colored dots). Although alternative projections like Multiple Correspondence Analysis or data transformations (e.g., log or square root) could be more suitable for count data, we apply PCA directly on raw counts to maintain comparability with \textit{scVelo} results. Note that, since \textit{BayVel} operates on discrete counts, its results are mapped to a distinct PCA space compared to \textit{scVelo}. According to Equation \eqref{eq:nb}, for each cell $c$ assigned to group $k$ and subgroup $r$, $\lambda_c s^\sim_{krg}$ is the the expected value of the spliced counts. \textit{BayVel} estimates the posterior distribution of this quantity, we extract its MAP estimate and then project it onto the PCA projection (colored squares in Figure \ref{fig:pcaPosBayVel}) of the observed spliced counts, allowing for direct comparison with real data. The capture efficiency has an important impact in the projected space. Indeed all cells within the same subgroup $r$ share the same MAP estimate of the $s$-coordinate $s^\sim_{krg}$, for each gene $g$. When the estimated capture efficiency is incorporated to compute $\lambda_c s^\sim_{krg}$ for the different cells, these values settle along a line. Because PCA is a linear transformation, this structure is preserved, leading to stripe-like patterns for each subgroup in the projection (Figure \ref{fig:pcaPosBayVel}). Notably, this effect is also observed in the real data projection, where cells of the same type form linear structures in PCA space. This consistency supports the inclusion of capture efficiency in \textit{BayVel}. 
Figure \ref{fig:pcaPosBayVel} presents PCA projections of the first against the second PCA component for the simplest model ($K = 1, R = 8$, Figure \ref{fig:posBayVelGS1-T1-D4-12}) and for the best WAIC-selected model ($K = 8, R = 38$, Figure \ref{fig:posBayVelGS2-T3-D4-12}). In general, \textit{BayVel} estimates align well with the observed data in both settings, particularly for \textit{Ductal}, \textit{Ngn3 low EP}, and terminal cells. Just for \textit{Ngn3 high EP} cells, some discrepancies emerge. The simplest model ($K=1, R=8$) assigns all cells to the same subgroup, preventing it from capturing the two distinct directions along which these cells are distributed. Even the model with $R=9$ (not shown here) does not fully resolve this structure. One possible reason is that the clustering does not optimally separate \textit{Ngn3 high EP} cells, leading to inaccurate position estimates. Specifically, Figure \ref{fig:r9} shows that most cells at the branching point are assigned to the grey subgroup, while the other subgroup fails to fully capture the direction of its cells, particularly without clear terminal reference points. On the other hand, the model with $K=8, R=38$ over-clusters these cells, dispersing them into many subgroups and masking their expected linear structure. Comparing Figures \ref{fig:posBayVelGS1-T1-D4-12} and \ref{fig:posBayVelGS2-T3-D4-12}, the most notable differences in mean estimation between the two models occur for \textit{Ngn3 high EP} and \textit{Pre-endocrine} cells. For other cell types, both models place subgroup positions along the same lines. This suggests that subdividing these particular cell types may not be necessary, as the observed differences in their positions likely stem from capture efficiency rather than distinct biological subgroups. This observation appears to contradict the WAIC findings, which favor models with many clusters. The discrepancy may arises from our choice of subgroup structures analyzed, which may not be the optimal for the data, leading the most granular configuration to seem preferable. More sophisticated clustering approaches might eliminate irrelevant subgroups while still capturing the essential dynamics of the data (see Section \ref{sec:conclusionRNA} for further discussion).
Figures \ref{fig:velBayVelGS1-T1-D4-12} and \ref{fig:velBayVelGS2-T3-D4-12} display the PCA projection of velocity estimates from \textit{BayVel} in setting $K = 1, R = 8$, and $K = 8, R = 38$ respectively. Velocity vectors are computed plotting the future states using MAP estimates of \textit{BayVel} parameters. Due to the linear effect of capture efficiency, cells within the same subgroup exhibit identical velocity directions. To improve clarity, we display only one representative velocity arrow for subgroup. While \textit{BayVel} demonstrates strong performance on simulated data (Section \ref{sec:perfBayVel}), its velocity estimates on real data do not always align with expected cellular evolution. For example, the direction of arrows of \textit{Ductal} and \textit{Ngn3 low EP} fail to consistently point toward any cell types. Moreover, still for these two cellular types, we observe that the arrows are not totally consistent between the two presented scenarios (Figure \ref{fig:velBayVelGS1-T1-D4-12} and \ref{fig:velBayVelGS2-T3-D4-12}). This suggests that, while the model performs well in controlled settings, its biological interpretation may be more complex.\\
\indent \cite{bergen2020generalizing} identifies several putative driver genes, characterized by high inferred likelihoods and expected to exhibit strong dynamic behavior and influence cellular differentiation. In Section 6.1 of Supplementary Material \citep{supplementary}, we compare the gene-specific \emph{almond} inferred by \textit{BayVel} with those from \textit{scVelo} for these genes. However, the absolute value of the likelihood is not a reliable metric for comparing model fits across different genes, as it is not normalized and thus cannot be meaningfully compared across distinct genes. Instead, we turn to PCA to identify genes that best explain variability in the data. Specifically, we again perform PCA on the raw spliced counts $\{\boldsymbol{y}_{s, c} = (y_{s, c1}, \dots, y_{s, cG})\}_{c = 1, \dots, C}$ and select the top five genes with the highest loading values on each of the first three principal components. Figure \ref{fig:putative} presents \emph{almond} estimates for two representative genes, \textit{Cpe} and \textit{Fkbp2}, comparing results from \textit{scVelo} (Figure \ref{fig:putativeScVelo}) and \textit{BayVel} (Figures \ref{fig:putativeCpeBayVel} and \ref{fig:putativeFkbp2BayVel}). The scVelo results were generated using the original notebook provided in \cite{scvelo_notebooks}, where the colored dots correspond to observed, pre-processed data. The dots in the \textit{BayVel}'s figures \ref{fig:putativeCpeBayVel} and \ref{fig:putativeFkbp2BayVel} represent the MAP estimates of the group-specific positions. These two genes represent two distinct patterns observed among the PCA-selected genes. The first behavior, exemplified by \textit{Cpe}, includes genes for which the characteristic \emph{almond} structure is visibly present in the data. In these cases, expression levels vary across different cell types (although formal statistical testing would be needed to confirm the significance of these differences), and cells distribute across the full developmental trajectory, spanning both branches. For instance, \textit{Cpe} shows clearly higher expression in\textit{Alpha}, \textit{Beta}, and \textit{Ngn3 low EP} cells. In contrast, the second group, represented by \textit{Fkbp2}, comprises genes for which the almond pattern is difficult to detect. Their expression data appear scattered and noisy, lacking a clear evolutionary trajectory, and no notable differences between cell types emerge. Here, the estimated latent positions cluster near the lower steady state, suggesting that only a limited portion of the full dynamic process is captured. As a result, reconstructing the unobserved parts of the trajectory becomes challenging. The homogeneity in expression levels across different cell types implies that these genes may play a less significant role in cellular differentiation. This highlights the challenges of interpreting RNA velocity when expression patterns are not clearly observed. Overall, the difficulties in drawing clear biological conclusions from RNA velocity arrows raise concerns about the broader interpretability of RNA velocity. We discuss these challenges further in Section \ref{sec:conclusionRNA}.

\section{Conclusions and future perspectives} \label{sec:conclusionRNA}

In this work we introduced \textit{BayVel}, a novel Bayesian approach for RNA velocity's estimation. Developed in response to \textit{scVelo} criticisms, our method analyzes the identifiability issues of \textit{scVelo} and provides mathematically and biologically justified solutions to these problems. We extend the original method by incorporating group-specific switching points, that better capture the heterogeneous behaviors of the different cellular types. We also introduce a subgroups structure, that enables a more identifiable estimation of elapsed times in cellular dynamics and we eliminate the majority of the artificial pre-processing steps. Indeed, \textit{BayVel} works directly with discrete counts, integrating normalization of this data directly into the model through the capture efficiency parameters. Another important improvement of \textit{BayVel} over \textit{scVelo} is its natural quantification of the variability of the parameters of the model. Our simulation study demonstrates that \textit{BayVel} effectively reconstructs gene expression dynamics with high accuracy, in contrast with \textit{scVelo}, that in many settings struggles to recover the true underlying structure. In general \textit{BayVel} improves the estimations of \textit{scVelo} also on real data, but there are some velocity estimates on the pancreatic data that fail to align with biological expectations. This reinforces the idea that certain key model assumptions may not hold in real datasets and further strengthens broader concerns about RNA velocity’s applicability in capturing true cellular dynamics.
\cite{zheng2023pumping} has already questioned the reliability of RNA velocity in predicting cellular trajectories and differentiation. Our observations strongly support and align with these concerns. The large differences in mean gene expression between cell types challenge the interpretability of RNA velocity. Indeed, Bergen and colleagues present RNA velocity as a continuous vector field, suggesting smooth transitions between cell types. However, this representation contradicts the observed abrupt changes in mean gene expression and the fundamental assumptions of RNA velocity modeling, which relies on a sharp ON/OFF model of regulation of genes. In reality, cellular transitions are often driven by the activation or repression of key regulatory genes, leading to sudden shifts rather than gradual changes. This inherent discreteness in gene regulation conflicts with the continuous framework of RNA velocity plotted in \textit{scVelo}, making it difficult to accurately capture and represent cellular trajectories. \\
\indent One future improvement for \textit{BayVel} relies in how the subgroup structure is decided. Right now, when biological additional information are not available, the subgroups are chosen in an arbitrary way before the estimation. One potential advance would be to integrate subgroup identification directly into the model, eliminating the need for an external clustering step and reducing arbitrariness. This approach could also eliminate redundant subgroups, decreasing the number of unnecessary parameters and helping the model to better individuate subgroups that drive the dynamic. However, incorporating clustering into the model introduces additional complexity to it, and it remains to be determined whether this structure can be reliably identified without creating further identifiability issues.
To conclude, the limitations on RNA velocity we observe on real data are not specific to \textit{BayVel}, that represent a valid improvement over \textit{scVelo}, but rather indicative of broader challenges in RNA velocity approaches, particularly when applied to biological processes characterized by complex dynamics.

\newpage

\part{	\textbf{\large Supplemental Materials to "BayVel: A Bayesian Framework for  RNA Velocity Estimation in Single-Cell Transcriptomics"}}

\section{Solution of the ODE for $\beta = \gamma$}
When $\gamma = \beta$ the solution of the ODE system \eqref{eq:systemODE} can be derived as limit of \eqref{eq:ussol} for $\gamma$ approaching $\beta$. While the form of $u(t,\ton,\omega, \boldsymbol{\theta})$ remains unchanged from \eqref{eq:ussol}, $s(t,\ton,\omega, \boldsymbol{\theta})$ becomes
\footnotesize
\begin{equation} \label{eq:ussolUGUALI}
	s(t,\ton,\omega, \boldsymbol{\theta}) =
	\begin{cases}
		\frac{\aoff}{\beta} & \text{ if }  0 \leq  t < t_{0}^{\text{on}},\\
		\frac{\aoff}{\beta} e^{-\beta \tilde{t}(t)} + \frac{\alpha^{\text{on}}}{\beta}\left[1 - e^{-\beta \tilde{t}(t)} \right] - \left[\alpha^{\text{on}} - \alpha^{\text{off}}\right]\tilde{t}(t)e^{-\beta \tilde{t}(t)}   & \text{ if }   t_{0}^{\text{on}} \leq  t \leq t_{0}^{\text{on}} + \omega, \\
		s^\text{on}\,(t_{0}^{\text{on}} + \omega,\ton,\omega, \boldsymbol{\theta}) e^{-\beta \left[\tilde{t}(t) - \omega\right]} + \frac{\alpha^{\text{off}}}{\beta}\left[1-e^{-\beta \left[\tilde{t}(t) - \omega\right]} \right] +  & \text{ if }   t_{0}^{\text{on}} + \omega < t.\\
		\qquad  - \left[\alpha^{\text{off}} - \beta  u^\text{on}(t_{0}^{\text{on}} + \omega,\ton,\omega, \boldsymbol{\theta}) \right]\left[\tilde{t}(t) - \omega\right]e^{-\beta \left[\tilde{t}(t) - \omega\right]}
	\end{cases} 
\end{equation} 

\subsection{Analytic relationship between $\phi_{rg}$ and the coordinates on the $(s, u)$-plane}
For a fixed subgroup $r$ and gene $g$, we show here the relationship between the angular coordinate $\phi_{rg} \in [0, 2\pi]$, which, given the values of $(\upsilon_g^{\text{off}}, \upsilon_g^{\text{on}},
\sigma_g^{\text{on}})$ and of $u^\omega_{kg}$, determines the coordinates on the $(s, u)$-plane 
\begin{equation}
	\label{eq:fromPhi}
	\begin{aligned}
		&s(\dots) = 
		\begin{cases}
			s^{\text{on}}(\dots) = \left[1 - \frac{\phi_{kg}}{(2\pi-p)/2}\left(\frac{u_{rg}^\omega - \upsilon_g^{\text{off}}}{ \upsilon_g^{\text{on}} -\upsilon_g^{\text{off}}}\right)\right]^{\frac{\gamma_g}{\beta_g}} \frac{\beta_g(\sigma_g^{\text{on}} - \sigma_g^{\text{off}})}{\gamma_g - \beta_g} + \\
			\qquad \qquad   + \sigma_g^{\text{on}}+ \frac{\beta_g \left(\upsilon_g^{\text{off}} - \upsilon_g^{\text{on}}\right)}{\gamma_g - \beta_g}\left[1 - \frac{\phi_{kg}}{(2\pi-p)/2}\left(\frac{u_{rg}^\omega - \upsilon_g^{\text{off}}}{ \upsilon_g^{\text{on}} -\upsilon_g^{\text{off}}}\right)\right]\qquad	& \text{if } \phi_{rg} \in \left[0, \frac{2\pi-p}{2}\right),\\
			s^{\text{off}}(\dots) = \left[1 - \frac{\phi_{kg}-(2\pi-p)/2}{(2\pi-p)/2}\right]^{\frac{\gamma_g}{\beta_g}} \left(s^{\omega}_{kg} - \sigma_g^{\text{off}} + \frac{\beta_g(\upsilon_g^{\text{off}} - u^{\omega}_{kg})}{\gamma_g - \beta_g}\right) \\  
			\qquad \qquad +\sigma_g^{\text{off}} +  \frac{\beta_g \left( u_g^{\omega} - \upsilon_g^{\text{off}}\right)}{\gamma_g - \beta_g}\left[1 - \frac{\phi_{kg} -(2\pi-p)/2}{(2\pi-p)/2}\right]\qquad	& \text{if } \phi_{rg} \in \left[\frac{2\pi-p}{2}, 2\pi - p\right),\\
			\sigma_g^{\text{off}}& \text{if }\phi_{rg} \in [2\pi-p, 2\pi];\\
		\end{cases} \\
		&u(\dots) = 
		\begin{cases} 
			u^{\text{on}}(\dots) = \upsilon_g^{\text{off}} + \frac{\phi_{kg}}{(2\pi-p)/2}\left(u_{rg}^\omega - \upsilon_g^{\text{off}}\right)	& \text{if } \phi_{rg} \in \left[0, \frac{2\pi-p}{2}\right),\\
			u^{\text{off}}(\dots) = u_{kg}^\omega + \frac{\phi_{kg} - (2\pi-p)/2}{(2\pi-p)/2}\left(\upsilon_g^{\text{off}} - u_{rg}^\omega \right)    & \text{if }\phi_{rg} \in \left[ \frac{2\pi-p}{2}, 2\pi-p\right),\\
			\upsilon_g^{\text{off}}& \text{if }\phi_{rg} \in [2\pi-p, 2\pi].\\
		\end{cases}\\
	\end{aligned}
\end{equation}

\section{Induced distributions on the time parameters}
Priors \eqref{eq:priorSwitch} on the switching $u$-coordinate $u_{kg}^\omega$ and on the angular coordinate $\phi_{rg}$ induce the following priors respectively on the switching time $\omega_{kg}$ and on the elapsed time since the beginning of the ON phase $\tilde{t}_{rg}$
\begin{equation}
	\label{eq:priorTswitch}
	f_{\omega_{kg}} (\omega) |\beta_g \propto \beta_g e^{-\beta_g \omega}\mathbb{1}_{[0, +\infty)}(\omega),
\end{equation}
\begin{equation}
	\label{eq:priorT}
	f_{\tilde{t}_{rg}}(t)| \beta_g, \omega_{kg} \propto p\mathbb{1}_{0}(t) + \frac{1-p}{2}\frac{\beta_g e^{-\beta_g t}}{1-e^{\beta_g \omega_{kt}}} \mathbb{1}_{\left(0, \omega_{kg} \right)}(t) + \frac{1-p}{2} \beta_g e^{-\beta_g \left(t - \omega_{kg}\right)}\mathbb{1}_{\left[\omega_{kg}, + \infty \right)}(t).
\end{equation}
It is worth noting that \eqref{eq:priorTswitch} corresponds to an Exponential prior over the switching time (that depends on $\beta_g$ alone, which we fix to 1), while \eqref{eq:priorT} imposes a mixed distribution over $\tilde{t}_{rg}$, consisting of a point mass at the upper steady state (corresponding to the amplitude of the circular sector mapped to $(\upsilon_g^{\text{off}}, \sigma_g^{\text{off}})$ by the transformation \eqref{eq:fromPhi}), and two Exponential distributions (one truncated at $\omega_{kg}$, and the other one translated by $\omega_{kg}$ itself). 

\section{Details on the simulation strategy}
\label{sec:simRNAvel}

\subsection{Generation of the parameters for the synthetic datasets}
\begin{table}[t]
	\small
	\centering
	\begin{tabular}{cccccc}
		$K$ & L&$R$ & \textit{scVelo} & \textit{BayVel}& Cells for subgroup \\
		\hline
		$1$ & 10&$10$ &\checkmark & \checkmark& 300\\
		$1$ &10 &$30$ &\checkmark & \checkmark& 100\\
		$1$ & 10&$100$ &\checkmark & \checkmark&30\\
		$10$ &- &$10$ &$\times$ & \checkmark&300\\ 
		$10$ & -&$30$ &$\times$ & \checkmark&100\\
		$10$ &- &$100$ &$\times$ & \checkmark&30\\
	\end{tabular}
	\caption[Details of a simulated scenario highlighting whether \textit{scVelo} and \textit{BayVel} are applied or not.]{Each row provides the details of a simulated scenarios, highlighting whether \textit{scVelo} and \textit{BayVel} are applied (\checkmark) or not ($\times$) and the sample size that is available to the estimate of the parameters in each subgroup. A total of 3000 cells with 2000 genes are generated in each sample.}
	\label{tab:structSim}
\end{table}

\normalsize
\indent Under the assumptions of both \textit{BayVel} and \textit{scVelo}, the rate parameters governing the ODE are gene-specific. We simulate them, for every gene $g$, as $\alpha_g^{\text{off}}\sim \mathcal{U}(1, 5)$, $\alpha_g^{\text{on}} \sim \mathcal{U}(6, 10)$, and $\gamma_g \sim \mathcal{U}(0.5,1.5)$. Given the invariance $\eqref{eq:inv2}$, to allow a direct comparison between the values estimated by \textit{BayVel} and the ground true, we fix $\beta_g=1$. Note that we do not enforce that $\alpha_g^{\text{off}}=0$, hence the \textit{scVelo} tool will automatically apply the translation \eqref{eq:translation}. 
The assumptions regarding the switching time and the elapsed times differ between \textit{scVelo} and \textit{BayVel} and the parameters of the two approaches are $\boldsymbol{\tau}_{cg}^{\text{sc}} = (\omega_g, \tilde{t}_{cg})$ and $\boldsymbol{\tau}_{krg}^{\text{B2}}=(\omega_{kg}, \tilde{t}_{rg})$, respectively.  \\

\indent To evaluate the two methods on common samples, compatible with both the hypotheses, we simulate for every gene $g$ a unique parameter $\omega_g$. In the notation of \textit{BayVel} this choice correspond to set $K=1$ (only a single group). The durations of the ON phase, $\omega_{g}=\omega_{1g}$, are then chosen uniformly at random from a predefined, equi-spaced grid $\{w_{(1)}, w_{(2)}, \dots, w_{(M)}\}$. The specific choice of this grid is not crucial, as long as the corresponding coordinates $\{(s_g^{w_i}, u_g^{w_i})\}_{i=1}^M$ span the entire ON branch of the dotted \emph{almond}. However, for the sake of reproducibility, we detail the exact
construction used in our numerical experiments. 
The grid is generated iteratively, starting with $w_{(1)} = -\log(1 - 0.3)$. This initialization ensures that, for each gene $g$, the associated $u$-coordinate, $u_g^{w_1}$, is positioned at 30\% of the total range between the lower and upper steady-state $u$-coordinates for each gene $g$. Specifically, 
\begin{equation}
	u^{\omega_1}_g = 0.3(\upsilon_g^{\text{on}} - \upsilon_g^{\text{off}}) + \upsilon_g^{\text{off}}.
\end{equation}
Subsequent points are defined iteratively as $w_{(i)} = w_{(i-1)} + w_{(1)}$, where $i = 2, \dots, 20$. This results in uniformly spaced values along the temporal scale. Due to the nonlinear nature of the ODE solution in Equation \eqref{eq:ussol}, the uniform spacing in time does not translate into uniform spacing in the $u$-coordinate. Specifically, at high values of $\omega_{(i)}$ the corresponding $u$-coordinates become densely concentrated near the upper steady state. To ensure a more balanced distribution, we remove values $\omega_{(i)}$ where the associated $u$-coordinate is too close to the upper steady state, using the threshold
\begin{equation}
	u_g^{\omega_i} <  \upsilon_g^{\text{on}} - 0.005(\upsilon_g^{\text{on}} - \upsilon_g^{\text{off}}). 
\end{equation}
The remaining values after this filtering process constitute the final grid $\{w_{(1)}, w_{(2)}, \dots, w_{(M)}\}$.

Regarding the parameters $ \tilde{t}_{cg}$, the choice of generation mechanism is irrelevant for \textit{scVelo}, since they are estimated individually cell-by cell. In contrast, \textit{BayVel} requires cells to be clustered into groups and subgroups based on their expression levels, with the parameters $\tilde{t}_{rg}$ being specific to each subgroup. While the need for compatibility with \textit{scVelo} prevents us from introducing group-dependent $\omega_g$ parameters, we can still incorporate the group-subgroup structure into the generation of $\tilde{t}_{rg}$. Specifically, for each gene $g$, we generate the
$\tilde{t}_{rg}$ using the following hierarchical structure: we first sample $L= 10$ values $\mu_{lg}$ from the prior distribution of $\tilde{t}$ (Formula \eqref{eq:priorT}) and then, for each $\mu_{lg}$, we generate $R/L$ subgroup specific times $\tilde{t}_{rg}=\max(0, z_{rg})$ where the $z_{rg}\sim \mathcal{N}(\mu_{lg}, 0.5)$. We explore different values for $R$, creating three different sample with $R$ equal respectively to $10$, $30$ and $100$. We do not expect any significant difference in the quality of the performance of \textit{scVelo} under different values of $R$, since the group and subgroup annotation is just ignored by it.\\    
\indent As an additional benchmark, we generate three additional \textit{BayVel} dedicated datasets  with $R \in \{10, 30, 100\}$, where the hierarchical subdivision in groups and subgroups is fully exploited. We fix $K=10$ and sample $\omega_{kg}$ again uniformly at random on the above mentioned grid. For each gene, we generate the $\tilde{t}_{rg}$ in a similar manner as before, but now using the group label $k$ in place of $l$. Specifically, we sample $K$ values $\mu_{kg}$ from the prior distribution of $\tilde{t}$ and then, for each $\mu_{kg}$ we generate $R/K$ subgroup specific times $\tilde{t}_{rg}=\max(0, z_{rg})$, where the $z_{rg}$ are normally distributed around the $\mu_{kg}$, with variance 0.5..
In total, we generate six datasets, three of which are compatible with both \textit{scVelo} and \textit{BayVel}, while the remaining three are tailored explicitly for \textit{BayVel}.
Table \ref{tab:structSim} summarizes all the different explored simulation structures.
To complete the description of parameters simulated for \textit{BayVel}, we sample the overdispersion parameter $\eta_g \sim \mathcal{U}(0.5, 1)$, and the capture efficiency $\lambda_c \sim \mathcal{U}(0.5, 1)$ and to address the identifiability issue we normalize the generated values $\lambda_1, \dots, \lambda_C$ so that they have a sample mean of 1. Taking care to not
change the data likelihood, we simultaneously scale both $\alpha_g^{\text{off}}$ and $\alpha_g^{\text{off}}$ accordingly, see equation \eqref{eq:invCapture}.  \\

\subsection{Simulating the Independent Normal data}
The variances of $m_{s, cg}$ and $m_{u, cg}$ used during the simulation of the Independent Normal (IN) data are
\begin{equation} \label{eq:varIN}
	\sigma^2_s = \frac{0.8}{10} q_{0.99}\left( {s(\boldsymbol{\tau}_{cg}^{\text{sc}}, \boldsymbol{\theta}{_g}^{\text{sc}})}_{c, g}\right), \quad \sigma^2_u = \frac{0.8}{10} q_{0.99}\left( {u(\boldsymbol{\tau}_{cg}^{\text{sc}}, \boldsymbol{\theta}_{g}^{\text{sc}})}_{c, g}\right),
\end{equation}
with $q_\alpha$ denoting the quantile of order $\alpha$. It is straightforward to verify that the data generated in this way do not align with the statistical model used in \textit{scVelo}. 

\subsection{Simulating the Deming data}
To simulate Deming (\textit{Dem}) data, we leverage a key property of \textit{scVelo}'s likelihood. Specifically, we consider a circle centered at $(s^\sim_{krg}, u^\sim_{krg})$ with a fixed radius. All points on the right semicircle correspond to the same value of the associated Deming residual and, consequently, share the same likelihood. Similarly, points on the left semicircle have a Deming residual of the opposite sign but still exhibit the same likelihood. This symmetry implies that within each semicircle, all points are equally probable under the likelihood model, making their exact position along the semicircle irrelevant to the probability distribution. To generate the simulated data while preserving this structure, we first sample an angle $\psi_{rg}$ uniformly from $\left(\frac{\pi}{2}, \frac{3\pi}{2}\right)$, ensuring that the initial points lie within the right semicircle. We then introduce controlled variability by drawing a radial displacement $\rho_{rg} \sim \mathcal{N}(0, \sigma^2)$, allowing for deviations from the semicircle while maintaining a natural spread of the data. Importantly, because $\rho_{rg}$ follows a symmetric normal distribution centered at zero, the left semicircle is also explored, ensuring that points are generated on both sides of the circle in accordance with the likelihood properties. The final coordinates of the simulated data points are given by 
\begin{equation}
	\label{eq:demingsSim}
	m_{s, cg} = \rho_{rg} \cos(\psi_{rg}) + s^\sim_{krg}, \quad  
	m_{u, cg} = \rho_{rg} \sin(\psi_{rg}) + u^\sim_{krg}.
\end{equation}
This formulation ensures that the generated data naturally respects the likelihood structure. The variance $\sigma^2$ is set as the mean of $\sigma^2_s$ and $\sigma^2_u$ (see Equation \eqref{eq:varIN}), which were used in the simulation of \textit{IN} data.

\subsection{Details on the MCMC}
For every gene $g$ the parameters $\upsilon_g^{\text{off}}, \upsilon_g^{\text{on}}, \sigma_g^{\text{on}}$ are proposed altogether with a multivariate normal with an adaptive covariance matrix, following \cite{andrieu2008adaptiveMCMC} (algorithm 4). The other parameters are proposed independently with an adaptive variance, as suggested in \cite{Robert}. The adaptation of the variance and covariance matrix begins at the $100$-th iteration and ends after $\left\lfloor 0.9 n_\text{burn.in} \right\rfloor$ iterations, where $n_\text{burn.in}$ is the total number of burn-in iterations. The adaptation gradually diminishes according to the factor $\Gamma_k = \frac{2500}{k + 20000}$, where $k$ is the current iteration. The target
acceptance probability $\alpha^*$ set to $0.25$. Additionally, for univariate proposals, variances are adapted every $100$ iterations. 

\section{\textit{BayVel}'s results on simulated data}
Table \ref{tab:CI} of \textit{BayVel} presents the median length of the $95\%$ credible intervals (CI) obtained from the posterior distribution of parameters estimated by \textit{BayVel} on the different simulated datasets. 
\begin{table}[t]
	\centering
	\begin{tabular}{l|ccccccccccccc}
		\hline
		& $\upsilon_{g}^{\text{off}}$ & $\sigma_{g}^{\text{off}}$ &$\upsilon_{g}^{\text{on}}$  & $\sigma_{g}^{\text{on}}$ &  $u_{k g}^{\omega}$ & $s_{kg}^{\omega}$ &  $u_{krg}^\sim$ & $s^\sim_{krg}$ & $v_{krg}$ & $\eta_{g}$ &  $\lambda_{c}$ \\
		\hline
		$K = 1, R = 10$ & 0.59 & 0.46 & 3.13 & 4.33 & 1.02 & 0.96 & 0.75 & 0.59 & 0.91 & 0.08 & 0.06 \\     
		$K = 1, R = 30$& 0.46 & 0.42 & 2.51 & 3.79 & 0.83 & 0.88 & 1.13 & 0.89 & 1.18 & 0.08 & 0.06 \\     
		$K = 1, R = 100$ & 0.38 & 0.37 & 2.33 & 4.21 & 0.68 & 0.74 & 1.54 & 1.19 & 1.67 & 0.06 & 0.05 \\ 
		$K = 10, R = 10$ & 0.60 & 0.40 & 3.43 & 4.17 & 2.87 & 2.99 & 0.67 & 0.60 & 1.04 & 0.06 & 0.05 \\    
		$K = 10, R = 30$ & 0.40 & 0.35 & 2.14 & 2.80 & 1.65 & 1.82 & 1.01 & 0.88 & 1.10 & 0.06 & 0.05 \\
		$K = 10, R = 100$ & 0.38 & 0.34 & 1.92 & 2.81 & 1.31 & 1.36 & 1.54 & 1.27 & 1.56 & 0.06 & 0.05 \\
		\hline
	\end{tabular}
	\caption{Median length of the $95\%$ credible intervals obtained from the posterior distributions estimated by \textit{BayVel}. The rows correspond to different simulated datasets, while the columns represent the various parameters defining the underlying dynamics.}
	\label{tab:CI}
\end{table}

\section{\textit{BayVel}'s results on real data}
\begin{figure}[t]
	\centering
	\subfloat[\label{fig:velBayVelGS1-T1-D4-13}]{\includegraphics[scale = 0.25]{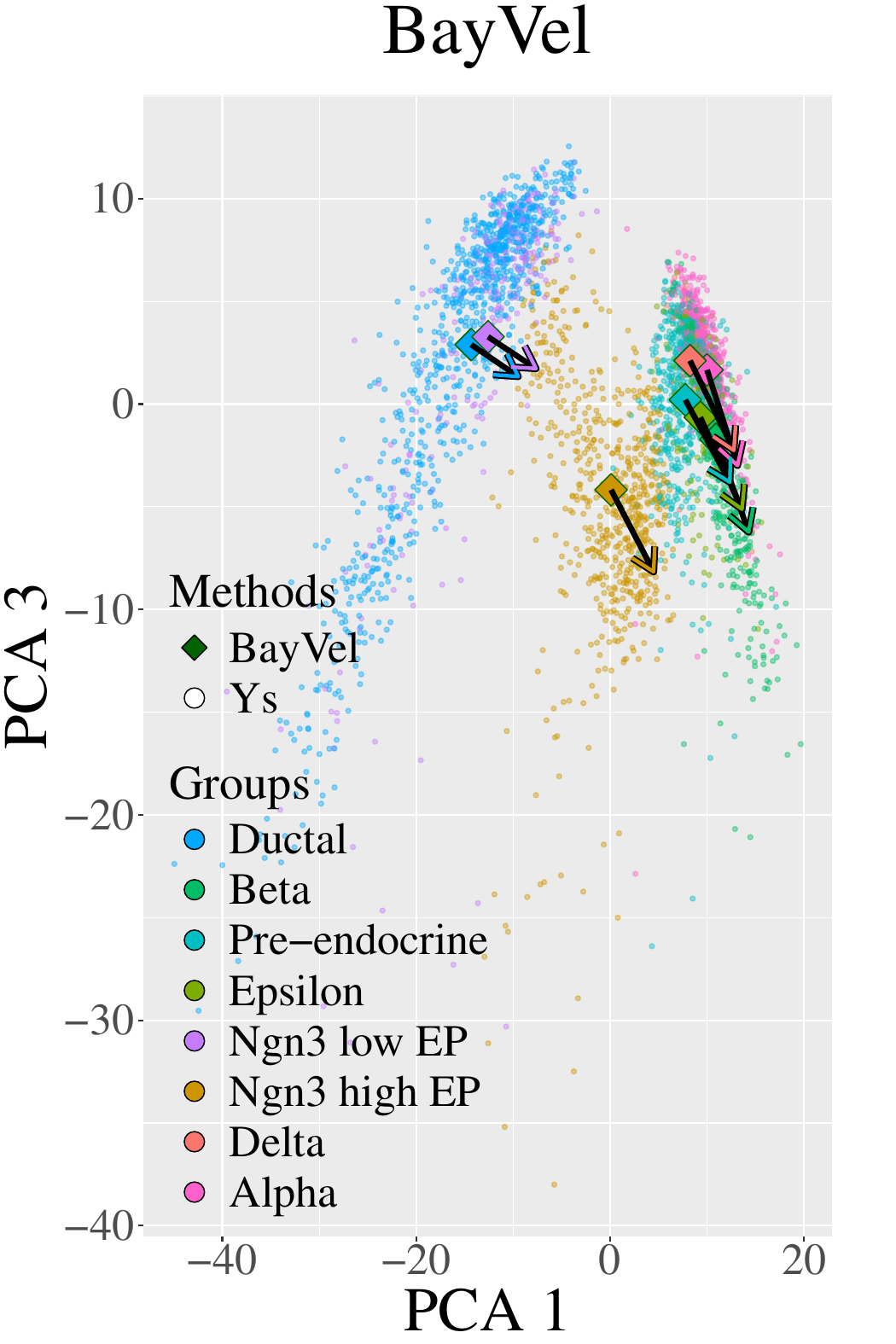}}
	\subfloat[\label{fig:velBayVelGS1-T1-D4-14}]{\includegraphics[scale = 0.25]{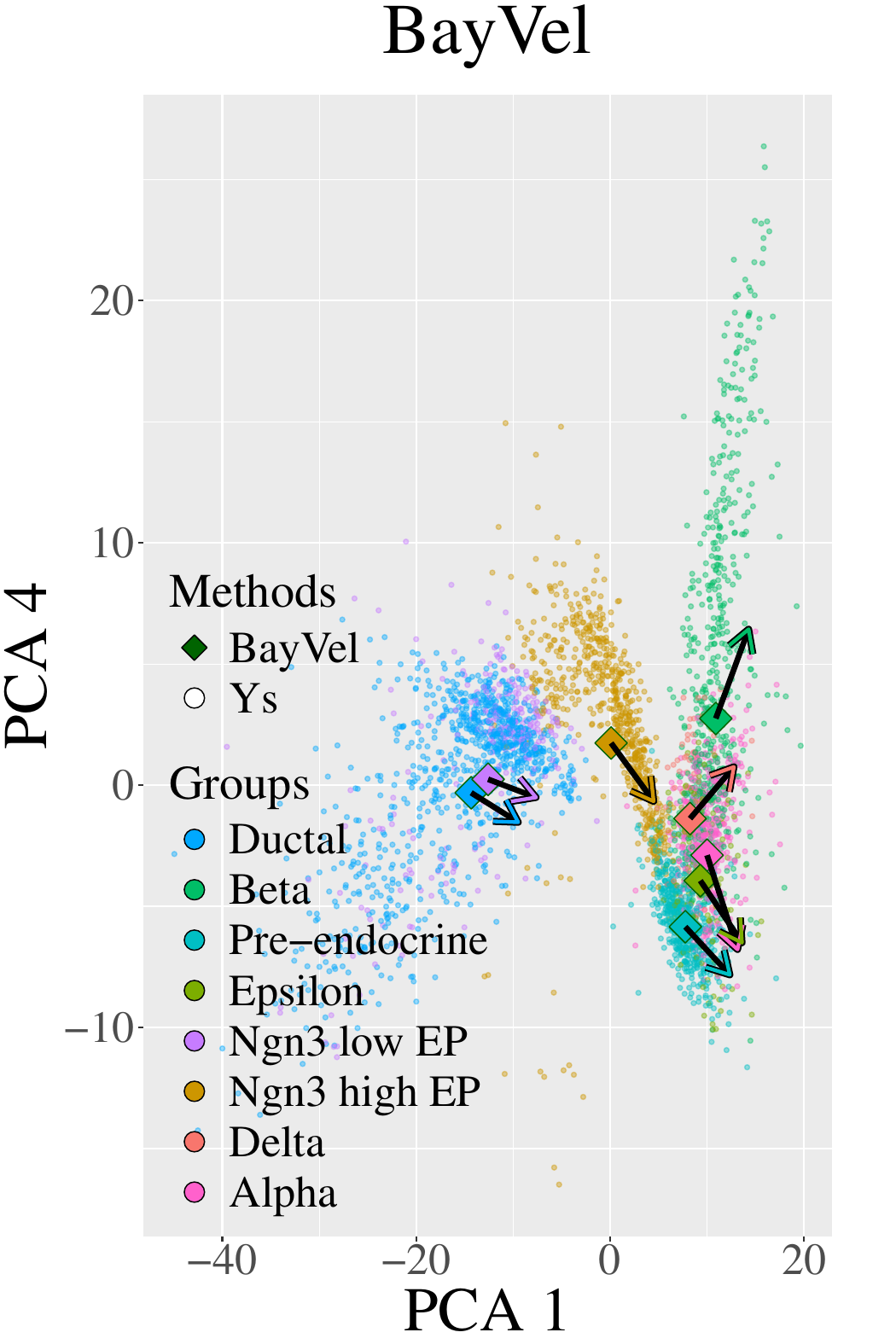}}\\
	\subfloat[\label{fig:velBayVelGS2-T3-D4-12}]{\includegraphics[scale = 0.25]{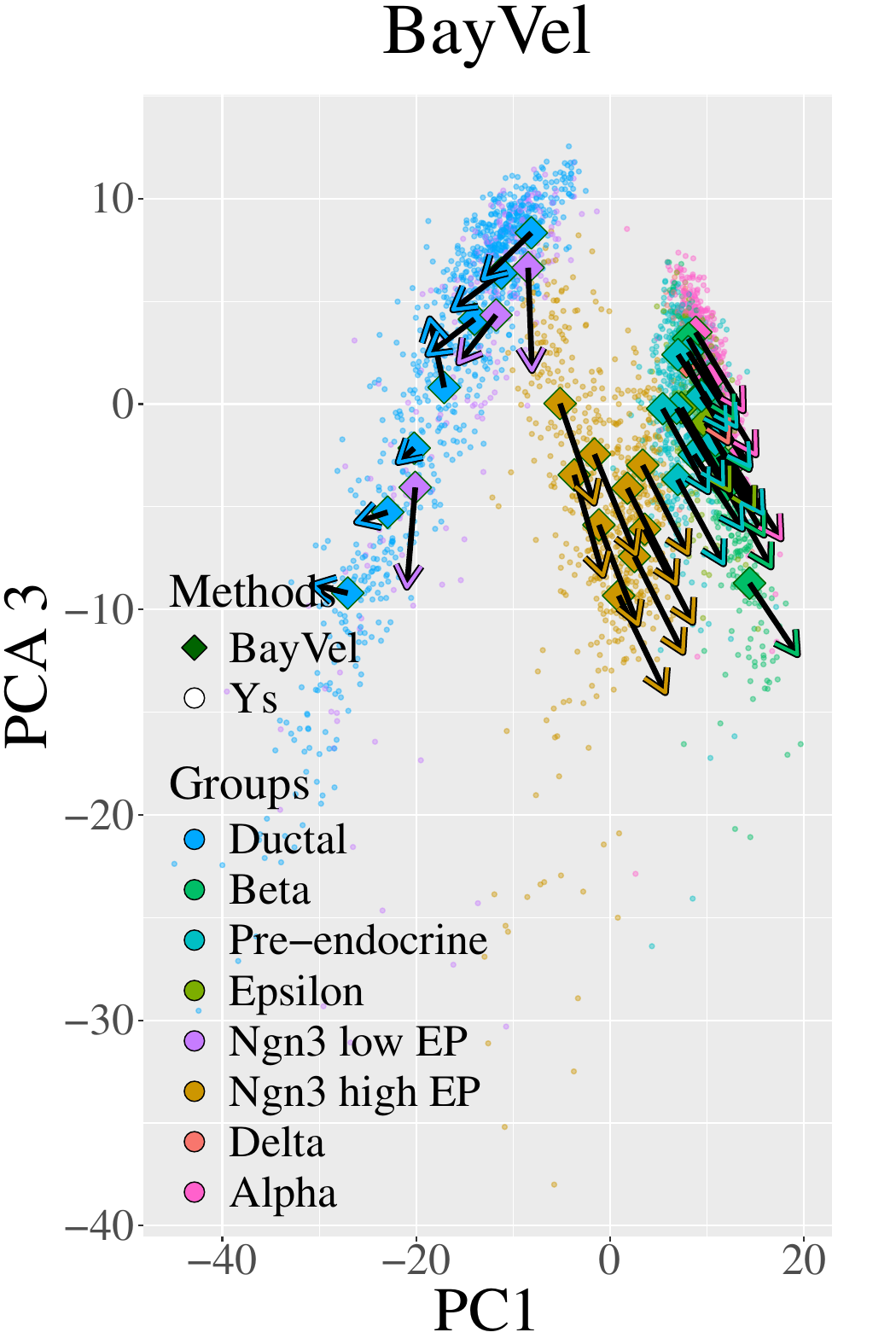}}
	\subfloat[\label{fig:velBayVelGS2-T3-D4-14}]{\includegraphics[scale = 0.25]{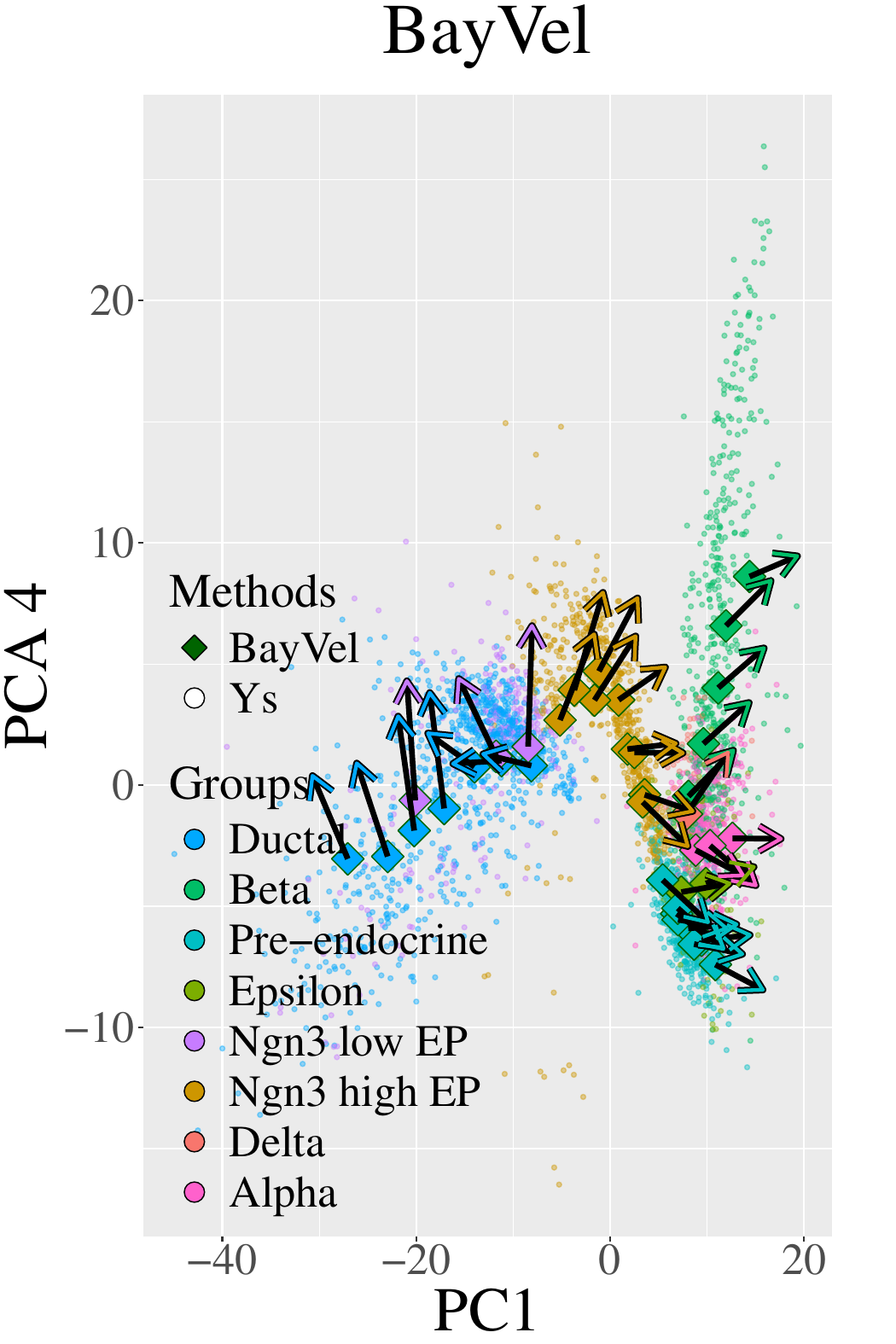}}
	\caption{PCA projection of the pancreatic cell data and the corresponding \textit{BayVel} estimates of RNA velocity for model with $K = 1, R = 8$ (first row) and for the best WAIC-selected model with $K = 8, R = 38$ (second row). We first project the discrete spliced count, ${\boldsymbol{y}_{s, c} = (y_{s, c1}, \dots, y_{s, cG})}_{c=1, \dots, C}$, from the pancreatic dataset into a low-dimensional space through PCA. We plot the first PCA component against the third (first column) and the fourth one (second column). Small dots represent individual cells, with colors indicating different cell types. Velocity vectors are computed plotting the future states using MAP estimates of \textit{BayVel} parameters. Due to the linear effect of capture efficiency, we display only one representative velocity arrow for subgroup.}
	\label{fig:pcaVelBayVel}
\end{figure}

Figure \ref{fig:pcaPosBayVel} represents the PCA projection of pancreatic real cells, comparing first and third PCA components (first column) and first and fourth components (second column). The first row presents the estimates of \textit{BayVel} of the expected spliced counts for model $K = 1, R = 8$ and the second for model $K = 8, R = 38$. Figure \ref{fig:pcaVelBayVel} represents \textit{BayVel} estimates of RNA velocity, again comparing first and third PCA components (first column) and first and fourth components (second column). 

\begin{figure}[t]
	\centering
	\subfloat[\label{fig:posBayVelGS1-T2-D4-13}]{\includegraphics[scale = 0.25]{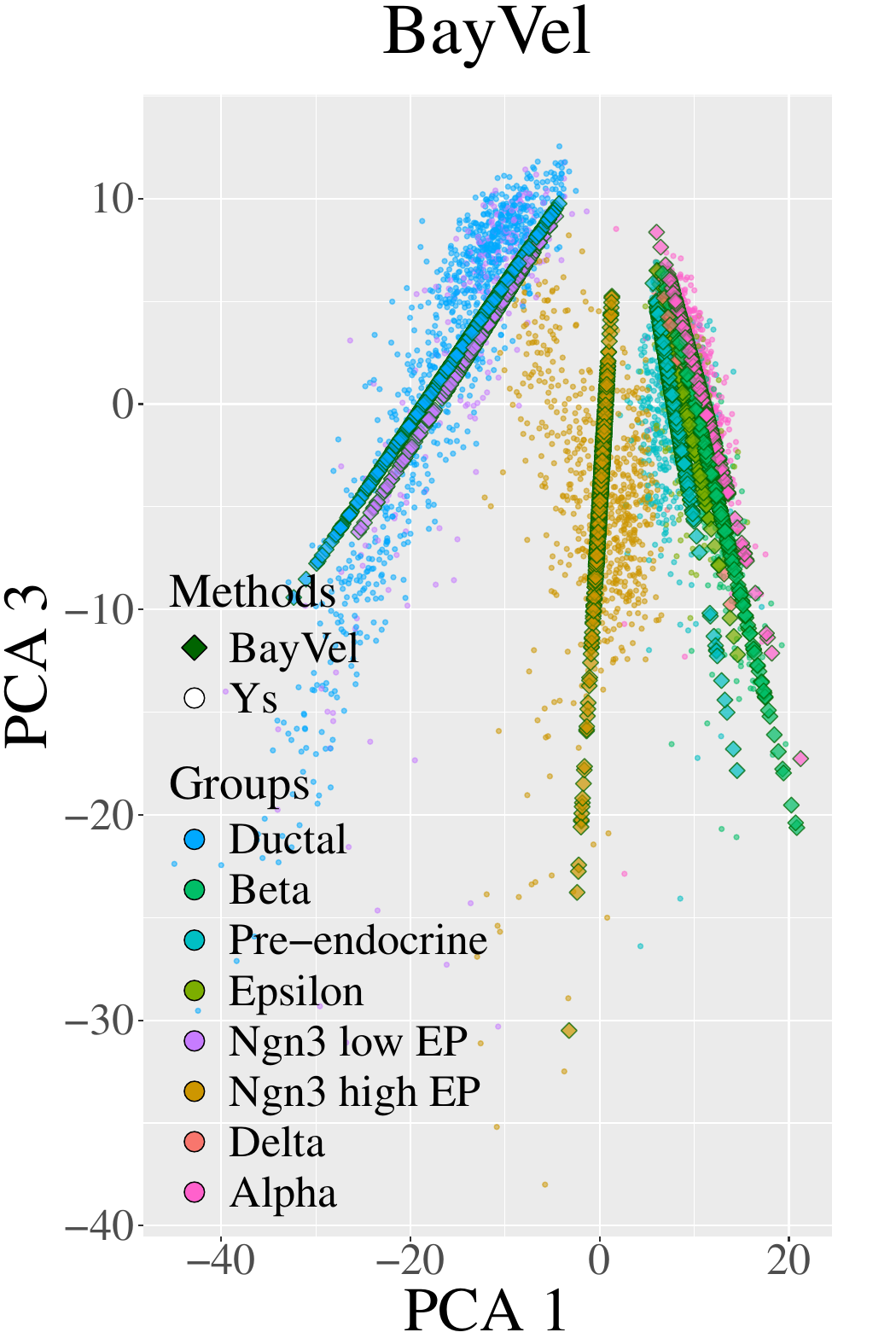}}
	\subfloat[\label{fig:posBayVelGS1-T3-D4-14}]{\includegraphics[scale = 0.25]{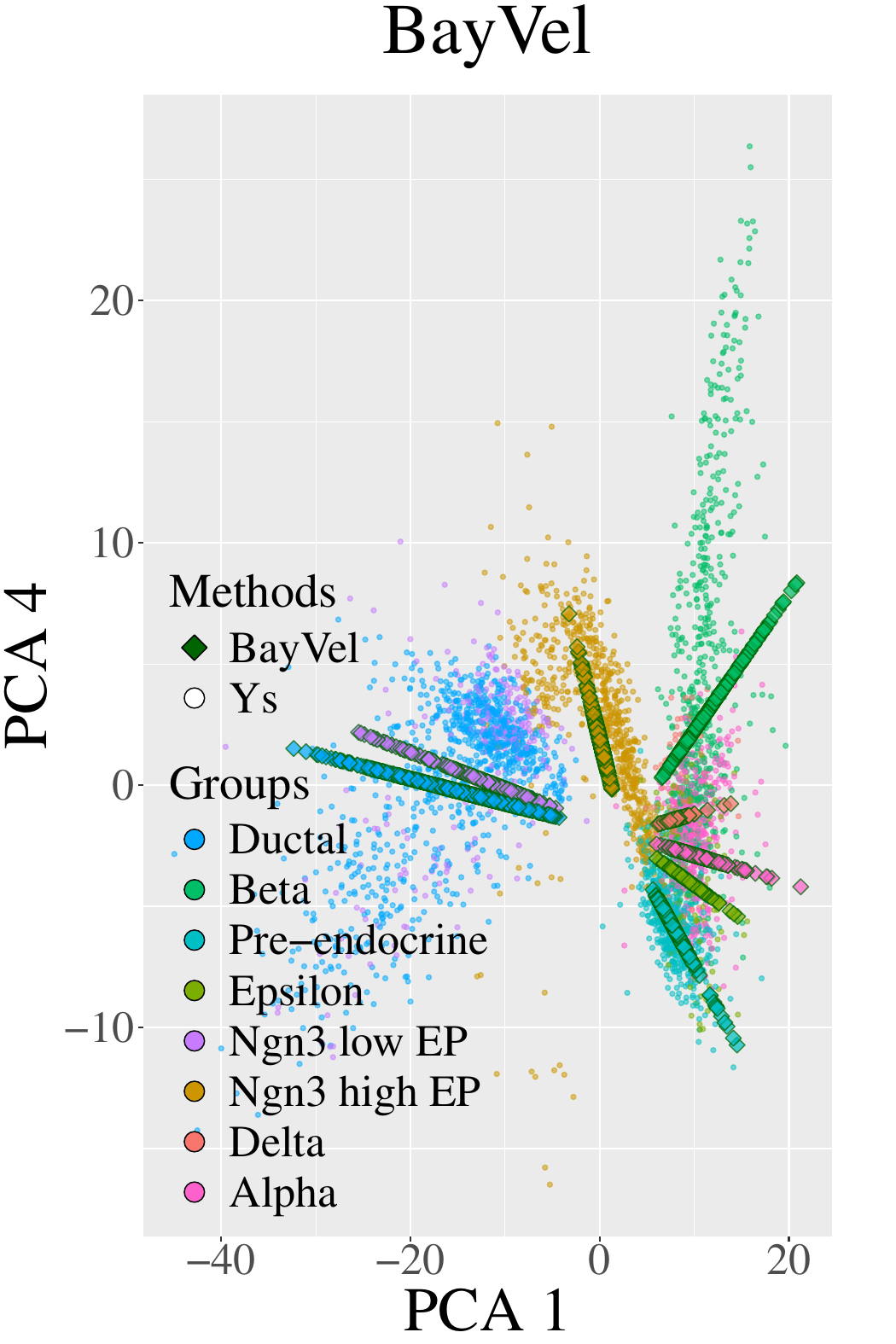}}\\
	\subfloat[\label{fig:posBayVelGS2-T3-D4-13}]{\includegraphics[scale = 0.25]{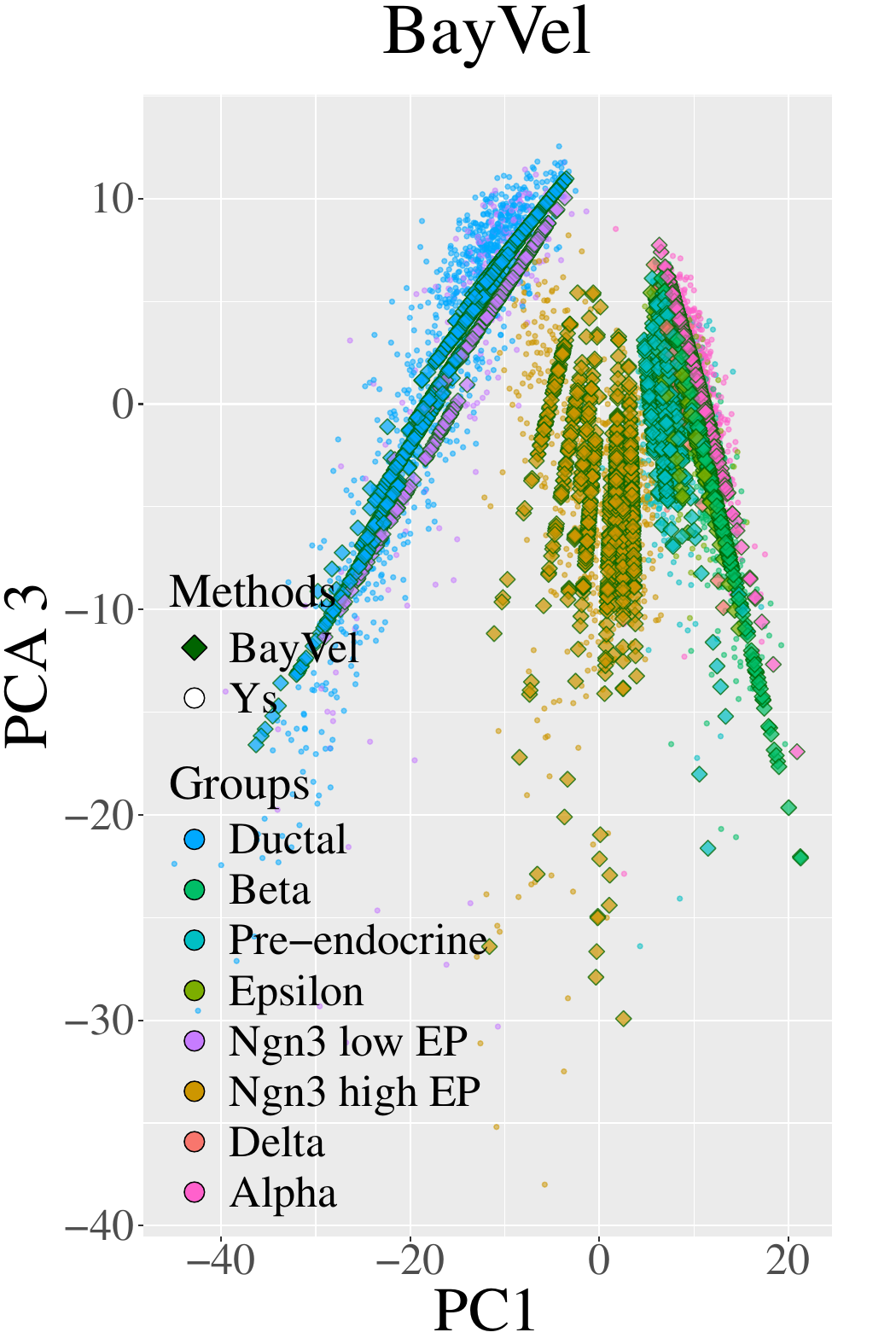}}
	\subfloat[\label{fig:posBayVelGS2-T3-D4-14}]{\includegraphics[scale = 0.25]{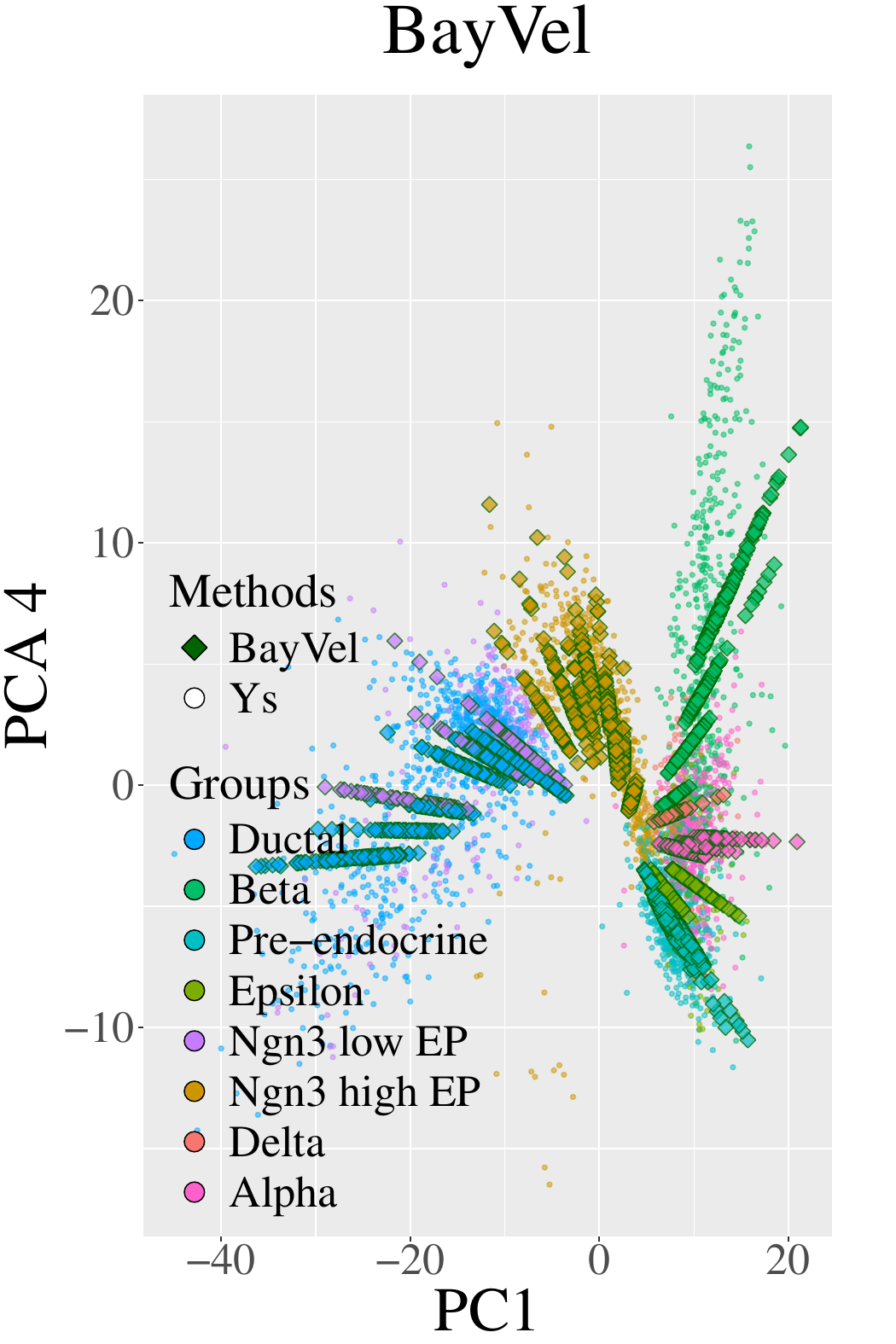}}
	\caption{PCA projection of the pancreatic cell data and the corresponding \textit{BayVel} estimates of spliced expected counts for model with $K = 1, R = 8$ (first row) and for the best WAIC-selected model with $K = 8, R = 38$ (second row). We first project the discrete spliced count, ${\boldsymbol{y}_{s, c} = (y_{s, c1}, \dots, y_{s, cG})}_{c=1, \dots, C}$, from the pancreatic dataset into a low-dimensional space through PCA. We plot the first PCA component against the third (first column) and the fourth one (second column). Small dots represent individual cells, with colors indicating different cell types. Then we project on the PCA embedding the MAP estimate of the expected mean of spliced count, $\lambda_c s^\sim_{krg}$, obtained with \textit{BayVel} (rectangles).}
	\label{fig:pcaPosBayVel}
\end{figure}

\clearpage

\subsection{Putative driver genes}

\begin{figure}[t]
	\centering 
	\subfloat[\label{fig:putativeScVelo}]{\includegraphics[scale = 0.6]{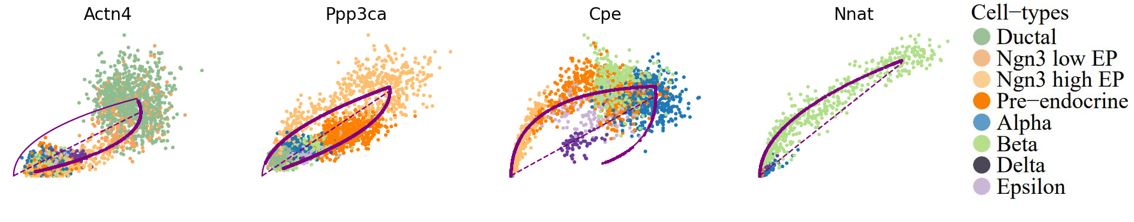}}
	\\
	\subfloat[\label{fig:putativeActn4BayVel}]{\includegraphics[scale = 0.15]{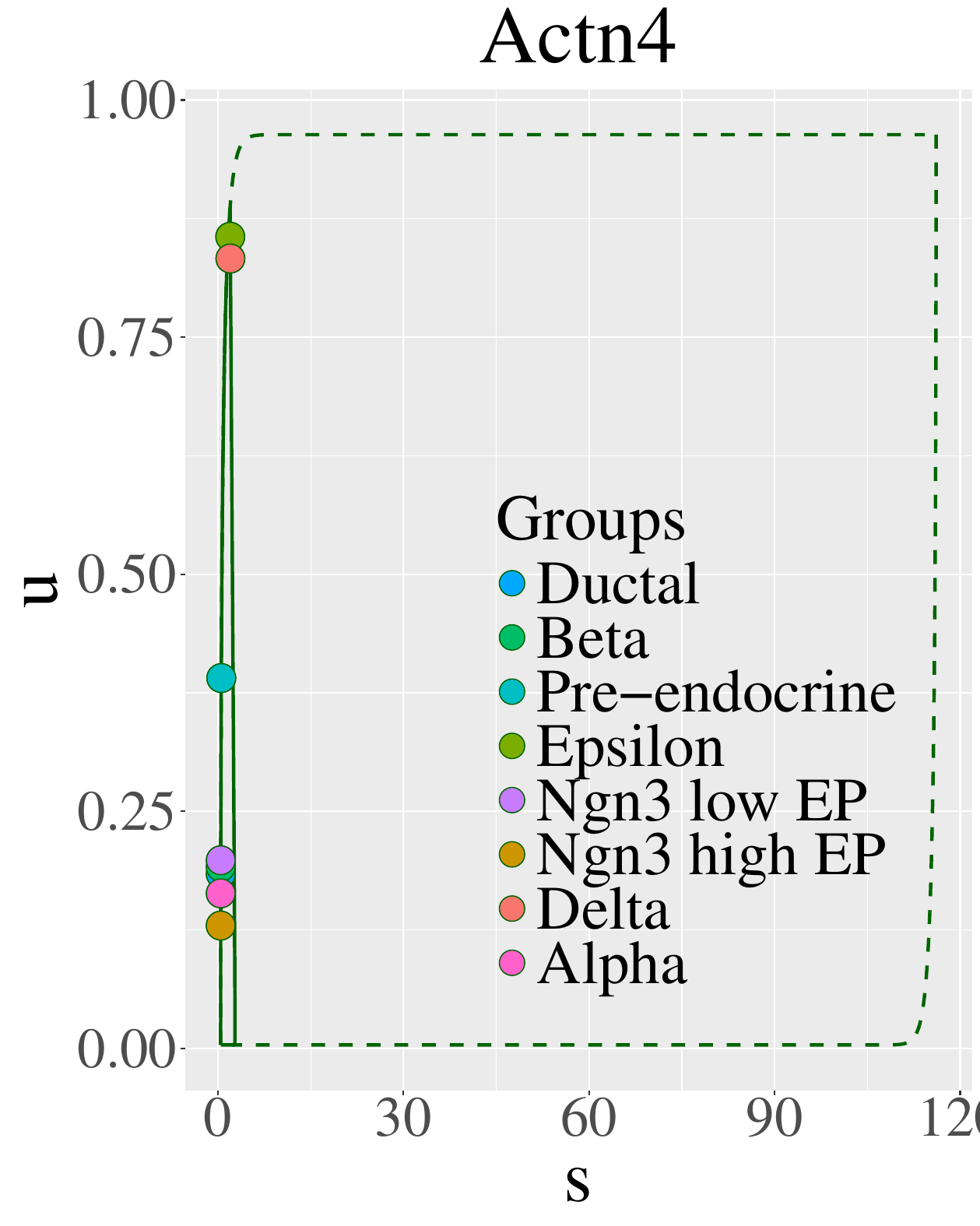}}
	\subfloat[\label{fig:putativePpp3caBayVel}]{\includegraphics[scale = 0.15]{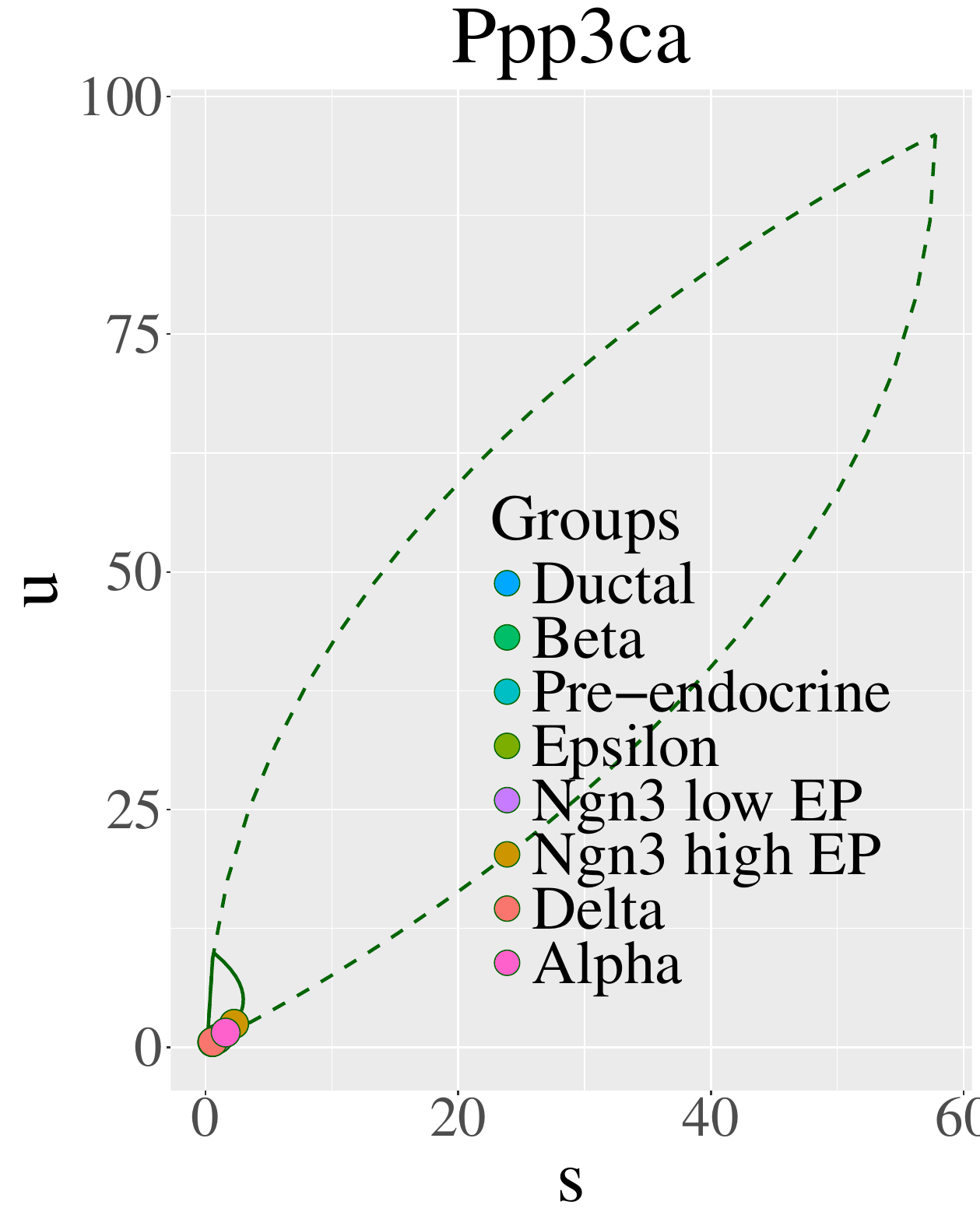}}
	\subfloat[\label{fig:putativeCpeBayVel}]{\includegraphics[scale = 0.15]{putativeCpe-GS1-T1-D4}}
	\subfloat[\label{fig:putativeNnatBayVel}]{\includegraphics[scale = 0.15]{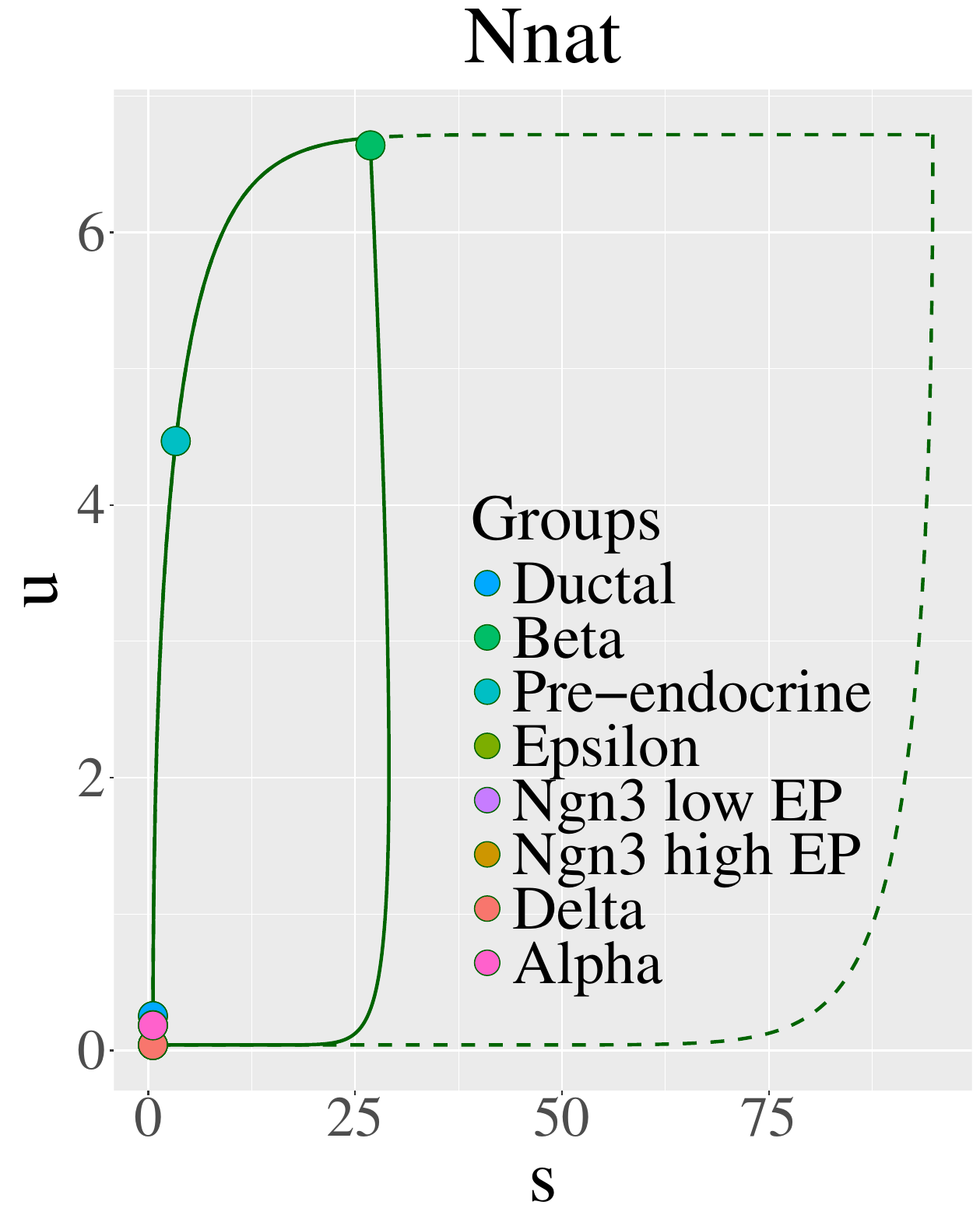}}
	\caption{Gene specific\emph{almonds} for \textit{putative driver genes} (Actn4, Ppp3ca, Cpe, Nnat) estimated by \textit{scVelo} (first row) and \textit{BayVel} (second row). The results of \textit{scVelo} are reproduced using the notebook at \cite{scvelo_notebooks}. The colored dots in this figure represents the observed pre-processed data for these genes, the gene dynamic represent \textit{scVelo} estimates. \textit{BayVel}'s \emph{almonds} are obtained using the MAP estimates of the parameters, with points representing the estimated subgroup specific positions.}
	\label{fig:putative}
\end{figure}

\begin{table}[t]
	\footnotesize
	\centering
	\begin{minipage}{.4\textwidth}
		\begin{tabular}{c|rrrr|}
			\hline
			& \multicolumn{4}{c}{$y_{s, cg}$}\\
			\hline
			& Actn4 & Ppp3ca &  Cpe & Nnat  \\
			\hline
			Ductal & 2.143 & 0.628 & 0.086 & 0.189 \\
			Ngn3 low EP & 1.966 & 0.519 & 0.084 & 0.176 \\
			Ngn3 high EP & 0.508 & 2.012 & 0.581 & 0.575 \\
			Pre-Endocrine & 0.372 & 1.620 & 7.123 & 0.583 \\
			Alpha & 0.422 & 0.836 & 13.229 & 1.605 \\
			Beta & 0.479 & 0.944 & 11.178 & 34.291 \\
			Delta & 0.800 & 0.714 & 5.729 & 3.014 \\ 
			Epsilon & 0.507 & 0.951 & 8.634 & 0.211 \\
			\hline
		\end{tabular}
	\end{minipage}
\hfill
	\begin{minipage}{.4\textwidth}
		\begin{tabular}{rrrr}
			\hline
			\multicolumn{4}{c}{$y_{u, cg}$}\\
			\hline
			Actn4 & Ppp3ca &  Cpe & Nnat\\
			\hline
			0.949 & 0.493 & 0.015 & 0.032 \\
			0.706 & 0.527 & 0.000 & 0.034 \\
			0.098 & 2.159 & 0.338 & 0.174 \\
			0.147 & 1.449 & 1.488 & 0.167 \\
			0.181 & 1.048 & 1.628 & 0.254 \\
			0.198 & 0.919 & 2.188 & 8.071 \\
			0.300 & 0.771 & 0.643 & 0.771 \\ 
			0.218 & 0.852 & 1.437 & 0.035 \\
			\hline
		\end{tabular}
	\end{minipage}
	\caption{Mean across different types for discrete counts $y_{s, cg}$ and $y_{u, cg}$ of the real pancreatic dataset for 4 specific putative driver genes.}
	\label{tab:meanYputative}
\end{table}
\normalsize
We compare here the \textit{putative driver genes} identified by \cite{bergen2020generalizing} through their high inferred likelihoods. For these genes, we compare the estimates of the \emph{almonds} estimated by\ \textit{BayVel} and \textit{scVelo}. The \textit{scVelo} results are obtained with the original notebook available at \cite{scvelo_notebooks} and are plotted in Figure \ref{fig:putativeScVelo}. The colored dots in this figure represents the observed pre-processed data for these genes. For some genes, such as \textit{Ppp3ca}, the \emph{almond} is difficult to discern from the data, as they appear more dispersed along a noisy linear trajectory. Moreover, further analysis of the data suggest \textit{Ppp3ca} does not exhibit substantial differences across cell types. Indeed, Table \ref{tab:meanYputative} presents the mean of spliced and unspliced counts for the analyzed genes across different cell types. Although a formal statistical test would be needed, the mean values for gene \textit{Ppp3ca} seem similar across types, with only a slight increase in \textit{Ngn3 high EP} cells. This suggests this gene may not play a major role in differentiation. In line with this observation, \textit{BayVel} estimates similar cellular positions across subgroups for \textit{Ppp3ca} (Figure \ref{fig:putativePpp3caBayVel}), consistent with the real data. In contrast, genes such as \textit{Cpe} and \textit{Nnat} show high expression levels in specific cell types, with pronounced differences in mean expression between groups. For example, \textit{Nnat} is more highly expressed in \textit{Beta} cells, and \textit{BayVel} accurately captures this behavior (Figure \ref{fig:putativeNnatBayVel}). The dynamic of \textit{Cpe} is more complex \ref{fig:putativeCpeBayVel}. While \textit{Alpha} and \textit{Beta} cells exhibit higher spliced expression levels, unspliced counts remain similar across cell types. 


\begin{thebibliography}{23}
	\providecommand{\natexlab}[1]{#1}
	\providecommand{\url}[1]{\texttt{#1}}
	\expandafter\ifx\csname urlstyle\endcsname\relax
	\providecommand{\doi}[1]{doi: #1}\else
	\providecommand{\doi}{doi: \begingroup \urlstyle{rm}\Url}\fi
	
\bibitem[Abdelaal et~al.(2024)]{abdelaal2024sirv}
T.~Abdelaal, L.~M. Grossouw, R.~J. Pasterkamp, B.~P. F. Lelieveldt, M.~J. T. Reinders, and A.~Mahfouz.
\newblock SIRV: Spatial inference of RNA velocity at the single-cell resolution.
\newblock \emph{NAR Genomics and Bioinformatics}, 6(3), 2024.


\bibitem[Aivazidis et~al.(2023)]{aivazidis2023cell2fate}
A.~Aivazidis, F.~Memi, V.~Kleshchevnikov, B.~Clarke, O.~Stegle, and O.~Ali Bayraktar.
\newblock Model-based inference of RNA velocity modules improves cell fate prediction.
\newblock \emph{BioRxiv}, 2023.
\newblock \doi{10.1101/2023.08.07.552680}.

\bibitem[Anderson et~al.(2010)]{ACK2010}
D.~F. Anderson, G.~Craciun, and T.~G. Kurtz.
\newblock Product-Form Stationary Distributions for Deficiency Zero Chemical Reaction Networks.
\newblock \emph{Bulletin of Mathematical Biology}, 72(8):\penalty0 1947--1970, 2010.
\newblock \doi{10.1007/s11538-010-9517-4}

\bibitem[Anderson and Kurtz(2015)]{anderson2015stochastic}
D.~F. Anderson and T.~G. Kurtz.
\newblock \emph{Stochastic analysis of biochemical systems}, volume 674.
\newblock Springer, 2015.

\bibitem[Andrieu and Thoms(2008)]{andrieu2008adaptiveMCMC}
C.~Andrieu and J.~Thoms.
\newblock A tutorial on adaptive {MCMC}.
\newblock \emph{Statistics and Computing}, 18:\penalty0 343--373, 2008.

\bibitem[Atta et~al.(2022)]{atta2022veloviz}
L.~Atta, A.~Sahoo, and J.~Fan.
\newblock VeloViz: RNA velocity-informed embeddings for visualizing cellular trajectories.
\newblock \emph{Bioinformatics}, 38(2):391--396, 2022.
\newblock \doi{10.1093/bioinformatics/btac771}.

\bibitem[Bastidas-Ponce et~al.(2019)]{bastidas2019edocrinogenesisDataset}
A.~Bastidas-Ponce, S.~Tritschler, L.~Dony, K.~Scheibner, M.~Tarquis-Medina, C.~Salinno, S.~Schirge, I.~Burtscher, A.~B{\"o}ttcher, F.~J. Theis, et~al.
\newblock Comprehensive single cell mRNA profiling reveals a detailed roadmap for pancreatic endocrinogenesis.
\newblock \emph{Development}, 146\penalty0 (12):\penalty0 dev173849, 2019.

\bibitem[Bergen et~al.(2020)]{bergen2020generalizing}
V.~Bergen, M.~Lange, S.~Peidli, F.~A. Wolf, and F.~J. Theis.
\newblock Generalizing RNA velocity to transient cell states through dynamical modeling.
\newblock \emph{Nature Biotechnology}, 38(12):\penalty0 1408--1414, 2020.

\bibitem[Bergen et~al.(2021)]{bergen2021rna}
V.~Bergen, R.~A. Soldatov, P.~V. Kharchenko, and F.~J. Theis.
\newblock RNA velocity—current challenges and future perspectives.
\newblock \emph{Molecular Systems Biology}, 17(8):\penalty0 e10282, 2021.


\bibitem[Bezanson et~al.(2017)]{bezanson2017julia}
J.~Bezanson, A.~Edelman, S.~Karpinski, and V.~B. Shah.
\newblock Julia: A fresh approach to numerical computing.
\newblock \emph{SIAM Review}, 59(1):\penalty0 65--98, 2017.

\bibitem[Burdziak et~al.(2023)]{burdziak2023sckinetics}
C.~Burdziak, C.~J. Zhao, D.~Haviv, D.~Alonso-Curbelo, S.~W. Lowe, and D.~Pe’er.
\newblock scKINETICS: inference of regulatory velocity with single-cell transcriptomics data.
\newblock \emph{Bioinformatics}, 39\penalty0 (Supplement\_1):\penalty0 i394--i403, 2023.


\bibitem[Cappelletti and Wiuf(2016)]{daniele_inv}
D.~Cappelletti and C.~Wiuf.
\newblock Product-Form Poisson-Like Distributions and Complex Balanced Reaction Systems.
\newblock \emph{SIAM Journal on Applied Mathematics}, 76(1):\penalty0 411--432, 2016.
\newblock \doi{10.1137/15M1029916}


\bibitem[Cui et~al.(2024)]{cui2024deepvelo}
H.~Cui, H.~Maan, M.~C. Vladoiu, J.~Zhang, M.~D. Taylor, and B.~Wang.
\newblock DeepVelo: deep learning extends RNA velocity to multi-lineage systems with cell-specific kinetics.
\newblock \emph{Genome Biology}, 25(1):27, 2024.

\bibitem[Gao et~al.(2022)]{gao2022unitvelo}
M.~Gao, C.~Qiao, and Y.~Huang.
\newblock UniTVelo: temporally unified RNA velocity reinforces single-cell trajectory inference.
\newblock \emph{Nature Communications}, 13(1):6586, 2022.
\newblock \doi{10.1038/s41467-022-34364-x}.

\bibitem[Gayoso et~al.(2024)]{gayoso2024veloVI}
A.~Gayoso, P.~Weiler, M.~Lotfollahi, D.~Klein, J.~Hong, A.~Streets, F.~J. Theis, and N.~Yosef.
\newblock Deep generative modeling of transcriptional dynamics for RNA velocity analysis in single cells.
\newblock \emph{Nature Methods}, 21\penalty0 (1):\penalty0 50--59, 2024.

\bibitem[Gorin et~al.(2022)]{gorin2022rna}
G.~Gorin, M.~Fang, T.~Chari, and L.~Pachter.
\newblock RNA velocity unraveled.
\newblock \emph{PLOS Computational Biology}, 18(9):\penalty0 e1010492, 2022.

\bibitem[Gorin et~al.(2020)]{gorin2020protaccel}
G.~Gorin, V.~Svensson, and L.~Pachter.
\newblock Protein velocity and acceleration from single-cell multiomics experiments.
\newblock \emph{Genome Biology}, 21(1):1--6, 2020.
\newblock \doi{10.1186/s13059-020-02074-2}.


\bibitem[Gu et~al.(2022)]{gu2022veloVAE}
Y.~Gu, D.~Blaauw, and J.~D. Welch.
\newblock Bayesian inference of RNA velocity from multi-lineage single-cell data.
\newblock \emph{BioRxiv}, 2022.
\newblock \doi{10.1101/2022.07.07.498592}.

\bibitem[Gupta et~al.(2022)]{gupta2022cytoPath}
R.~Gupta, D.~Cerletti, G.~Gut, A.~Oxenius, and M.~Claassen.
\newblock Simulation-based inference of differentiation trajectories from RNA velocity fields.
\newblock \emph{Cell Reports Methods}, 2(12):2022.
\newblock \doi{10.1016/j.crmeth.2022.100170}.

\bibitem[He et~al.(2022)]{he2022alevin}
D.~He, M.~Zakeri, H.~Sarkar, C.~Soneson, A.~Srivastava, and R.~Patro.
\newblock Alevin-fry unlocks rapid, accurate and memory-frugal quantification of single-cell RNA-seq data.
\newblock \emph{Nature Methods}, 19(3):\penalty0 316--322, 2022.

\bibitem[Jahnke and Huisinga(2007)]{Jahnke2007}
T.~Jahnke and W.~Huisinga.
\newblock Solving the chemical master equation for monomolecular reaction systems analytically.
\newblock \emph{Journal of Mathematical Biology}, 54(1):\penalty0 1--26, 2007.
\newblock \doi{10.1007/s00285-006-0034-x}

\bibitem[Jia and Chen(2023)]{jia2023velde}
J.~Jia and L.~Chen.
\newblock Velde: constructing cell potential landscapes by RNA velocity vector field decomposition.
\newblock \emph{arXiv preprint arXiv:2311.10403}, 2023.

\bibitem[Kouadri Boudjelthia et~al.(2023)]{kouadri2023neurovelo}
I.~Kouadri Boudjelthia, S.~Milite, N.~El Kazwini, J.~Fernandez-Mateos, N.~Valeri, Y.~Huang, A.~Sottoriva, and G.~Sanguinetti.
\newblock NeuroVelo: interpretable learning of cellular dynamics from single-cell transcriptomic data.
\newblock \emph{bioRxiv}, pages 2023--11, 2023.

\bibitem[La Manno et~al.(2018)]{laManno2018rna}
G.~La Manno, R.~Soldatov, A.~Zeisel, E.~Braun, H.~Hochgerner, V.~Petukhov, K.~Lidschreiber, M.~E. Kastriti, P.~Lönnerberg, A.~Furlan, et~al.
\newblock RNA velocity of single cells.
\newblock \emph{Nature}, 560(7719):\penalty0 494--498, 2018.

\bibitem[Lange et~al.(2022)]{lange2022cellrank}
M.~Lange, V.~Bergen, M.~Klein, M.~Setty, B.~Reuter, M.~Bakhti, H.~Lickert, M.~Ansari, J.~Schniering, H.~B. Schiller, et~al.
\newblock CellRank for directed single-cell fate mapping.
\newblock \emph{Nature Methods}, 19(2):159--170, 2022.

\bibitem[Lederer et~al.(2024)]{lederer2024veloCycle}
A.~R. Lederer, M.~Leonardi, L.~Talamanca, A.~Herrera, C.~Droin, I.~Khven, H.~J. F. Carvalho, A.~Valente, A.~Dominguez Mantes, P.~Mulet Arabí, et~al.
\newblock Statistical inference with a manifold-constrained RNA velocity model uncovers cell cycle speed modulations.
\newblock \emph{BioRxiv}, 2024--01, 2024.

\bibitem[Li et~al.(2023)]{li2023multiVelo}
C.~Li, M.~C. Virgilio, K.~L. Collins, and J.~D. Welch.
\newblock Multi-omic single-cell velocity models epigenome--transcriptome interactions and improves cell fate prediction.
\newblock \emph{Nature Biotechnology}, 41(3):387--398, 2023.
\newblock \doi{10.1038/s41587-022-01206-x}.

\bibitem[Li et~al.(2024)]{li2024tfvelo}
J.~Li, X.~Pan, Y.~Yuan, and H.-B. Shen.
\newblock TFvelo: gene regulation inspired RNA velocity estimation.
\newblock \emph{Nature Communications}, 15(1):1387, 2024.

\bibitem[Li(2023)]{li2023sctour}
Q.~Li.
\newblock scTour: a deep learning architecture for robust inference and accurate prediction of cellular dynamics.
\newblock \emph{Genome Biology}, 24(1):149, 2023.

\bibitem[Li et~al.(2024)]{li2024cellDancer}
S.~Li, P.~Zhang, W.~Chen, L.~Ye, K.~W. Brannan, N.~T. Le, J.~Abe, J.~P. Cooke, and G.~Wang.
\newblock A relay velocity model infers cell-dependent RNA velocity.
\newblock \emph{Nature Biotechnology}, 42(1):99--108, 2024.

\bibitem[Liu et~al.(2022)]{liu2022rnaOde}
R.~Liu, A.~O. Pisco, E.~Braun, S.~Linnarsson, and J.~Zou.
\newblock Dynamical systems model of RNA velocity improves inference of single-cell trajectory, pseudo-time and gene regulation.
\newblock \emph{Journal of Molecular Biology}, 434(15):167606, 2022.
\newblock \doi{10.1016/j.jmb.2022.167606}.

\bibitem[Marot-Lassauzaie et~al.(2022)]{marot2022kvelo}
V.~Marot-Lassauzaie, B.~J. Bouman, F.~D. Donaghy, Y.~Demerdash, M.~A. G. Essers, and L.~Haghverdi.
\newblock Towards reliable quantification of cell state velocities.
\newblock \emph{PLoS Computational Biology}, 18(9):e1010031, 2022.


\bibitem[McInnes et~al.(2018)]{mcinnes2018umap}
L.~McInnes, J.~Healy, and J.~Melville.
\newblock UMAP: Uniform manifold approximation and projection for dimension reduction.
\newblock \emph{arXiv preprint arXiv:1802.03426}, 2018.

\bibitem[Melsted et~al.(2021)]{melsted2021modular}
P.~Melsted, A.~S. Booeshaghi, L.~Liu, F.~Gao, L.~Lu, K.~H. Min, E.~da Veiga Beltrame, K.~E. Hj{\"o}rleifsson, J.~Gehring, and L.~Pachter.
\newblock Modular, efficient and constant-memory single-cell RNA-seq preprocessing.
\newblock \emph{Nature Biotechnology}, 39(7):\penalty0 813--818, 2021.

\bibitem[Peng et~al.(2023)]{peng2023storm}
Q.~Peng, X.~Qiu, and T.~Li.
\newblock Storm: incorporating transient stochastic dynamics to infer the RNA velocity with metabolic labeling information.
\newblock \emph{BioRxiv}, 2023.
\newblock \doi{10.1101/2023.06.14.543892}.

\bibitem[Putri et~al.(2022)]{putri2022analysing}
G.~H.~Putri, S.~Anders, P.~T.~Pyl, J.~E.~Pimanda, and F.~Zanini.
\newblock Analysing high-throughput sequencing data in Python with HTSeq 2.0.
\newblock \emph{Bioinformatics}, 38(10):2943--2945, 2022.


\bibitem[Qiao and Huang(2021)]{qiao2021veloAE}
C.~Qiao and Y.~Huang.
\newblock Representation learning of RNA velocity reveals robust cell transitions.
\newblock \emph{Proceedings of the National Academy of Sciences}, 118(49):e2105859118, 2021.
\newblock \doi{10.1073/pnas.2105859118}.

\bibitem[Qin et~al.(2022)]{qin2022pyro}
Q.~Qin, E.~Bingham, G.~La Manno, D.~M. Langenau, and L.~Pinello.
\newblock Pyro-Velocity: Probabilistic RNA Velocity inference from single-cell data.
\newblock \emph{BioRxiv}, 2022.
\newblock \doi{10.1101/2022.09.12.506498}.

\bibitem[Qiu et~al.(2020)]{qiu2020massively}
Q.~Qiu, P.~Hu, X.~Qiu, K.~W. Govek, P.~G. C{\'a}mara, and H.~Wu.
\newblock Massively parallel and time-resolved RNA sequencing in single cells with scNT-seq.
\newblock \emph{Nature Methods}, 17(10):991--1001, 2020.
\newblock \doi{10.1038/s41592-020-00992-3}.

\bibitem[Qiu et~al.(2022)]{qiu2022dynamo}
X.~Qiu, Y.~Zhang, J.~D. Martin-Rufino, C.~Weng, S.~Hosseinzadeh, D.~Yang, A.~N. Pogson, M.~Y. Hein, K.~H. J. Min, L.~Wang, and others.
\newblock Mapping transcriptomic vector fields of single cells.
\newblock \emph{Cell}, 185(4):690--711, 2022.


\bibitem[Riba et~al.(2022)]{riba2022deepCycle}
A.~Riba, A.~Oravecz, M.~Durik, S.~Jim{\'e}nez, V.~Alunni, M.~Cerciat, M.~Jung, C.~Keime, W.~M. Keyes, and N.~Molina.
\newblock Cell cycle gene regulation dynamics revealed by RNA velocity and deep-learning.
\newblock \emph{Nature Communications}, 13(1):2865, 2022.
\newblock \doi{10.1038/s41467-022-30465-3}.

\bibitem[Robert and Casella(2009)]{Robert}
C.~P. Robert and G.~Casella.
\newblock \emph{Introducing Monte Carlo Methods with R}.
\newblock Springer-Verlag, Berlin, Heidelberg, 1st edition, 2009.
\newblock ISBN 1441915753.

\bibitem[Sabbioni et~al.(2023)]{sisNonIdentifiability}
E.~Sabbioni, E.~Bibbona, G.~Mastrantonio, and G.~Sanguinetti.
\newblock Analyzing RNA data with scVelo: identifiability issues and a Bayesian implementation.
\newblock 2023.

\bibitem[Sabbioni et~al.(2025)]{supplementary}
E.~Sabbioni, E.~Bibbona, G.~Mastrantonio, and G.~Sanguinetti.
\newblock Supplement to "BayVel: A Bayesian Framework for RNA velocity estimation in single-cell transcriptomics".
\newblock 2025.

\bibitem[Shah et~al.(2024)]{shah2024systematic}
N.~Shah, Q.~Meng, Z.~Zou, and X.~Zhang.
\newblock Systematic analysis on the horse-shoe-like effect in PCA plots of scRNA-seq data.
\newblock \emph{Bioinformatics Advances}, 4(1):\penalty0 vbae109, 2024.


\bibitem[Singh et~al.(2024)]{singh2024velorama}
R.~Singh, A.~P. Wu, A.~Mudide, and B.~Berger.
\newblock Causal gene regulatory analysis with RNA velocity reveals an interplay between slow and fast transcription factors.
\newblock \emph{Cell Systems}, 15(5):462--474, 2024.
\newblock \doi{10.1016/j.cels.2023.08.007}.


\bibitem[Soneson et~al.(2021)]{soneson2021preprocessing}
C.~Soneson, A.~Srivastava, R.~Patro, and M.~B. Stadler.
\newblock Preprocessing choices affect RNA velocity results for droplet scRNA-seq data.
\newblock \emph{PLoS Computational Biology}, 17(1):e1008585, 2021.
\newblock \doi{10.1371/journal.pcbi.1008585}.

\bibitem[Su et~al.(2024)]{su2024hodge}
Z.~Su, Y.~Tong, and G.-W. Wei.
\newblock Hodge decomposition of single-cell RNA velocity.
\newblock \emph{Journal of chemical information and modeling}, 64(8):3558--3568, 2024.

\bibitem[Tang et~al.(2023)]{tang2023captureEfficiency}
W.~Tang, A.~C. S{\o}lvsten J{\o}rgensen, S.~Marguerat, P.~Thomas, and V.~Shahrezaei.
\newblock Modelling capture efficiency of single-cell RNA-sequencing data improves inference of transcriptome-wide burst kinetics.
\newblock \emph{Bioinformatics}, 39(7):btad395, 2023.


\bibitem[Tedesco et~al.(2022)]{tedesco2022chromatinVelocity}
M.~Tedesco, F.~Giannese, D.~Lazarevi{\'c}, V.~Giansanti, D.~Rosano, S.~Monzani, I.~Catalano, E.~Grassi, E.~Zanella, O.~A. Botrugno, et~al.
\newblock Chromatin Velocity reveals epigenetic dynamics by single-cell profiling of heterochromatin and euchromatin.
\newblock \emph{Nature Biotechnology}, 40(2):235--244, 2022.
\newblock \doi{10.1038/s41587-021-00471-5}.

\bibitem[TheisLab(2025)]{scvelo_notebooks}
TheisLab.
\newblock scVelo Notebooks - Pancreas, 2025.
\newblock \url{https://github.com/theislab/scvelo_notebooks/blob/master/Pancreas.ipynb}.
\newblock Accessed: 2025-03-27.

\bibitem[Van der Maaten and Hinton(2008)]{van2008tsne}
L.~Van der Maaten and G.~Hinton.
\newblock Visualizing data using t-SNE.
\newblock \emph{Journal of Machine Learning Research}, 9(11), 2008.


\bibitem[Wang and Zheng(2021)]{VeloPredictor}
X.~Wang and J.~Zheng.
\newblock Velo-Predictor: an ensemble learning pipeline for RNA velocity prediction.
\newblock \emph{BMC bioinformatics}, 22:1--14, 2021.
\newblock \doi{10.1186/s12859-021-04209-4}.

\bibitem[Watanabe and Opper(2010)]{watanabe2010asymptotic}
S.~Watanabe and M.~Opper.
\newblock Asymptotic equivalence of Bayes cross validation and widely applicable information criterion in singular learning theory.
\newblock \emph{Journal of Machine Learning Research}, 11(12), 2010.

\bibitem[Weng et~al.(2021)]{weng2021vetra}
G.~Weng, J.~Kim, and K.~J. Won.
\newblock VeTra: a tool for trajectory inference based on RNA velocity.
\newblock \emph{Bioinformatics}, 37(20):3509--3513, 2021.
\newblock \doi{10.1093/bioinformatics/btab377}.



\bibitem[Xie et~al.(2024)]{xie2024symVelo}
C.~Xie, Y.~Yang, H.~Yu, Q.~He, M.~Yuan, B.~Dong, L.~Zhang, and M.~Yang.
\newblock RNA velocity prediction via neural ordinary differential equation.
\newblock \emph{Iscience}, 27(4), 2024.


\bibitem[Zeisel et~al.(2011)]{zeisel2011coupled}
A.~Zeisel, W.~J. K{\"o}stler, N.~Molotski, J.~M. Tsai, R.~Krauthgamer, J.~Jacob-Hirsch, G.~Rechavi, Y.~Soen, S.~Jung, Y.~Yarden, et~al.
\newblock Coupled pre-mRNA and mRNA dynamics unveil operational strategies underlying transcriptional responses to stimuli.
\newblock \emph{Molecular Systems Biology}, 7(1):529, 2011.
\newblock \doi{10.1038/msb.2011.53}.

\bibitem[Zhang et~al.(2024)]{zhang2024consensusVelo}
H.~Zhang, N.~Bochkina, and S.~Wade.
\newblock Quantifying uncertainty in RNA velocity.
\newblock \emph{bioRxiv}, pages 2024--05, 2024.

\bibitem[Zhang and Zhang(2021)]{CellPath}
Z.~Zhang and X.~Zhang.
\newblock Inference of high-resolution trajectories in single-cell RNA-seq data by using RNA velocity.
\newblock \emph{Cell Reports Methods}, 1(6):100095, 2021.
\newblock \doi{https://doi.org/10.1016/j.crmeth.2021.100095}.

\bibitem[Zhang and Zhang(2021)]{zhang2021velosim}
Z.~Zhang and X.~Zhang.
\newblock VeloSim: Simulating single cell gene-expression and RNA velocity.
\newblock \emph{BioRxiv}, 2021.
\newblock \doi{10.1101/2021.01.18.427649}.

\bibitem[Zheng et~al.(2023)]{zheng2023pumping}
S.~C. Zheng, G.~Stein-O’Brien, L.~Boukas, L.~A. Goff, and K.~D. Hansen.
\newblock Pumping the brakes on RNA velocity by understanding and interpreting RNA velocity estimates.
\newblock \emph{Genome Biology}, 24(1):\penalty0 246, 2023.

\end{thebibliography}
\end{document}